\documentclass[reprint, preprintnumbers, amsmath, amssymb, superscriptaddress, nofootinbib, aps, pra, floatfix]{revtex4-2}

\usepackage[english]{babel}
\usepackage[T1]{fontenc}
\usepackage[utf8]{inputenc}

\usepackage{amsmath}
\usepackage{amsfonts}
\usepackage{amssymb}
\usepackage[
]{graphicx}
\usepackage[colorinlistoftodos]{todonotes}
\usepackage[utf8]{inputenc}
\usepackage{bbold}
\usepackage{cancel}
\usepackage{comment}
\usepackage{mathtools}
\usepackage{hyperref}
\usepackage{physics}
\usepackage{hyperref}

\newcounter{protocol}
\newenvironment{protocol}[1]{\refstepcounter{protocol} \textbf{Protocol~\theprotocol: #1} \begin{enumerate}}{\end{enumerate}}

\begin{document}

\title{Analysis of Multipartite Entanglement Distribution using a Central Quantum-Network Node}
\date{\today}

\author{Guus Avis}
\email{guusavis@hotmail.com}
\affiliation{QuTech, Delft University of Technology, Lorentzweg 1, 2628 CJ Delft, The Netherlands}
\affiliation{Kavli Institute of Nanoscience, Delft University of Technology, Lorentzweg 1, 2628 CJ Delft, The Netherlands}
\author{Filip Rozp\k{e}dek}
\affiliation{QuTech, Delft University of Technology, Lorentzweg 1, 2628 CJ Delft, The Netherlands}
\affiliation{Kavli Institute of Nanoscience, Delft University of Technology, Lorentzweg 1, 2628 CJ Delft, The Netherlands}
\affiliation{Pritzker School of Molecular Engineering, University of Chicago, Chicago, IL 60637, USA}
\author{Stephanie Wehner}
\email{s.d.c.wehner@tudelft.nl}
\affiliation{QuTech, Delft University of Technology, Lorentzweg 1, 2628 CJ Delft, The Netherlands}
\affiliation{Kavli Institute of Nanoscience, Delft University of Technology, Lorentzweg 1, 2628 CJ Delft, The Netherlands}

\begin{abstract}

We study the performance (rate and fidelity) of distributing multipartite entangled states in a quantum network through the use of a central node.
Specifically, we consider the scenario where the multipartite entangled state is first prepared locally at a central node,
and then transmitted to the end nodes of the network through quantum teleportation.
As our first result, we present leading-order analytical expressions and lower bounds for both the rate and fidelity at which a specific class of multipartite entangled states,
namely Greenberger-Horne-Zeilinger (GHZ) states,
are distributed.
Our analytical expressions for the fidelity accurately account for time-dependent depolarizing noise encountered by individual quantum bits while stored in quantum memory,
as verified using Monte Carlo simulations.
As our second result,
we compare the performance to the case where the central node is an entanglement switch and the GHZ state is created by the end nodes in a distributed fashion.
Apart from these two results,
we outline how the teleportation-based scheme could be physically implemented using trapped ions or nitrogen-vacancy centers in diamond.

\end{abstract}

\maketitle

\section{Introduction} \label{sec:introduction}

A quantum network is capable of distributing entangled quantum states between end nodes
that are possibly separated by large distances \cite{castelvecchiQuantumInternetHas2018, wehnerQuantumInternetVision2018a, kimbleQuantumInternet2008, caleffiRiseQuantumInternet2020}.
The development of quantum networks is an active field of research,
with recent milestones including
the distribution of entanglement over 1203 kilometers using a satellite \cite{yinSatellitebasedEntanglementDistribution2017},
quantum teleportation without using a preshared entangled state \cite{langenfeldQuantumTeleportationRemote2021},
the generation of light-matter entanglement over 50 kilometers of optical fiber through the use of quantum frequency conversion \cite{krutyanskiyLightmatterEntanglement502019},
and the creation of the first three-node quantum network \cite{pompiliRealizationMultinodeQuantum2021}.
\\

Much research focuses on the distribution of bipartite entangled states, or Bell states, which are shared only between two nodes.
Bell states allow for many interesting applications, such as quantum key distribution \cite{ekertQuantumCryptographyBased1991, bennettQuantumCryptographyUsing1992, bennettQuantumCryptographyBell1992, pirandolaAdvancesQuantumCryptography2020} and blind quantum computation \cite{feigenbaumEncryptingProblemInstances1986, fitzsimonsUnconditionallyVerifiableBlind2017, leichtleVerifyingBQPComputations2021}.
Some quantum-network applications, however, require the distribution of multipartite entangled states.
One class of multipartite entangled states is formed by graph states.
Graph states are states that can be represented using mathematical graphs, with each node corresponding to a qubit, and each edge corresponding to an entangling operation \cite{heinEntanglementGraphStates2006}.
An example of a state that is equivalent to a graph state up to single-qubit operations is the Greenberger-Horne-Zeilinger (GHZ) state \cite{greenbergerGoingBellTheorem1989},
which is equivalent to graph states both corresponding to the complete graph and the star graph.
Distributed GHZ states can be used for, among others, conference-key agreement \cite{murtaQuantumConferenceKey2020, grasselliRobustAnonymousConference2021, hahnAnonymousQuantumConference2020, thalackerAnonymousSecretCommunication2021}, distributed quantum computing \cite{groverQuantumTelecomputation1997, ciracDistributedQuantumComputation1999}, secret sharing \cite{qinDynamicQuantumSecret2017}, clock synchronization \cite{komarQuantumNetworkClocks2014}, and two-dimensional quantum-repeater schemes \cite{wallnoferTwodimensionalQuantumRepeaters2016}.
A multipartite state that is not equivalent to a graph state is the W state \cite{durThreeQubitsCan2000},
which can be used for e.g. anonymous transmission \cite{lipinskaAnonymousTransmissionNoisy2018}.\\

Various investigations have been performed into how specific multipartite entangled states can best be distributed in a quantum network \cite{
meterRecursiveQuantumRepeater2011a, wallnoferTwodimensionalQuantumRepeaters2016, pirkerModularArchitecturesQuantum2018, capraravivoliHighfidelityGreenbergerHorneZeilingerState2019, pirkerQuantumNetworkStack2019, % network design
benjaminBrokeredGraphstateQuantum2006, campbellAdaptiveStrategiesGraphstate2007, % growing graph states
kruszynskaQuantumCommunicationCost2006, deboneProtocolsCreatingDistilling2020, % distillation
coopmansImprovedAnalyticalBounds2022,coopmansNetSquidNETworkSimulator2021,nainAnalysisMultipartiteEntanglement2020,nainAnalysisTripartiteEntanglement2021,vardoyanCapacityRegionBipartite2021,vardoyanStochasticAnalysisQuantum2021, % switch
cuquetGrowthGraphStates2012, eppingLargescaleQuantumNetworks2016, dahlbergTransformingGraphStates2018a, yamasakiMultipartiteEntanglementOutperforming2018, meignantDistributingGraphStates2019, bugalhoDistributingMultipartiteEntanglement2021, fischerDistributingGraphStates2021}. % protocols
A recurring theme that can be discerned in prior work is the use of a central node that establishes bipartite entanglement with a number of end nodes,
and then executes local operations to transform the bipartite states into a single multipartite entangled state between those end nodes \cite{kruszynskaQuantumCommunicationCost2006, cuquetGrowthGraphStates2012, meignantDistributingGraphStates2019, fischerDistributingGraphStates2021, pirkerModularArchitecturesQuantum2018, bugalhoDistributingMultipartiteEntanglement2021, vardoyanStochasticAnalysisQuantum2021, nainAnalysisMultipartiteEntanglement2020}.
Notably, such a scheme is a key ingredient for different efficient protocols and network architectures for distributing multipartite entanglement \cite{cuquetGrowthGraphStates2012, meignantDistributingGraphStates2019, fischerDistributingGraphStates2021, pirkerModularArchitecturesQuantum2018, bugalhoDistributingMultipartiteEntanglement2021}.\\

\begin{figure}[h]
	\includegraphics[width=0.8\linewidth]{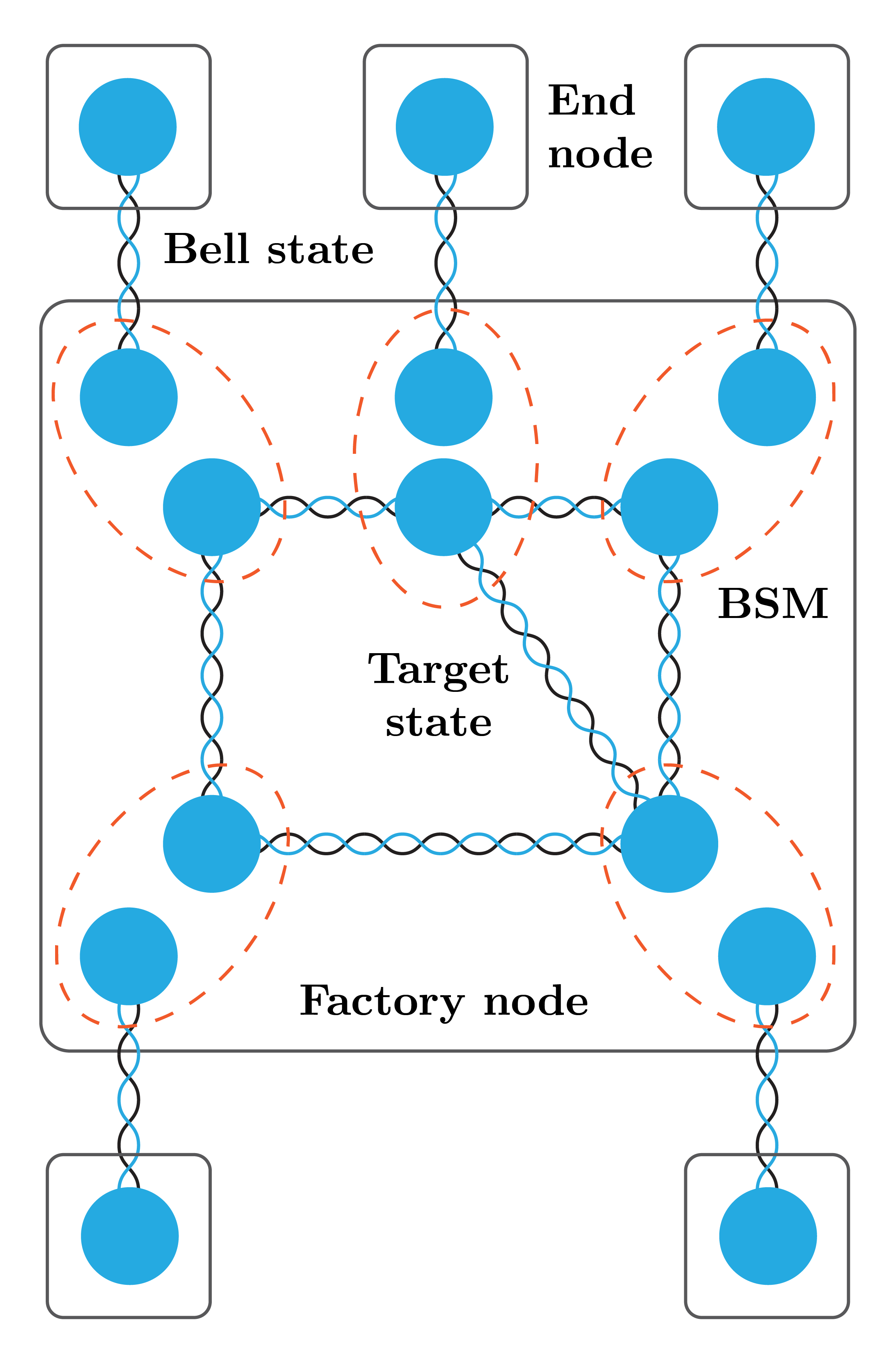}
	\caption[
		Distribution of some entangled target state between end nodes through a factory node.
	]
	{
        A factory node can be used to distribute some multipartite entangled target state (for example, a graph state) between a set of end nodes.
        This is done by preparing the target state locally at the factory node and teleporting it.
        Quantum teleportation of the target state is realized using Bell states shared between the factory node and the end nodes and Bell-state measurements (BSMs).
	}
	\label{fig:intro:general_factory}
\end{figure}

In this paper,
we consider the case where a multipartite entangled state is distributed in a quantum network by first creating the target state locally at the central node,
and then transmitting the qubits of the state to the end nodes through quantum teleportation using preshared Bell states \cite{bennettTeleportingUnknownQuantum1993}.
Teleportation is realized by executing a Bell-state measurement (BSM) on the to-be teleported qubit and a qubit in a Bell state.
Here, we refer to a node capable of creating and teleporting multipartite entangled states as a \textit{factory node}.
The function of a factory node is illustrated in Figure \ref{fig:intro:general_factory}.\\

Understanding the performance of factory nodes in the presence of hardware imperfections allows for the assessment of the different proposed protocols and network architectures that incorporate such central nodes.
Metrics that quantify the performance of multipartite entanglement distribution are the rate at which states can be distributed, and the fidelity of distributed states to the target state.
Developing a good understanding of the rate and fidelity is of special relevance to the work done in \cite{bugalhoDistributingMultipartiteEntanglement2021}.
Here, the authors present a protocol to decide which node in a larger network to select as the central node for the distribution of GHZ states.
This protocol relies on an analytical model of the rate and fidelity with which the states can be distributed for different possible placements of the central node.
We contribute to understanding the rate and fidelity in Section \ref{sec:estimates}.
Furthermore, we remark that it is not only of interest to quantify the performance of factory nodes in an absolute sense.
It is also of interest to understand how the performance of factory nodes compares to other schemes that also allow for distributing multipartite entangled states,
such that statements about their relative performance can be made.
We contribute to this by considering different types of central nodes in Sections \ref{subsec:different_central_nodes} and \ref{sec:comparison}.
\\

In this work, we specifically study the use of factory nodes to distribute GHZ states in a symmetric star-shaped network.
In such a network, depicted in Figure \ref{fig:intro:star_network}, a central node is connected to $N$ end nodes through, in total, $N$ identical quantum connections.
These quantum connections can be used to distribute Bell states.
We will model the distribution of Bell states using quantum connections as a series of attempts of constant duration and success probability.
When such an attempt is successful, the series terminates and a Bell state is created.
When a quantum connection creates a Bell state, it is shared between the central node and the corresponding end node,
and can be stored in quantum memory.
These Bell states can be used as a resource to create multipartite entangled states shared by the end nodes.
\\

\subsection{Summary of results}

In this paper, we present two main results.
As our first result,
in Section \ref{sec:estimates},
we provide analytical leading-order expressions and lower bounds for both the rate and fidelity of GHZ-state distribution in a symmetric star-shaped network using a factory node,
and additionally an exact expression for the rate.
The leading-order expressions become exact in the limit when
the success probability of a single attempt at Bell-state distribution using a quantum connection is small,
and the probability of losing a qubit due to memory decoherence during the time span of a single such attempt is small.
As our second result,
in Section \ref{sec:comparison},
we provide a comparison between the performance of GHZ-state distribution on a symmetric star-shaped network when the central node is a factory node,
and when the central node is instead a ``2-switch'' capable of performing BSMs to create Bell states shared between end nodes \cite{vardoyanCapacityRegionBipartite2021}.
A key advantage to the use of factory nodes is an increased resilience to noise in Bell-state distribution.
However, a disadvantage is reduced resilience to noise in BSMs.
Additionally, the factory node is typically outperformed by the 2-switch in terms of rate.
\\

\subsection{Comparison of analytical results to prior work} \label{subsec:prior_work}

Here, we compare the analytical results for the rate and fidelity that we present in Section \ref{sec:estimates} to existing results.
First, we note that we are aware of only one prior analytical result for the fidelity of distributed GHZ states in a similar scheme,
which is found in \cite{bugalhoDistributingMultipartiteEntanglement2021}.
However, the authors make the simplifying assumption that Bell states cannot be stored in quantum memory between attempts at Bell-state distribution.
Therefore, all connections need to be successful simultaneously.
When the success probability for distributing Bell states is small,
this is a very inefficient scheme.
In contrast, we assume entangled qubits are stored within the factory node until all Bell states are in place and the GHZ state can be teleported.
Here, we are able to accurately account for the time-dependent noise due to qubits being stored in noisy quantum memory for random periods of time.
Additionally, it is assumed in \cite{bugalhoDistributingMultipartiteEntanglement2021} that local operations are always noiseless,
which is not an assumption made in this paper.  \\

Second, we compare our results with the study of the ``entanglement switch''.
An entanglement switch, first defined in \cite{vardoyanCapacityRegionBipartite2021},
is a quantum-network node capable of generating and storing Bell states with $k$ end nodes,
and executing GHZ-state measurements on $n$ local qubits,
thereby creating GHZ states shared by $n$ out of $k$ end nodes.
From this perspective, a factory node that distributes GHZ states, as studied in this paper, can be described as an $n=k$ entanglement switch.
An entanglement switch for which $n = 2$ is referred to as ``2-switch'' throughout this paper.
\\

In \cite{vardoyanCapacityRegionBipartite2021, vardoyanExactAnalysisIdealized2020, vardoyanStochasticAnalysisQuantum2021, nainAnalysisMultipartiteEntanglement2020, nainAnalysisTripartiteEntanglement2021},
the entanglement switch is studied analytically using Markov-chain techniques.
In \cite{vardoyanStochasticAnalysisQuantum2021},
it is discussed that a minimum fidelity can be guaranteed by incorporating a cutoff time after which qubits are discarded from memory in the protocol,
and the effects of the cutoff time on the rate are studied for $n=2$.
However, there are no expressions for the actual fidelity (with or without cutoff time),
and in case there is no cutoff time there is also no lower bound.
Additionally, none but \cite{nainAnalysisMultipartiteEntanglement2020} consider the case $n>3$,
where the only result that is presented for $n=k$ is that no steady-state solution exists in case the switch is able to store an infinite number of entangled qubits.
This is in contrast to the present paper, where we present analytical results for the fidelity in the absence of a cutoff time,
the parameter $n$ can take any value, and we assume there is only one qubit of buffer memory available per end node.
Our results are limited to $n=k$, but we discuss in Section \ref{sec:conclusion} how the results can be extended to $n < k$.
\\

A paper that does derive results for an entanglement switch of general $n=k$ with only a single qubit of buffer memory is \cite{coopmansImprovedAnalyticalBounds2022}.
The authors provide analytical tools for understanding and bounding the rate,
but do not consider the fidelity.
Finally, numerical results for the fidelity obtained from Monte Carlo simulations can be found in \cite{coopmansNetSquidNETworkSimulator2021}.
While Monte Carlo simulations can be used to study a larger range of setups than our analytical results
(e.g., they can be used to study asymmetric star-shaped networks),
they may need to be evaluated many times in order to obtain results with small error bars.
Doing so can be computationally expensive.
This is especially the case when there is a large number of end nodes,
as quantum states in the system will be large and therefore hard to simulate.
On the other hand, our analytical results are computationally cheap to evaluate and have no error bars.
Furthermore, analytical results are often more suited to understand how a quantity scales and gain intuition.
\\

\subsection{Different central nodes} \label{subsec:different_central_nodes}

\begin{figure}
	\includegraphics[width=0.8\linewidth]{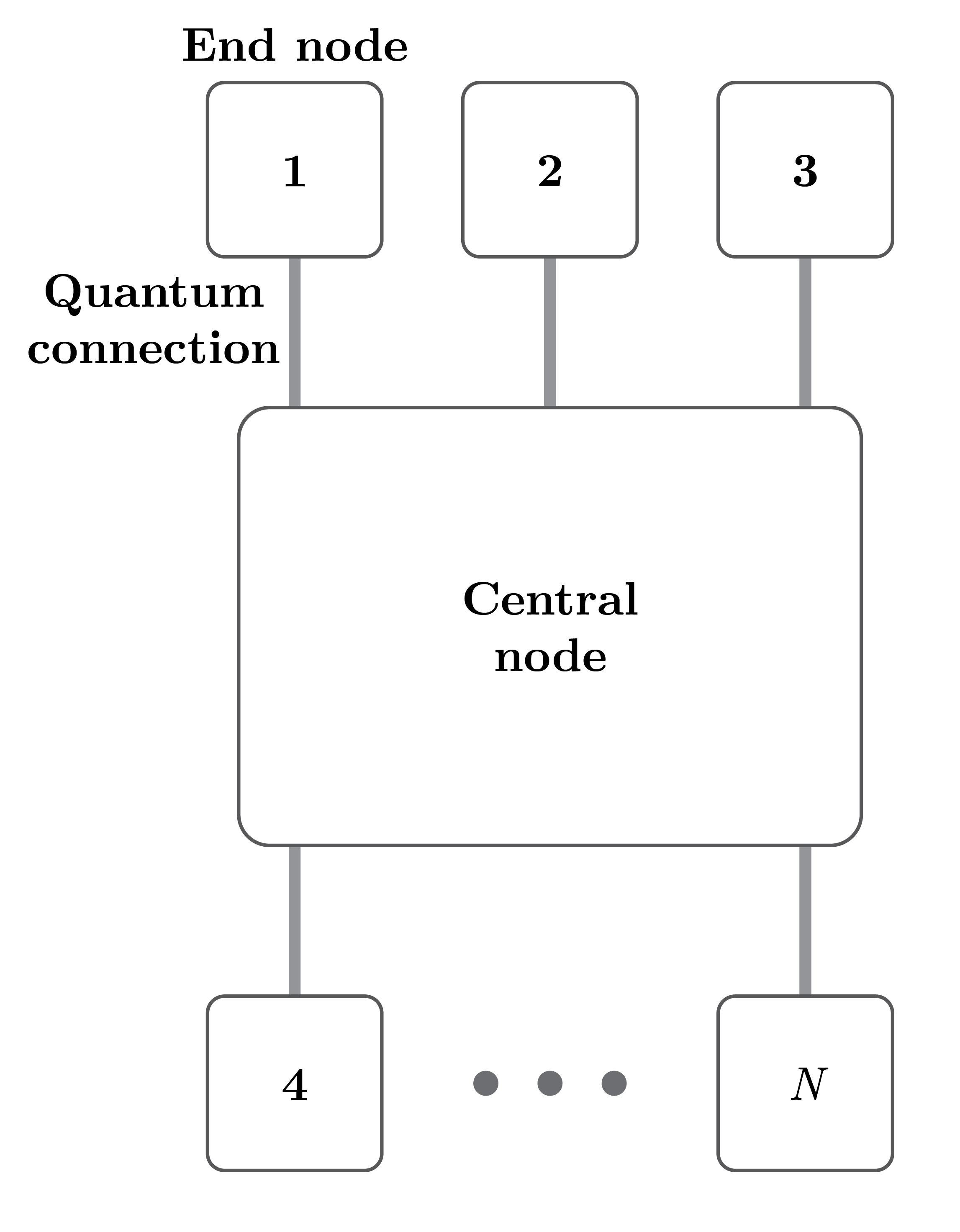}
	\caption[
		Symmetric star-shaped network studied in this paper.
	]
	{
		Symmetric star-shaped network studied in this paper.
        $N$ identical end nodes are each connected to a central node through one of, in total, $N$ identical quantum connections.
        These quantum connections can be used to distribute Bell states,
        which can be stored in quantum memory and provide a resource to create a multipartite entangled states shared by the end nodes.
        An example of a possible central node is a factory node.
	}
    \label{fig:intro:star_network}
\end{figure}

In order to understand how well factory nodes perform relative to other schemes that allow for the distribution of multipartite entangled states,
a comparison needs to be performed.
This allows us to put the rate and fidelity that factory nodes can achieve into context,
and can help determine under what circumstances it is best to use a factory node,
and under what circumstances it may be better to consider a different scheme.
Here, we provide a non-exhaustive comparison by discussing two alternative strategies for distributing multipartite entangled states on the symmetric star-shaped network depicted in Figure \ref{fig:intro:star_network}.
The first of these utilizes a central node without quantum memory,
while the second uses a 2-switch as central node.
\\

\begin{figure}
	\includegraphics[width=\linewidth]{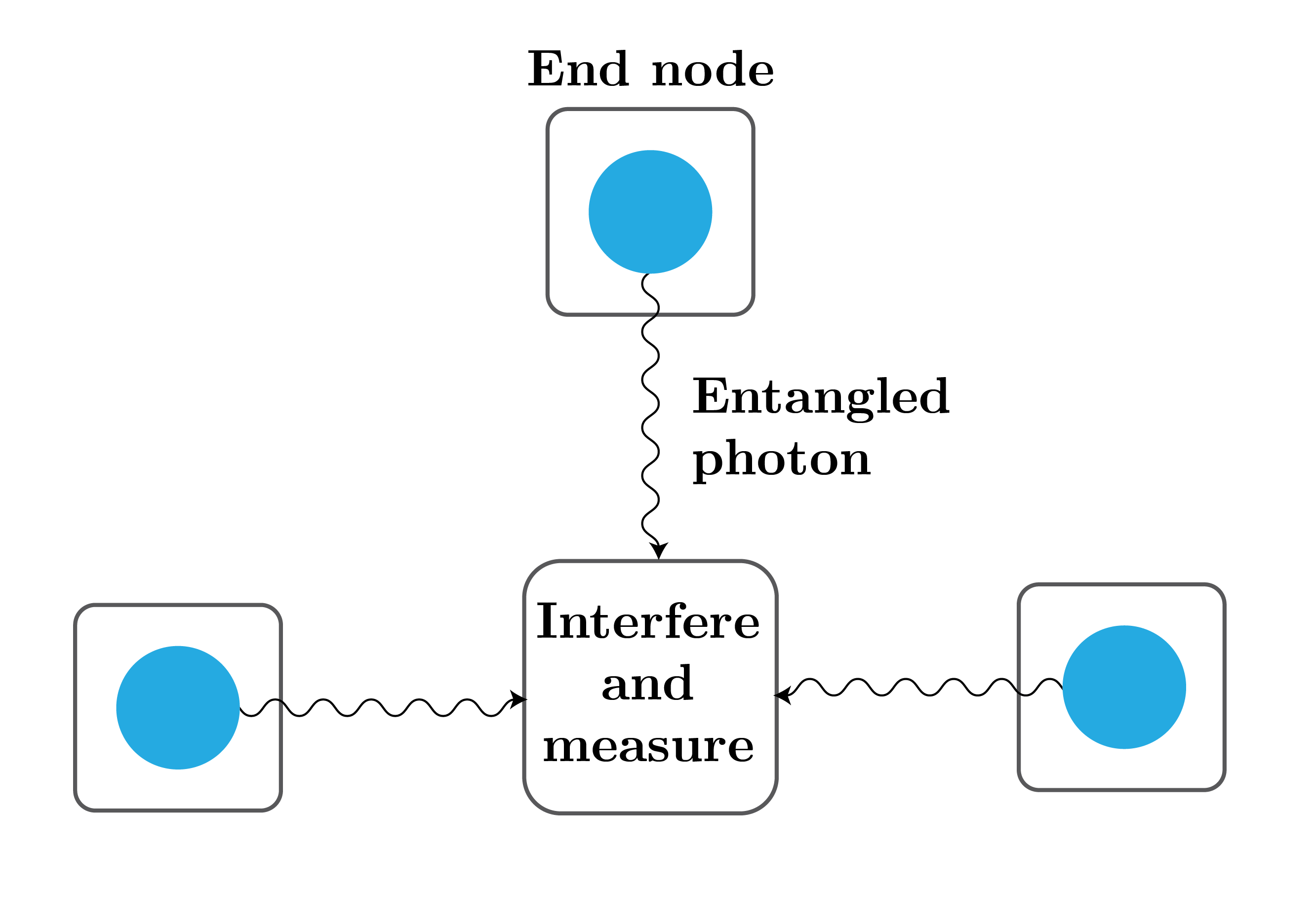}
	\caption[
		Distribution of some entangled target state between end nodes through a factory node.
	]
	{
        Some multipartite entangled states, such as GHZ states and W states, can be distributed between end nodes through the interference and measurement of entangled photons.
        Each of the end nodes needs to emit a photon that is entangled to a qubit held in local quantum memory,
        and transmit it to a central node.
        At this node, the photons originating from all the different nodes are interfered.
	}
	\label{fig:intro:interference}
\end{figure}

The first alternative method to factory nodes for the distribution of multipartite entanglement in a star-shaped network
is to utilize a central node that does not have any quantum memory.
This memoryless scheme requires connections through which photons can be directly transmitted, e.g. they can be optical fibers.
To distribute a multipartite entangled state, the end nodes emit entangled photons that are sent through the connections to the central node.
Here, the photons are interfered and measured, resulting in the creation of the target state on the end nodes.
Such schemes exist for the distribution of GHZ states \cite{wangSchemesGenerationMultipartite2009a, capraravivoliHighfidelityGreenbergerHorneZeilingerState2019}
and W states \cite{grasselliConferenceKeyAgreement2019, kalbDiamondbasedQuantumNetworks2018},
and they are illustrated in Figure \ref{fig:intro:interference}.
\\

An advantage of these schemes is that the central node can be very simple, requiring only linear-optics components and single-photon detectors.
A downside however, when distributing GHZ states, is that all photons need to arrive at the central station simultaneously, making it very sensitive to photon losses;
if each of the $N$ connections transmits photons successfully with probability $\eta$ (the transmittance of the connection),
the distribution rate will scale as $\eta^N$.
On the other hand, a factory node could be used to distribute states with a rate that falls only logarithmically with $N$,
and linearly with the success probability of Bell-state distribution (see Section \ref{sec:rate}).
How this success probability scales with $\eta$ depends on the nature of the connection and the specific method used to distribute Bell states.
When using direct transmission of entangled photons, the scaling will be linear in $\eta$,
but schemes with better scaling exist.
For example, single-click heralded entanglement generation \cite{cabrilloCreationEntangledStates1999} can be used for $\sqrt{\eta}$ scaling,
and the scaling could be further improved using quantum repeaters,
with the exact scaling depending on how they are implemented \cite{inside_quantum_repeaters}.
% When distributing W states, the rate scales only linearly with $\eta$ and is independent of $N$.
% While it scales better than the factory node in terms of $N$,
% the factory node can potentially scale more favourably with $\eta$.
No further comparison between memoryless schemes and the use of a factory node is performed in this paper.
\\

The second alternative method to using factory nodes for the distribution of multipartite entanglement in a star-shaped network,
is to use a 2-switch as a central node.
The 2-switch functions as an intermediary, allowing the end nodes to share Bell states with one another even though they are not directly connected.
By executing the appropriate local operations at the end nodes, these Bell states can be transformed into the target multipartite entangled state.
One downside to this option is that it imposes the requirement that end nodes must be able to store multiple qubits within their quantum memory,
and that they must be able to execute multipartite entangling operations.
An additional downside is that, even if each end node is able to store and exert full control over two qubits,
there still exist multipartite entangled states that the nodes would be able to store but cannot create in their limited quantum memory using only bipartite entangled states shared between them \cite{yamasakiMultipartiteEntanglementOutperforming2018}.
On the other hand, when utilizing a factory node,
any multipartite entangled state
that the end nodes have enough quantum memory to store
can be distributed among them.
Generally, when using a factory node,
advanced quantum capabilities are required only of the dedicated network device,
not of the end nodes.
\\

In section \ref{sec:comparison}, we present our second main result.
This result is a comparison, based on Monte Carlo simulations,
of the rate and fidelity of GHZ-state distribution on the symmetric star-shaped network using a factory node and using a 2-switch.
Here, we assume the 2-switch follows a specific protocol under which BSMs are not executed whenever possible,
but only when they result in a Bell state that directly contributes to the creation of a GHZ state.
\\

\subsection{Outline}

The remainder of this paper is set up as follows.
First, in Section \ref{sec:setup}, we introduce the exact factory-node setup and noise model we study.
Next, in Section \ref{sec:estimates}, we provide analytical results for the rate and fidelity with which GHZ states can be distributed on this setup.
In Section \ref{sec:comparison}, we use Monte Carlo simulations to compare the performance of GHZ-state distribution using a factory node and using a 2-switch.
We provide examples of how a factory node could be physically implemented using trapped ions or nitrogen-vacancy centers in diamond in Section \ref{sec:implementation}.
Finally, we conclude in Section \ref{sec:conclusion},
where we discuss how the results presented in this paper could be generalized and used for further study.

\section{Setup, Protocol and Model}\label{sec:setup}

In this section, we discuss in detail the factory-node setup that we study in this paper.
Additionally, we introduce the exact protocol used to distribute GHZ states on this setup,
and the model that we use to account for noise and losses.
% In this section, we introduce the exact setup and protocol that we study in this paper.
% Furthermore, we detail the exact model that we use to account for noise and losses.
\\

\begin{figure*}
    \includegraphics[width=\textwidth]{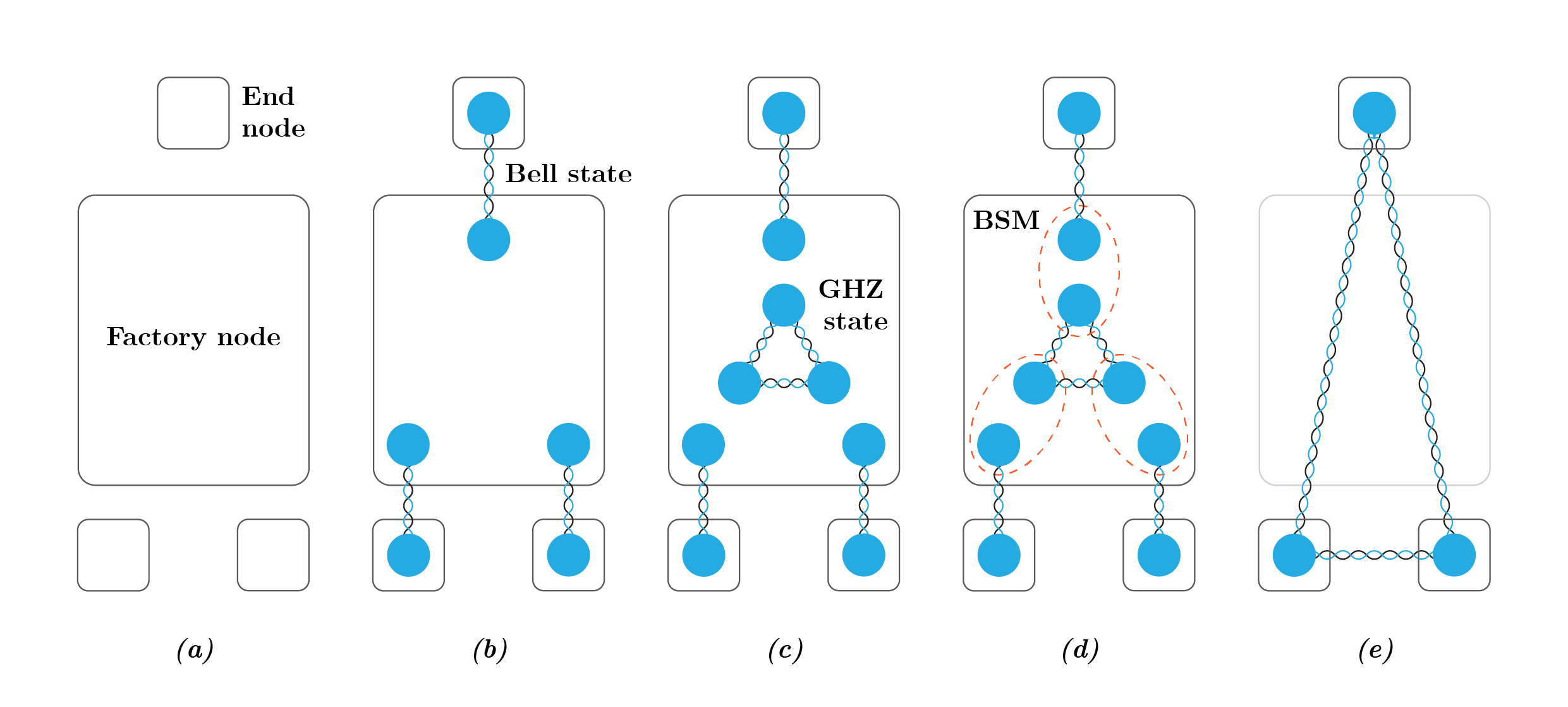}
	\caption[
        Illustration of Protocol \ref{prot:factory}.
	]
	{
        Illustration of GHZ-state distribution through a factory node, using Protocol \ref{prot:factory}.
        \textbf{(a)} There is one factory node, and there are $N=3$ end nodes.
        \textbf{(b)} Bell states are distributed between the factory node and each of the end nodes (Step 1 of Protocol \ref{prot:factory}).
        \textbf{(c)} After all Bell states are in place, a GHZ state is created locally (Step 2 of Protocol \ref{prot:factory}).
        \textbf{(d)} BSMs are executed between qubits in Bell states and qubits in the GHZ state (Step 3 of Protocol \ref{prot:factory}).
        \textbf{(e)} If all BSMs were successful and the corresponding Pauli corrections have been applied, the end nodes share a GHZ state (Steps 4 and 5 of Protocol \ref{prot:factory}).
	}
	\label{fig:setup:factory}
\end{figure*}

We consider a symmetric star-shaped quantum network.
Such a network, depicted in Figure \ref{fig:intro:star_network}, consists of $N$ end nodes, and one central node that shares a single quantum connection with each of the end nodes.
For the factory-node setup discussed in this section, this central node is a factory node.
The quantum connections can be used to distribute Bell states of the form
\begin{equation}
\ket{\phi_{00}}= \frac 1 {\sqrt2} \left( \ket {00} +  \ket {11} \right).
\end{equation}
Each end node contains a single qubit.
On the other hand, the factory node contains $2N$ qubits.
$N$ of these can be used to store the local halves of Bell states that are distributed using the quantum connections.
The other $N$ can be used to prepare and store a target quantum state to be distributed among the end nodes.
Furthermore, for each of the first $N$ qubits, the node is able to execute a BSM with exactly one of the second $N$ qubits.
In our modeling, we allow for probabilistic BSMs.
A BSM is probabilistic e.g. when it is implemented using linear optics \cite{calsamigliaMaximumEfficiencyLinearoptical2001, griceArbitrarilyCompleteBellstate2011}.
When a BSM has success probability $q_\text{BSM}$, we model this as raising a ``fail'' flag with probability $1 - q_\text{BSM}$,
and executing a perfect BSM otherwise.
On this setup, any $N$-partite target state can be distributed between the end nodes by creating the target state locally,
and then teleporting it to the end nodes using Bell states.
Specifically, we consider the distribution of an $N$-partite GHZ state using Protocol \ref{prot:factory},
which is illustrated in Figure \ref{fig:setup:factory}.
Such a state is defined by
\begin{equation}
\ket{\text{GHZ}} = \frac 1 {\sqrt 2} \left( \ket{0}^{\otimes N} + \ket {1}^{\otimes N}\right).
\end{equation}
\\

\begin{protocol}{GHZ-State Distribution Using Factory Node} \label{prot:factory}
	\item
	Repeatedly attempt Bell-state distribution over each of the $N$ quantum connections shared between the factory node and the $N$ different end nodes,
    until the factory node shares a Bell state with each end node.
	
	\item
	Create an $N$-partite GHZ state on the $N$ remaining free memory qubits in the factory node.
	
	\item
	Perform $N$ BSMs at the factory node, each between one qubit that holds part of the GHZ state, and one qubit that holds part of a Bell state.
	
	\item
	Send a classical message from the factory node to each of the end nodes containing the results of the BSMs.
	
	\item
	If any of the BSMs was unsuccessful, all end nodes reset their memory qubits.
	Return to Step 1.
    Otherwise, the end nodes perform Pauli corrections based on the outcomes of the BSMs,
    such that, in the absence of noise,
    the end nodes now share a GHZ state.
\end{protocol}
Each step in the protocol is performed after the previous step has been concluded.
In case the BSMs are all successful, the last three steps of Protocol \ref{prot:factory} implement quantum teleportation of the $N$ qubits sharing a GHZ state from the factory node to the end nodes.
Therefore, in the absence of noise, this results in the $N$ end nodes sharing an $N$-partite GHZ state.\\

In this study, we assume the time it takes to distribute a Bell state over a quantum connection follows a geometric distribution.
That is, Bell-state distribution is a series of attempts, where each attempt is of constant duration $\Delta t$,
and where the probability that an attempt is successful is described by the constant $q_\text{link}$.
To be more precise, $\Delta t$ is the time it takes after starting an attempt until both the end node and factory node know whether it was successful or not.
Only after they have obtained this knowledge, they can decide whether they want to reset their local qubits and start again, or whether they should instead keep the created quantum state stored in memory.
We use this time, i.e. $\Delta t$ after the start of the attempt, as the start of the storage time of the Bell state that is generated if the attempt is successful.
Describing Bell-state distribution as a sequence of independent attempts is accurate when the quantum connection consists of,
for example,
heralded entanglement generation through either
direct transmission \cite{linHeraldedQuantumMemory2009,langenfeldQuantumTeleportationRemote2021}
or photon interference \cite{barrettEfficientHighfidelityQuantum2005,bernienHeraldedEntanglementSolidstate2013,cabrilloCreationEntangledStates1999,humphreysDeterministicDeliveryRemote2018,kalbEntanglementDistillationSolidState2017,pompiliRealizationMultinodeQuantum2021,slodickaAtomAtomEntanglementSinglePhoton2013,stephensonHighRateHighFidelityEntanglement2020,yinEntanglementbasedSecureQuantum2020,yuEntanglementTwoQuantum2020,zippilliEntanglementDistantAtoms2008,northupQuantumInformationTransfer2014},
or a quantum-repeater chain with fixed-time quantum memory \cite{sinclairSpectralMultiplexingScalable2014, guhaRatelossAnalysisEfficient2015}.
\\

Another assumption made here is that all quantum connections are identical, i.e. $\Delta t$ and $q_\text{link}$ are the same for each of the $N$ connections between the factory node and the end nodes.
Therefore, $\Delta t$ is used as the standard time unit throughout the rest of this paper,
and one time step of duration $\Delta t$ during which attempts at Bell-state distribution take place is sometimes referred to as a ``round''.
\\

The time that it takes to send a classical message between the factory node and any of the end nodes is denoted $t_\text{cl}$.
Since Step 4 of Protocol \ref{prot:factory} consists of sending classical messages, it will take $t_\text{cl}$ to finish that step.
How large $t_\text{cl}$ is compared to $\Delta t$ depends on how the quantum connections are implemented.
For example, in the case of heralded entanglement generation through photon interference,
$\Delta t$ includes the time required to send photons to a midpoint station,
and the time required to send back the measurement outcome to the nodes.
Assuming classical signals travel at the same speed of light (in fiber) as the photons used to generate entanglement,
this time is exactly equal to $t_\text{cl}$.
$\Delta t$ may be further limited by, among others, the rate at which entangled photons can be emitted
and by classical overhead due to e.g. synchronizing emission times \cite{pfaffUnconditionalQuantumTeleportation2014, pompiliExperimentalDemonstrationEntanglement2021, pompiliRealizationMultinodeQuantum2021}.
In that case, $t_\text{cl} < \Delta t$.
In this paper, we focus on the case $q_\text{link} \ll 1$.
In that regime, the number of attempts required to successfully distribute a Bell state is typically very large.
Then, as long as $t_\text{cl}$ is not much larger than $\Delta t$,
classical communication will only take up a negligibly small part of both the time required to distribute one GHZ state and qubit storage times.
Therefore, we use $t_\text{cl} = 0$ throughout the rest of this paper.
Additionally, we assume that all local operations executed at the factory node and the end nodes are instantaneous.
These operations do not suffer from any speed-of-light delay,
and their execution time will always become comparatively small for small enough $q_\text{link}$.
Because both classical communication and local operations are modeled as instantaneous,
Step 1 is the only step of Protocol \ref{prot:factory} with nonzero duration.
\\

All noise in the network is modeled by depolarizing channels, described by the action  \cite{nielsenQuantumComputationQuantum2011}
\begin{equation}
\mathcal D _{\mathcal H_A, p} (\rho) = p \rho + (1 - p)\Tr_{\mathcal H_A}(\rho) \otimes \frac{\mathbb 1_{\mathcal H_A}}{\Tr \mathbb 1_{\mathcal H_A}}.
\end{equation}
Here, $\rho$ is a density matrix in the Hilbert space $\mathcal H = \mathcal H_A \otimes \mathcal H_B$,
$\mathcal H_A$ is the subspace of $\mathcal H$ that describes the system that the depolarizing channel acts on,
$\mathbb 1_{\mathcal H_A}$ is the identity operator of $\mathcal H_A$,
$\Tr_{\mathcal H_A}$ is the partial trace over $\mathcal H_A$,
and $p$ is the so-called depolarizing parameter.
It can be interpreted as losing all information about the system described by $\mathcal H_A$ with probability $1 - p$.
Specifically, we consider the following sources of noise:
\begin{itemize}
	\item Noisy connections.
	Whenever a Bell state is created, a depolarizing channel with parameter $p_\text{link}$ acts on the two qubits that hold the Bell state
    (i.e. $\mathcal H_A$ has dimension $4$).
    We note that, because of the symmetry of the Bell state,
    this is equivalent to a single-qubit depolarizing channel acting with parameter $p_\text{link}$ on either of the individual qubits.
	
	\item Noisy memory.
	For every time unit $\Delta t$ that a quantum state is stored in a memory qubit,
    a depolarizing channel with parameter $p_\text{mem}$ acts on that qubit
    (i.e. $\mathcal H_A$ has dimension 2).
	
	\item Noisy BSMs.
	Whenever a BSM is executed, it is preceded by two depolarizing channels with parameter $p_\text{BSM}$, one on each of the participating qubits
    (i.e. $\mathcal H_A$ has dimension 2).
	This measurement itself, following the depolarizing channels, is then modeled as being noiseless.
	
	\item Noisy GHZ states.
	Whenever a GHZ state is created, a depolarizing channel with parameter $p_\text{GHZ}$ acts on the $N$ qubits that hold the GHZ state
    (i.e. $\mathcal H_A$ has dimension $2^N$.)
\end{itemize}
Local Pauli corrections are modeled as noiseless.

\section{Analytical Results} \label{sec:estimates}

Here, we present analytical results for the rate and fidelity of GHZ-state distribution using Protocol \ref{prot:factory}.
For the rate, we provide three analytical results: an exact expression, a lower bound, and a leading-order expression.
For the fidelity, we present two analytical results: a lower bound and a leading-order expression.
The accuracy of the leading-order expression for the rate, and of both the leading-order expression and the lower bound for the fidelity,
is verified against a numerical model built using the quantum-network simulator NetSquid \cite{coopmansNetSquidNETworkSimulator2021} in Appendix \ref{app:verification}.

\subsection{Rate} \label{sec:rate}

We denote the time required to distribute a single GHZ state using Protocol \ref{prot:factory} by $T$, which is a random variable.
The (average) rate at which GHZ states are distributed is then defined by
\begin{equation}\label{key}
R = 1 / \expectationvalue{T}.
\end{equation}
Thus, to calculate the rate, we need to know the expected value of the distribution time.
To this end, we decompose the distribution time as
\begin{equation}
T = n_\text{teleport} T_\text{teleport}.
\end{equation}
Here, $n_\text{teleport}$ is the number of attempts at teleporting a GHZ state until such an attempt is successful.
That is, it is the number of times Steps 1 through 4 of Protocol \ref{prot:factory} need to be executed for the protocol to finish.
Such an attempt at teleportation may fail in case the BSMs are probabilistic, i.e. $q_\text{BSM} < 1$.
On the other hand, $T_\text{teleport}$ is
the time required to perform Steps 1 through 4 once.
Both these quantities are random variables.
Because under the present assumptions only Step 1 of Protocol \ref{prot:factory} has a nonzero duration,
$T_\text{teleport}$ can be further dissected into
\begin{equation}
T_\text{teleport} = n_\text{all} \Delta t,
\end{equation}
where $n_\text{all}$ is again a random variable, corresponding to the number of rounds of Bell-state distribution required to share Bell states between the factory node and all of the end nodes.
That is, it is the number of rounds required to finish Step 1 of Protocol \ref{prot:factory}.
Combining the two expressions yields
\begin{equation}
T = n_\text{teleport} n_\text{all} \Delta t.
\end{equation}
Because the expected value of a product of two independent random variables is the product of their expected values, we find
\begin{equation}
\expectationvalue{T} = \expectationvalue{n_\text{teleport}} \expectationvalue{n_\text{all}} \Delta t.
\end{equation}
Since each teleportation attempt succeeds with a fixed success probability of $q_\text{BSM}^N$
(teleportation succeeds if and only if all $N$ BSMs are successful),
$n_\text{teleport}$ is geometrically distributed with $\expectationvalue{n_\text{teleport}} = 1 / q_\text{BSM} ^ N$.
Thus,
\begin{equation} \label{eq:rate:rate_in_terms_of_n_all}
R = \frac{q_\text{BSM}^N}{\expval{n_\text{all}} \Delta t}.
\end{equation}

The probability distribution of $n_\text{all}$ is more complicated:
the number of rounds required to distribute Bell states with all $N$ end nodes is the number of rounds required to distribute the Bell state that takes the longest.
Writing $n_i$ for the number of attempts required to distribute a Bell state with end node $i$, we have
\begin{equation} \label{eq:rate:n_all}
n_\text{all} = \max \{n_1, n_2, ..., n_N\}.
\end{equation}
Each of the $n_i$ is geometrically distributed with $\expectationvalue{n_i} = 1 / q_\text{link}$.
It can be evaluated exactly using \cite{bernardesRateAnalysisHybrid2011}
\begin{equation}
\expval{n_\text{all}} = \sum_{j=1}^N (-1)^{j+1} \binom {N} {j} \frac 1 {1 - (1 - q_\text{link})^j}.
\end{equation}
This can be substituted into Eq. \eqref{eq:rate:rate_in_terms_of_n_all} to obtain an exact expression for the rate.
However, we also report here a known leading-order expression
\cite{coopmansImprovedAnalyticalBounds2022, shchukinWaitingTimeQuantum2019, schmidtMemoryassistedLongdistancePhasematching2020},
\begin{equation} \label{eq:rate:max_of_geoms}
\expval{n_\text{all}} \approx \frac {H_N}{q_\text{link}},
\end{equation}
where $H_N$ is the $N^\text{th}$ harmonic number,
\begin{equation}
H_N \equiv \sum_{i = 1}^{N} \frac 1 i = \gamma + \ln{N} + \mathcal O \Big(\frac 1 N \Big).
\end{equation}
Here, $\gamma \approx 0.5772$ is the Euler-Mascheroni constant.
Substituting this into Equation \eqref{eq:rate:rate_in_terms_of_n_all} yields
\begin{equation} \label{eq:factory_rate}
R \approx \frac{q_\text{BSM}^N q_\text{link}}{H_N \Delta t},
\end{equation}
which is valid up to leading order in $q_\text{link}$.
\\

There are two reasons why we report the leading-order approximation \eqref{eq:factory_rate} even though an exact expression is available.
First, in the regime $q_\text{link} \ll 1$,
Eq. \eqref{eq:factory_rate} is accurate and easier to evaluate.
Second, Eq. \eqref{eq:factory_rate} more clearly shows how the rate scales with $q_\text{link}$, $N$ and $q_\text{BSM}$,
thereby providing more intuition.
We additionally note that there exists an upper bound \cite{eisenbergExpectationMaximumIID2008, coopmansImprovedAnalyticalBounds2022},
\begin{equation}\label{eq:rate:bound}
\expectationvalue{n_\text{all}} < 1 + \frac{H_N}{- \ln(1 - q_\text{link})}.
\end{equation}
Therefore, Eq. \eqref{eq:factory_rate} is a lower bound on the actual rate if
\begin{equation}
\frac{H_N}{q_\text{link}} >  1 + \frac{H_N}{- \ln(1 - q_\text{link})}.
\end{equation}
This is the case for any $N > 3$.
Additionally, it is true for $N = 3$ if $q_\text{link} \gtrapprox 0.42$.
Therefore, using the simpler leading-order expression usually does not lead to overestimating the performance of Protocol \ref{prot:factory}.
In Appendix \ref{app:verification}, for $N = 5$,
we find that Eq. \eqref{eq:factory_rate} is indeed a tight lower bound for small values of $q_\text{link}$,
while underestimating the rate up to a factor of two for $q_\text{link} \sim 1$.

\subsection{Fidelity} \label{sec:fidelity}

In this section, we calculate the fidelity of the state shared by the end nodes after a successful execution of Protocol \ref{prot:factory}.
This fidelity is defined with respect to the perfect GHZ state.
The first step is to determine the density matrix of that state, which we denote $\rho$.
In the absence of noise, $\rho$ would simply be a perfect GHZ state.
However, due to the depolarizing noise in the creation of the local GHZ state within the factory node,
the performance of BSMs,
the distribution of Bell states
and the storage of qubits,
$\rho$ is generally not a GHZ state
and is a function of the noise parameters $p_\text{GHZ}$, $p_\text{BSM}$, $p_\text{link}$ and $p_\text{mem}$.
Additionally, we note that each individual execution of the protocol is characterized by the values that the random variables $n_1, n_2, ..., n_N$ take.
Just like above, the random variable $n_i$ represents the number of rounds it takes to distribute a Bell state between the factory node and end node $i$.
How much decoherence due to the storage of qubits in quantum memories is suffered,
will depend on the value that each $n_i$ takes.
Therefore, $\rho$ is additionally a function of the random variables $n_1, n_2, ..., n_N$.
\\

We derive $\rho$ as a function of the noise parameters and random variables in Appendix \ref{app:dm}.
Here, we briefly summarize how this derivation is performed.
First, we note that there are single-qubit depolarizing channels acting on three groups of qubits.
First, there are the qubits that are part of the locally created GHZ state in the factory node.
Second, there are the qubits stored at the GHZ factory that are entangled to those at the end nodes and partake in BSMs together with the GHZ-state qubits.
Finally, there are the qubits stored at the end nodes.
Because of the symmetry of Bell states, and by extension of BSMs,
it is possible to ``move'' all these single-qubit depolarizing channels to only the qubits stored at the end nodes.
That is, the state $\rho$ can be derived correctly by pretending that as the protocol is executed,
there is no single-qubit depolarizing noise within the factory node,
but instead there are only single-qubit depolarizing channels acting at the end nodes.
Because the composition of depolarizing channels is itself a depolarizing channel,
each end node $i$ only undergoes a single depolarizing channel with parameter
\begin{equation} \label{eq:fidelity:p_i}
p_i = p_\text{link} \; p_\text{BSM}^2 \; p_\text{mem} ^ {2 \Delta n_i},
\end{equation}
where
\begin{equation}
\Delta n_i \equiv n_\text{all} - n_i
\end{equation}
is the number of rounds the Bell state shared with end node $i$ is stored until it partakes in a BSM.
Describing the protocol in this way is very convenient,
because it then amounts to performing perfect quantum teleportation of a noisy GHZ state to the end nodes,
followed by depolarizing channels on each of the $N$ individual qubits of the state.
Resolving all these depolarizing channels gives the result
\begin{equation} \label{eq:fidelity:state}
\begin{aligned}
\rho &= \frac{1 - p_{\text{GHZ}}}{2^N} \mathbb 1_{\mathcal N} \\
&+ p_{\text{GHZ}} \Bigg[\prod_{i\in \mathcal N} p_i \big( \ketbra{\text{GHZ}} \big)_{\mathcal N} + \prod_{i \in \mathcal N} \frac{1 - p_i}{2} \mathbb 1 _{\mathcal N} \\
&+ \frac 1 2 \sum_{\substack{U \subset \mathcal N \\ 1 < |U| < N}} \left( \prod_{i \in U } \frac{1 - p_i}{2} \prod_{j \in \mathcal N \setminus U} p_j \right) \mathbb 1_U \otimes \mathcal P_{\mathcal N \setminus U} \Bigg].
\end{aligned}
\end{equation}
Here, we have defined $\mathcal N = \{1, 2, ..., N\}$,
and $\mathcal P$ is the classically correlated, unnormalized state
\begin{equation} \label{eq:fidelity:P}
\mathcal P_{1, 2, \dots, k} \equiv \big( \ketbra{0}\big)^{\otimes k} + \big( \ketbra{1}\big)^{\otimes k}.
\end{equation}
The different terms in the density matrix correspond to all different combinations of some of the qubits being lost due to single-qubit depolarizing noise,
and some being unscathed.
\\

Using Eq. \eqref{eq:fidelity:state},
the fidelity can be efficiently written as
\begin{equation} \label{eq:fidelity:F_rand}
\begin{aligned}
F_\text{rand} \equiv& \expval{\rho}{\text{GHZ}} \\
=& \frac{1 - p_\text{GHZ}}{2^N}  \\
+& p_\text{GHZ} \sum_{U \subseteq \mathcal N} 2^{\delta_{|U|, 0} + \delta_{|U|, N} - 1} \prod_{i \in U}  \Big ( \frac{1 - p_i}{2} \Big) \prod_{j \in \mathcal N \setminus U} p_j ,
\end{aligned}
\end{equation}
where $|U|$ is the cardinality of set $U$ and $\delta_{i,j}$ denotes the Kronecker delta function.
As the fidelity is a function of the random variables $\Delta n_i$,
it is itself a random variable: it depends on how quickly one after another the different Bell states are distributed.
This is the reason why the fidelity above is denoted with the subscript ``rand''.
The delta functions are there to account for the fact that there is ``one less'' factor of $\tfrac 1 2$ in the fidelity when no qubits are lost,
and when all qubits are lost.
The reason for this is that losing a single qubit (i.e. tracing that qubit out and then replacing it by a maximally mixed state) in a GHZ state does not only destroy the information held by that qubit,
but also reduces the correlation between the remaining qubits to classical correlation instead of quantum correlation.
Therefore, the first qubit that is lost accounts for a larger drop in fidelity than subsequent qubits.
Additionally, the last qubit that is lost does not account for any drop in fidelity,
as losing $N-1$ qubits of the GHZ state will already result in an $N$-qubit maximally mixed state,
the fidelity of which cannot be further decreased by depolarizing noise.
\\

Here, we are assuming no post-selection on distributed GHZ states takes place.
Therefore, we can describe the state produced by execution of Protocol \ref{prot:factory} as a mixture between all $\rho$'s corresponding to different values of $\Delta n_i$.
This state is then independent of the random variables, and the same for each execution of the protocol.
The mixed state is the expected value of the density matrix $\rho$,
and its fidelity is the expected value of $F_\text{rand}$,
which can be written as
\begin{equation} \label{eq:fidelity:fidelity_not_yet_rearranged}
\begin{aligned}
F &= \expectationvalue{F_\text{rand}} =  \frac{1 - p_\text{GHZ}}{2^N} \\
&+ p_\text{GHZ} \sum_{U \subseteq \mathcal N} 2^{\delta_{|U|, 0} + \delta_{|U|, N} - 1}  \expectationvalue{\prod_{i \in U}  \frac{1 - p_i}{2} \prod_{j \in \mathcal N \setminus U} p_j}.
\end{aligned}
\end{equation}
In appendix \ref{app:coefficients} we work out the combinatorics to rewrite the fidelity as
\begin{equation} \label{eq:factory_fidelity}
F = \frac{1 - p_\text{GHZ}}{2^N} + p_\text{GHZ} \sum_{U \subseteq \mathcal N}  A_{|U|} \expval{\prod_{i \in U} (p_\text{mem}^2)^{\Delta n_i}},
\end{equation}
where
\begin{equation} \label{eq:fidelity:coefficients}
A_{|U|} =
\begin{cases}
\left( p_\text{link} p_\text{BSM}^{2} \right)^{|U|} \left( \frac 1 {2 ^ N} + \frac 1 2 \delta_{|U|, N}\right) \hspace{0.1 cm} &\text{if $|U|$ is even,} \\
\frac 1 2 \left( p_\text{link} p_\text{BSM}^{2} \right)^{|U|} \delta_{|U|, N}  &\text{if $|U|$ is odd.}
\end{cases}
\end{equation}
\\

Now, we note that after Bell states have been distributed between the factory node and all end nodes,
it is possible to order the end nodes based on the order in which they were connected to the factory node.
That is, to each end node $i \in \mathcal N$ we assign $d_i \in \mathcal N$ such that if $d_i > d_j$,
then end node $i$ shared a Bell state with the factory node at the same time as or later than end node $j$.
For example, if end node 4 shared a Bell state first, we assign $d_4 = 1$.
If such an ordering is given, it is possible to use the results from Appendix \ref{app:exp_val} to evaluate expressions like Eq. \eqref{eq:factory_fidelity}.
However, in general, such an ordering cannot be imposed a priori; it is only well-defined after executing the protocol.
Because the order in which Bell states are shared is random, each $d_i$ is a random variable.
Therefore, to apply the results from the appendix, an average should be taken over all possible orders in which Bell states can be distributed.
Because of the symmetry of the setup under consideration, however, we need not worry about that.
The success probability is $q_\text{link}$ for all quantum connections, so all orderings are equally likely.
Furthermore, since the effective depolarizing probability per round is $p_\text{mem}^2$ for all end nodes,
the fidelity is invariant under changes in the ordering
(it does not matter if end node 4 shares a Bell state first and end node 6 last, or the other way around).
Therefore, we can safely pretend the order in which Bell states are distributed is fixed.
Furthermore, we set our labeling to coincide with this order.
That is, we set it such that $d_i = i$.
\\

It follows from Eq. \eqref{eq:app:G_final} in Appendix \ref{app:exp_val} that, to leading order in $q_\text{link}$ and $(1 - p_\text{mem}^2)$,
\begin{equation} \label{eq:factory_fidelity_approx}
\expectationvalue{\prod_{i \in U} (p_\text{mem}^2)^{\Delta n_i}} \approx \prod_{k=1}^N \frac{(N + 1 - k) q_\text{link} }{|U_k| (1 - p_\text{mem}^2)+ (N + 1 - k) q_\text{link}},
\end{equation}
where
\begin{equation}
U_k \equiv \{u \in U | u < k \}.
\end{equation}
For example, if $U = \{1, 3\}$, then $U_1 = \emptyset$, $U_2 = U_3 = \{1\}$ and $U_4 = U$.
Since the expression is to leading order in $1 - p_\text{mem}^2$
and $1 - p_\text{mem}^2 \geq 1 - p_\text{mem}$, we consider the approximation to be valid up to leading order in $1 - p_\text{mem}$.
A leading-order expression for the fidelity is then obtained by combining Eq. \eqref{eq:factory_fidelity} with Eq. \eqref{eq:factory_fidelity_approx}.
\\

The main reason why working to leading order in $q_\text{link}$ and $1 - p_\text{mem}^2$ allows us to derive Eq. \eqref{eq:factory_fidelity_approx},
is that in this approximation we can neglect the possibility of multiple Bell states being generated at the same time.
For $q_\text{link} \ll 1$, the probability of more than one Bell state being generated during a single round is very small;
most likely, there are many rounds between one success and the next.
Additionally, when $1 - p_\text{mem}^2 \ll 1$, the drop in fidelity per extra round that qubits have to wait in memory is small.
If that were not the case, the fidelity can be still high in case all Bell states succeed in quick succession, including some at the same time,
while the fidelity would already be small in case there is some waiting time between different successes.
Therefore, the contribution to the average fidelity of cases with multiple simultaneous successes would be relatively large despite them occurring with small probability,
and neglecting their contribution would be inaccurate.
\\

We see in Appendix \ref{app:verification} that the real fidelity of Protocol \ref{prot:factory} is typically larger than the leading-order expression given by Eq. \eqref{eq:factory_fidelity_approx}.
This is explained by the fact that we ignore cases where multiple Bell states are generated simultaneously:
we are effectively calculating the average of $F_\text{rand}$ over a sub-normalized probability distribution.
However, this does not prove Eq. \eqref{eq:factory_fidelity_approx} is a lower bound on the fidelity.
The reason for this is that, in Appendix \ref{app:exp_val}, in order to work consistently at leading order in $q_\text{link}$ and $1 - p_\text{mem}$ we have also neglected terms that would lower the calculated fidelity if they were included,
and we do not know if these neglected terms generally outweigh the terms corresponding to multiple simultaneously distributed Bell states.
When not throwing these higher-order terms out, a strict lower bound is obtained.
However, it typically approximates the real fidelity (far) worse than the leading-order expression, as discussed below.
The bound is calculated in Appendix \ref{app:exp_val} (Eq. \eqref{eq:app:G_bound}) and yields
\begin{equation} \label{eq:factory_fidelity_bound}
\begin{aligned}
&\expectationvalue{\prod_{i \in U} (p_\text{mem}^2)^{\Delta n_i}} \geq \\
&\prod_{k=1}^N \frac{(N + 1 - k) q_\text{link} (1 - q_\text{link})^{N-k} (1 - p_\text{mem}^2)^{|U_k|} }{1 - (1 - q_\text{link})^{N+1-k} (1 - p_\text{mem}^2)^{|U_k|}}.
\end{aligned}
\end{equation}
The lower bound on the fidelity is obtained by using Eq. \eqref{eq:factory_fidelity_bound} to evaluate Eq. \eqref{eq:factory_fidelity}.
\\

In Appendix \ref{app:verification},
we compare the analytical results to a Monte Carlo simulation of Protocol \ref{prot:factory}.
One such comparison figure is also included here, see Figure \ref{fig:fidelity:verification}.
In Appendix \ref{app:verification}, we find that both the leading-order expression and lower bound closely approximate simulation results for small values of $q_\text{link}$ and $1 - p_\text{mem}$.
Remarkably, the leading-order expression remains reasonably accurate all the way up to $q_\text{link} \sim 1$, where deviations are on the percent level.
This can be explained by the fact that as $q_\text{link}$ grows,
the effect of memory decoherence slowly becomes negligible in case $1 - p_\text{mem} \ll 1$,
and the leading-order expression happens to be accurate up to the point where the fidelity becomes approximately constant.
The lower bound however becomes very loose for larger values of $q_\text{link}$.
When instead $1 - p_\text{mem}$ is increased,
we find that the leading-order expression stays accurate and the lower bound remains tight until the fidelity becomes close to that of a maximally mixed state.
\\

\begin{figure}[h]
    \includegraphics[width=\linewidth]{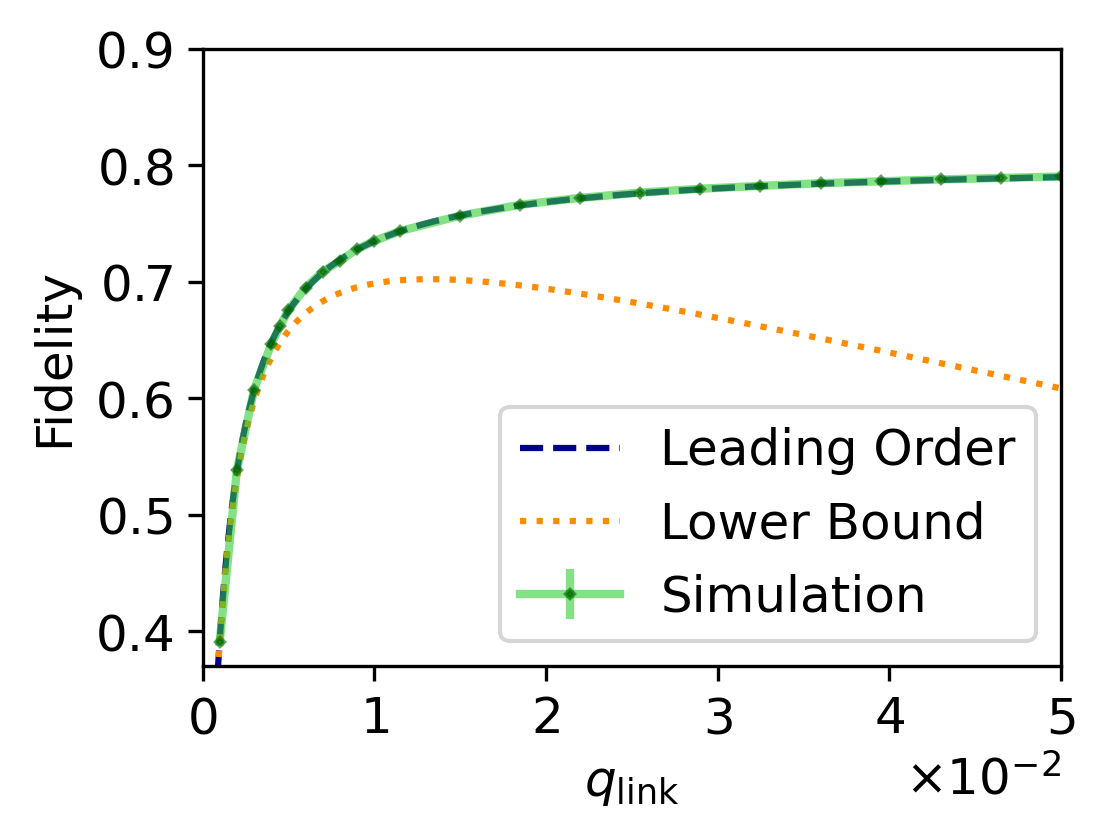}
	\caption
	{
        Comparison between simulation result and analytical expressions for the fidelity of Protocol \ref{prot:factory}.
        The parameters are $N=5$, $q_\text{BSM} = 0.95$, $p_\text{BSM} = p_\text{link} = 1 - 10^{-2}$ and $p_\text{mem} = 1 - 10^{-4}$.
        GHZ states are locally prepared with a fidelity of 0.9,
        which corresponds to $p_\text{GHZ} \approx 0.872$.
        The lower bound is tight for small values of $q_\text{link}$, but not for larger values.
        The leading-order expression on the other hand stays accurate also for larger values of $q_\text{link}$.
        Each data point represents the average over 10,000 simulated executions of Protocol \ref{prot:factory}.
        Error bars represent the standard deviation of the mean and are smaller than the markers.
        Note that the lines showing the leading-order result and the simulation result can be hard to distinguish because of their overlap.
	}
	\label{fig:fidelity:verification}
\end{figure}

To calculate both the approximate and bounded values of $F$,
we use a Python script that evaluates Eq. \eqref{eq:factory_fidelity} using either Eq. \eqref{eq:factory_fidelity_approx} (for an approximation)
or Eq. \eqref{eq:factory_fidelity_bound} (for a lower bound).
This script has been made public and can be found in our repository \cite{netsquid-factory}.
\\

\section{Comparison} \label{sec:comparison}

\begin{figure*}
    \includegraphics[width=\textwidth]{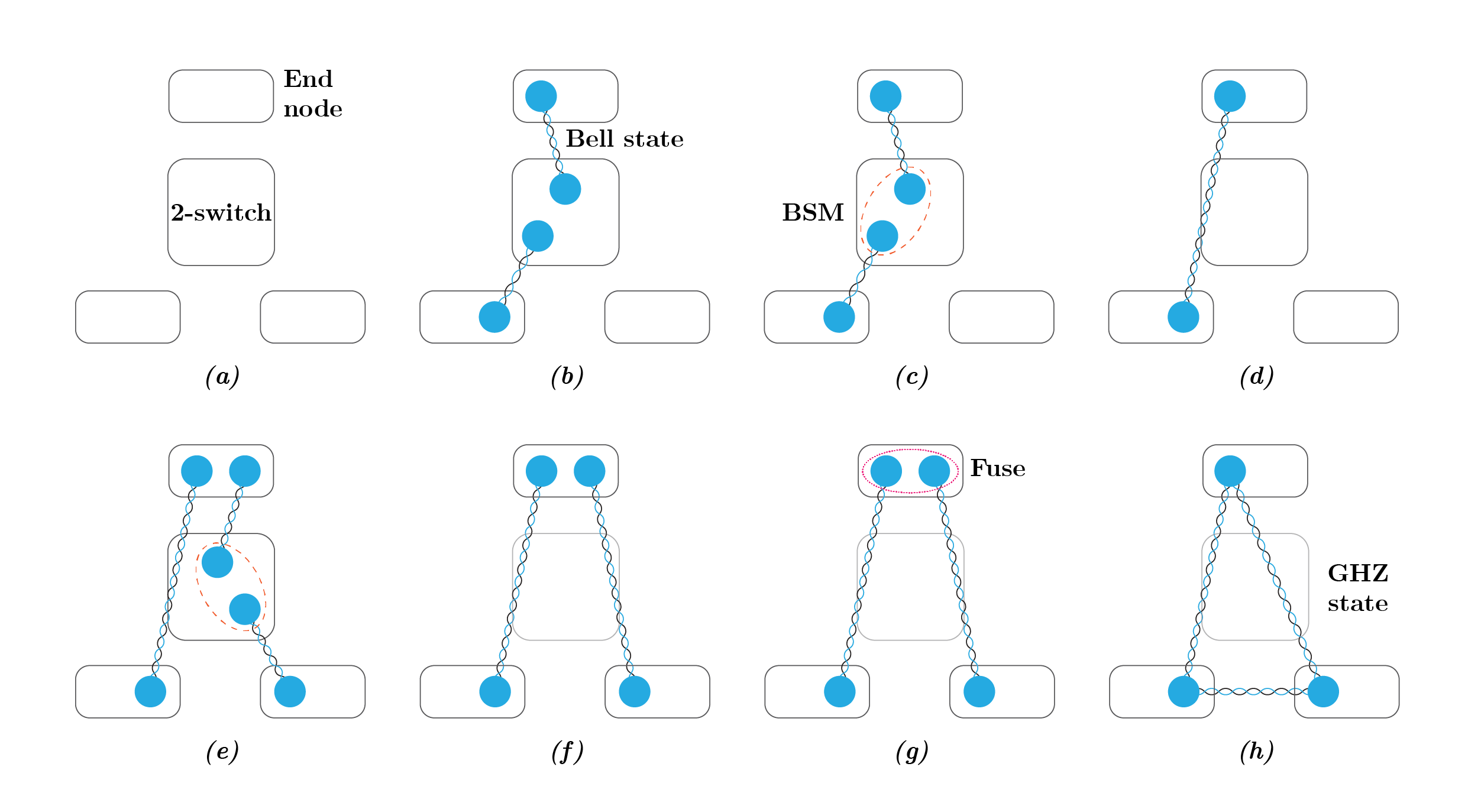}
	\caption[
        Illustration of Protocol \ref{prot:bipartite}.
	]
	{
        Illustration of GHZ-state distribution through a 2-switch, using Protocol \ref{prot:bipartite}.
        \textbf{(a)} There is one 2-switch, and there are $N=3$ end nodes.
        \textbf{(b)} Bell states are distributed between the 2-switch and end nodes (Step 1 of Protocol \ref{prot:bipartite}).
        \textbf{(c)} When there are two Bell states, a BSM is executed (Step 2 of Protocol \ref{prot:bipartite}).
        \textbf{(d)} If the BSM was successful and the corresponding Pauli corrections have been applied, the two end nodes now share a Bell state (Steps 3 and 4 of Protocol \ref{prot:bipartite}).
        \textbf{(e)} Bell states are distributed until the 2-switch is entangled to two end nodes that are not themselves already entangled.
        A BSM is executed on the corresponding entangled qubits (Steps 1 - 3 of Protocol \ref{prot:bipartite}).
        \textbf{(f)} If the BSM was successful and the corresponding Pauli corrections have been applied, one end node is now entangled to the two other end nodes, but those other end nodes are not themselves entangled to each other (Step 4 of Protocol \ref{prot:bipartite}).
        \textbf{(g)} A fusion operation (which involves a CNOT gate and Z-basis measurement) is executed in the end node holding two qubits (Steps 5 and 6 of Protocol \ref{prot:bipartite}).
        \textbf{(h)} As a consequence of the fusion operation, the three end nodes now share a GHZ state together.
	}
	\label{fig:comparison:bipartite}
\end{figure*}

In this section, we compare the performance of GHZ-state distribution on a symmetric star-shaped network (depicted in Figure \ref{fig:intro:star_network})
in case the central node is a factory node to the performance in case the central node is not a factory node.
Specifically, we will compare the performance of Protocol \ref{prot:factory} as described in Section \ref{sec:setup}
to the performance of Protocol \ref{prot:bipartite}, which requires the central node to be a 2-switch.
The 2-switch serves as an intermediary in the creation of Bell states between end nodes by performing BSMs on pairs of entangled qubits.
Protocol \ref{prot:bipartite} is illustrated in Figure \ref{fig:comparison:bipartite}.
\\

There are two differences between the factory-node setup discussed in Section \ref{sec:setup},
and the 2-switch setup considered here.
The first difference is in the central node.
The central node is the 2-switch,
and it is able to store a maximum of $N$ qubits in quantum memory (one per end node).
The only way this node can manipulate qubits,
is through the execution of BSMs on any pair of the qubits in its memory.
When the node executes a BSM between a qubit that is entangled to one end node and a qubit that is entangled to another end node,
this results in a Bell state shared between the two end nodes.
The second difference is in the end nodes.
As discussed in Section \ref{sec:introduction},
end nodes that only have access to bipartite entangled resource states among themselves cannot create multipartite entangled states if they can only store a single qubit.
Therefore, in order to enable the distribution of GHZ states through the use of a 2-switch,
end nodes in the 2-switch setup have a quantum memory of two qubits each.
Additionally, they are able to execute CNOT gates and Z-basis measurements.
\\

We model the 2-switch setup largely the same as the factory-node setup.
Each attempt at Bell-state distribution takes a time $\Delta t$.
Exchanging a classical message between the central node and an end node takes time $t_\text{cl}$,
which we assume to be zero.
An attempt at Bell-state distribution succeeds with probability $q_\text{link}$,
a BSM succeeds with probability $q_\text{BSM}$.
Whenever a Bell state is distributed by a quantum connection, the qubits are depolarized with parameter $p_\text{link}$.
Qubits stored in memory undergo depolarization with parameter $p_\text{mem}$ once during each time unit $\Delta t$.
Finally, whenever a BSM is executed,
both qubits first undergo depolarization with parameter $p_\text{BSM}$.
We model CNOT gates and Z-basis measurements as noiseless.
\\

\begin{protocol}{Bipartite GHZ-state distribution.} \label{prot:bipartite}

	\item
    Repeatedly attempt Bell-state distribution over all quantum connections for which there is a free qubit at the 2-switch
    until the first success occurs.
	
	\item
    At the 2-switch, execute BSMs randomly between pairs of entangled qubits,
    on the condition that
    the end nodes that are entangled to those qubits are not yet part of the same (noisy) GHZ state.
    If no BSMs are executed, go back to Step 1.
	
	\item
	Send a classical message from the 2-switch to each of the end nodes,
    informing them about which BSMs have been executed, and what the results of the measurements are.
	
	\item
    Each end node that was entangled to a qubit that has partaken in a BSM,
    checks the result of that BSM.
    If the BSM failed, the qubit is reset.
    If it succeeded, a Pauli correction (chosen based on the outcome of the BSM)
    is applied to the qubit to ensure this qubit and the qubit it is entangled with are in the $\ket{\phi_{00}}$ Bell state (in the absence of noise).

    \item
    Each end node that now holds two qubits in its quantum memory
    executes a CNOT gate between those qubits followed by a Z-basis measurement on the target qubit.

    \item
    Each end node that has executed a Z-basis measurement sends a classical message with the result to all other end nodes.
    These end nodes then perform single-qubit Pauli corrections, chosen based on the measurement outcomes,
    to transform each entangled state that is shared between end nodes into a GHZ state (in the absence of noise).

	\item
    If there is a GHZ state shared between all end nodes, the protocol has finished.
    Otherwise, go back to Step 1.  \end{protocol}
We now make some remarks about Protocol \ref{prot:bipartite}.
\begin{itemize}

    \item
    In Step 1 of Protocol \ref{prot:factory},
    Bell-state distribution is attempted until there has been one success for each of the $N$ quantum connections.
    In contrast, in Step 1 of Protocol \ref{prot:bipartite},
    Bell-state distribution is only attempted until there is a round during which as least one success occurs.

    \item
    Steps 5 and 6 together implement a fusion operation \cite{deboneProtocolsCreatingDistilling2020}.
    Such an operation combines two GHZ states into one,
    at the cost of measuring out a single qubit.
    Here, the $\ket{\phi_{00}}$ Bell state is considered a two-qubit GHZ state.
    Each time a fusion operation is executed,
    a larger GHZ state is created,
    until eventually all $N$ end nodes share in the GHZ state.

    \item
    For each time Step 1 is executed, classical communication takes up a time $3t_\text{cl}$
    (one $t_\text{cl}$ to send BSM results from the 2-switch to the end nodes, one $t_\text{cl}$ to send Z-basis-measurement results from the end nodes to the 2-switch, and one $t_\text{cl}$ to forward those measurement results from the 2-switch to the end nodes).
    When $q_\text{link} \ll 1$, Step 1 requires many rounds and therefore both the completion time and the qubit storage times are dominated by entanglement distribution, assuming $t_\text{cl}$ is not much larger than $\Delta t$.
    The classical communication time can then be safely neglected, just as for Protocol \ref{prot:factory}.
    This motivates the choice to consistently set $t_\text{cl} = 0$ throughout the paper.

    \item
    Protocol \ref{prot:bipartite} is inefficient in terms of the amount of classical communication it requires.
    Specifically, the protocol could be altered such that all Pauli corrections are only performed after creating a GHZ-like state shared between all end nodes.
    Additionally, in the case of deterministic BSMs,
    the 2-switch does not need to inform the end nodes about the success of the measurements.
    In this paper, however, we make the assumption that the exchange of classical messages is instantaneous ($t_\text{cl}=0$).
    Therefore, any inefficiency with respect to classical communication does not affect the results presented here.

\end{itemize}

We have studied the performance of Protocol \ref{prot:bipartite} numerically using quantum-network simulator NetSquid \cite{coopmansNetSquidNETworkSimulator2021}.
NetSquid is able to track time-dependent noise accurately by jumping through a timeline consisting of discrete events, at which quantum states are acted upon to account for errors.
On top of NetSquid, our simulations utilize user-contributed NetSquid snippets \cite{netsquid-magic, netsquid-netconf}.
Apart from using NetSquid to study Protocol \ref{prot:bipartite}, we also set up a NetSquid simulation to study Protocol \ref{prot:factory}.
This simulation model serves two purposes.
First, it is used to verify the accuracy of the analytical results presented in Section \ref{sec:estimates}.
This verification is described in Appendix \ref{app:verification}.
Second, simulations of Protocol \ref{prot:factory} are used in this section to compare the performance of Protocols \ref{prot:factory} and \ref{prot:bipartite}.
Note that it would also have been possible to compare simulations of Protocol \ref{prot:bipartite} to our leading-order expressions for Protocol \ref{prot:factory}.
Instead, we are comparing simulations to simulations.
This makes the results of this section independent of the importance of subleading terms that are not included in the leading-order expressions.
\\

Every numerical value that is reported in this paper,
either for Protocol \ref{prot:factory} or for Protocol \ref{prot:bipartite},
is based on the simulation of 10,000 protocol executions.
Error bars on the rate and fidelity represent the standard deviation of the mean,
and are sometimes smaller than the marker size.
Additionally, we remark that when simulating Protocol \ref{prot:bipartite},
the network state is not reset between executions of the protocol.
It can happen that there are Bell states in the network, generated during Step 1, that never feed into a BSM during Step 2 and are thus not used to create a GHZ state.
Then, there are already Bell states present in the network at the start of the next protocol execution.
This entanglement is used as a resource to create the next GHZ state.
\\

While comparing Protocols \ref{prot:factory} and \ref{prot:bipartite},
we observe the relative sensitivity of their performance to the various parameters describing their setups.
This comparison can help us understand in what parameter regimes the use of a factory node can be beneficial.
Throughout the comparison, we use $\Delta t = 1$ to make the results independent of specific time scales.
As a result, the rate is a dimensionless quantity, and can be interpreted as ``average number of GHZ states distributed per round''.
Our comparison will focus on the regime $q_\text{link} \ll 1$.
Only at the end of this Section will we briefly study what happens for $q_\text{link} \sim 1$.
\\

\begin{figure}[!]
    \includegraphics[width=\columnwidth]{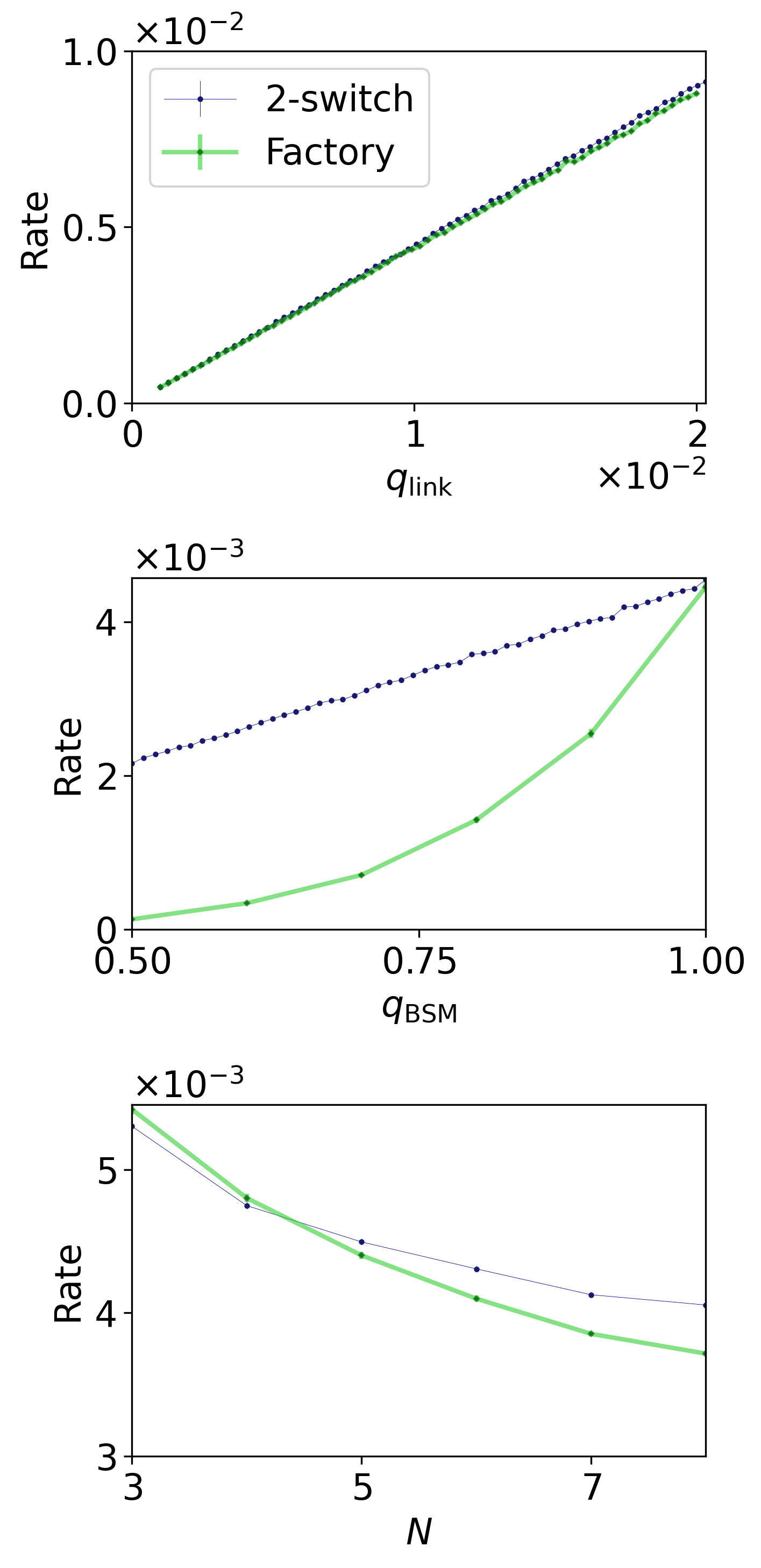}
    \caption{
    Sensitivity of the rate of Protocols \ref{prot:factory} (``Factory'') and \ref{prot:bipartite} (``2-switch'')
    to
    the success probability of Bell-state distribution $q_\text{link}$,
    the number of end nodes $N$, and
    the BSM success probability $q_\text{BSM}$.
    When the parameters are not varied over, their values are $q_\text{link}=0.01$, $N=5$ and $q_\text{BSM} = 1$.
    % For Protocol \ref{prot:bipartite}, we show the results of Monte Carlo simulations (10,000 samples per data point).
    We see that for small values of $q_\text{link}$,
    the rates are of similar magnitude for $q_\text{BSM} = 1$ and $N = 5$, with Protocol \ref{prot:bipartite} slightly outperforming Protocol \ref{prot:factory}.
    If either $q_\text{BSM}$ is decreased or $N$ is increased, this difference becomes more pronounced.
    Note that the lines in the top figure can be hard to distinguish because of their overlap.
    The rate is dimensionless as the round time $\Delta t$ has been set to 1.
    }
    \label{fig:comparison:rate}
\end{figure}

First, we compare the rates of the two protocols.
Since noise parameters of the setups cannot affect the rate at which GHZ states are distributed (only the fidelity),
we limit our attention to the effects of the success probability of Bell-state distribution $q_\text{link}$,
the BSM success probability $q_\text{BSM}$,
and the number of end nodes $N$.
Their effects are shown in Figure \ref{fig:comparison:rate}.
From this figure, we must conclude that for small $q_\text{link}$ Protocol \ref{prot:bipartite} typically has a higher rate than Protocol \ref{prot:factory}.
It is notable that the difference in rate becomes large especially for probabilistic BSMs,
as the rate of Protocol \ref{prot:factory} drops exponentially as $q_\text{BSM}$ is decreased.
However, also for deterministic BSMs Protocol \ref{prot:factory} tends to be slower than Protocol \ref{prot:bipartite}, especially for larger values of $N$.
This can be surprising, considering that Protocol \ref{prot:bipartite} requires a larger total number of Bell states to be distributed than Protocol \ref{prot:factory}
($2(N-1)$, as opposed to $N$ for Protocol \ref{prot:factory}).
The reason for this is that, as discussed above,
Bell states that are generated but not used during one execution of Protocol \ref{prot:bipartite} can still be used during the next execution.
In Protocol \ref{prot:bipartite}, BSMs are executed continuously at the central node, thereby freeing up qubits.
This allows quantum connections to generate multiple Bell states during a single execution of Protocol \ref{prot:bipartite},
which is not the case for Protocol \ref{prot:factory}.
Combining this with the possibility to distribute Bell states ahead of time for the next GHZ state
allows Protocol \ref{prot:bipartite} to use its quantum connections more efficiently than Protocol \ref{prot:factory},
to such a degree that the larger number of Bell states can be distributed in a smaller amount of time.
\\

\begin{figure}[h]
    \includegraphics[width=\columnwidth]{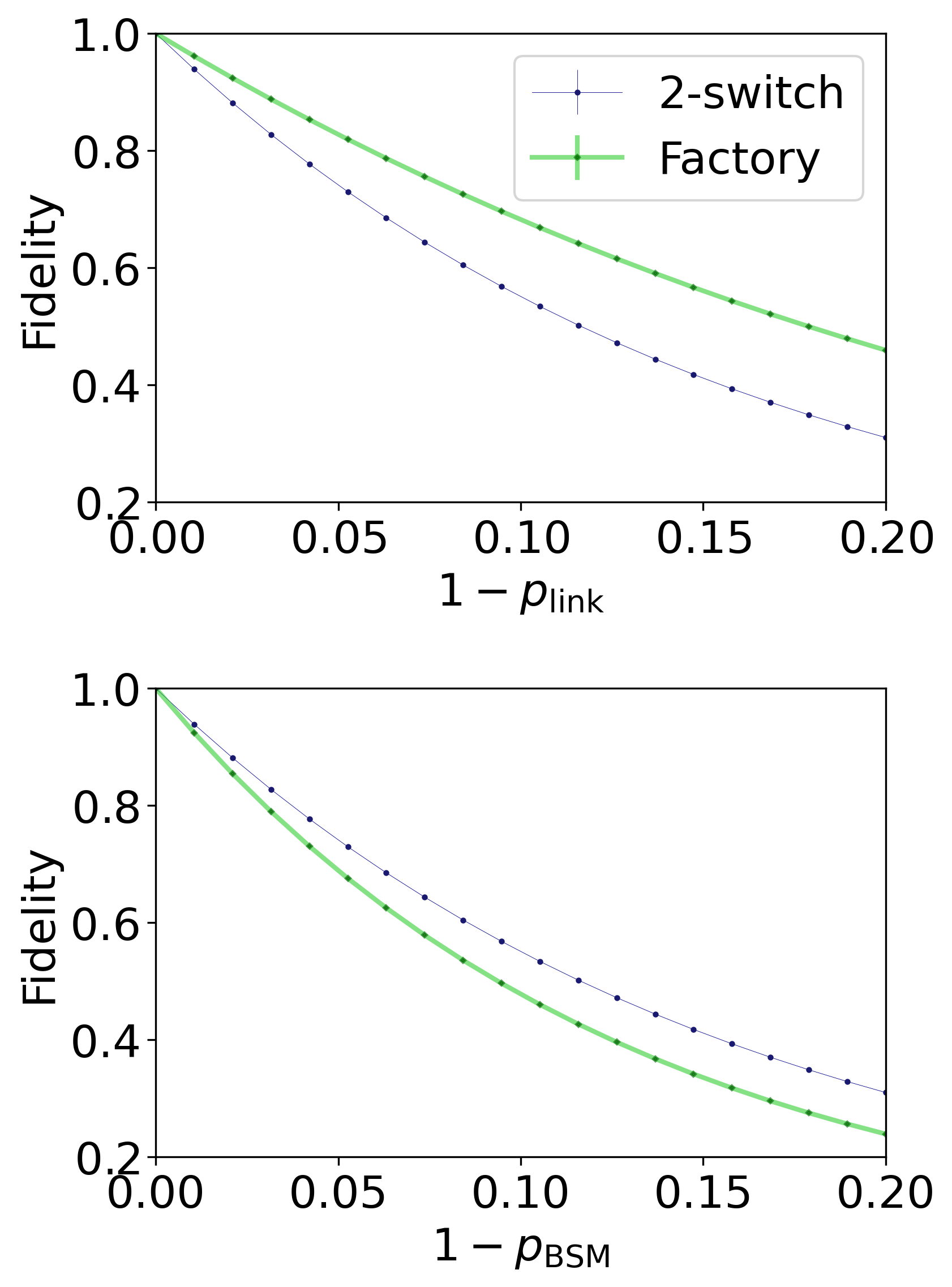}
    \caption{
    Sensitivity of the fidelity of Protocols \ref{prot:factory} (``Factory'') and \ref{prot:bipartite} (``2-switch'') to
    the noise in Bell states shared between the central node and the end nodes ($p_\text{link}$) and
    the noise in BSMs ($p_\text{BSM}$).
    Apart from the parameter varied over,
    there are no sources of noise ($p_\text{link} = p_\text{BSM} = p_\text{mem} = p_\text{GHZ} = 1$).
    The other parameters have the values $q_\text{link} = 0.01$, $N = 5$ and $q_\text{BSM} = 1$.
    % For Protocol \ref{prot:bipartite}, we show the results of Monte Carlo simulations (10,000 samples per data point).
    While Protocol \ref{prot:factory} is more resilient against noise in Bell states,
    Protocol \ref{prot:bipartite} is more resilient against noise in BSMs.
    }
    \label{fig:comparison:plink_pbsm}
\end{figure}

Now, we compare the fidelities of the two protocols.
From Figure \ref{fig:comparison:plink_pbsm},
we see that Protocol \ref{prot:bipartite} is more sensitive to the noise parameter $p_\text{link}$.
This is explained by the fact that it requires more Bell states between the central node and end nodes to distribute a single GHZ state
($2(N-1)$ instead of $N$).
Additionally, we see that Protocol \ref{prot:factory} is more sensitive to $p_\text{BSM}$.
The reason for this, is that the protocol executes more successful BSMs per GHZ state than Protocol \ref{prot:bipartite}
($N$ vs $N - 1$).
We note though that Protocol \ref{prot:bipartite} also requires the execution of fusion operations at the end nodes,
consisting of a CNOT gate and one Z-basis measurement.
As a deterministic BSM can be implemented using a CNOT gate, a Hadamard gate, and two Z-basis measurements,
it could very well be the case that the noise in the fusion operations is of similar magnitude as the noise in the BSMs.
If we would have modeled the fusion operation as also inflicting depolarizing channels with parameter $p_\text{BSM}$ on the involved qubits,
we would likely instead have found that Protocol \ref{prot:bipartite} is more sensitive to $p_\text{BSM}$,
as it requires $N-1$ successful BSMs and $N-2$ fusions, giving a total of $2N-3$ instances at which the noise is suffered.
\\

\begin{figure}[!]
    \includegraphics[width=\columnwidth]{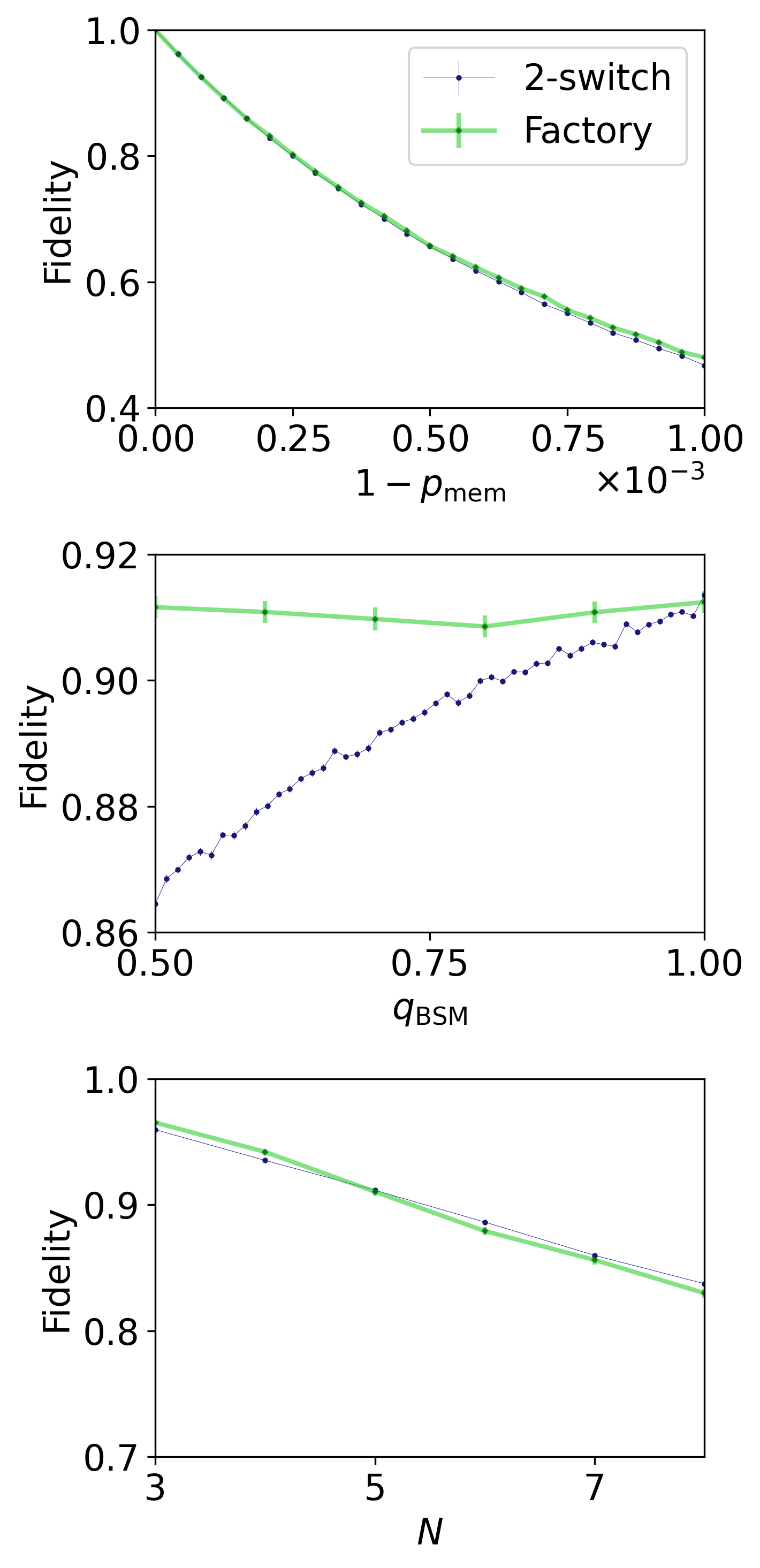}
   \caption{
       Sensitivity of the fidelity of Protocols \ref{prot:factory} (``Factory'') and \ref{prot:bipartite} (``2-switch'') to
       the memory depolarizing parameter $p_\text{mem}$,
       the number of end nodes $N$, and
       the BSM success probability $q_\text{BSM}$,
       when the only source of noise is memory decoherence ($p_\text{link} = p_\text{BSM} = p_\text{GHZ} = 1$).
       When the parameters are not varied over,
       their values are $p_\text{mem}=1-10^{-4}$, $q_\text{link} = 0.01$, $N=5$ and $q_\text{BSM} = 1$.
       % For Protocol \ref{prot:bipartite}, we show the results of Monte Carlo simulations (10,000 samples per data point).
       We see that when both $q_\text{link}$ and $1 - p_\text{mem}$ are small,
       the fidelities are approximately equal for $q_\text{BSM} = 1$ and $N = 5$.
       When $q_\text{BSM}$ is decreased, this is in favor of Protocol \ref{prot:factory}.
       However, if $N$ is increased, this is slightly in favor of Protocol \ref{prot:bipartite}.
       Note that the lines in the top (and to lesser degree, the bottom) figure can be hard to distinguish because of their overlap.
   }
   \label{fig:comparison:fid_memory}
\end{figure}

The final source of noise that the two setups have in common is the memory decoherence, $p_\text{mem}$.
How much decoherence enters into the final GHZ state depends on the amount of time qubits are stored while executing the protocol.
Therefore, it is reasonable to expect that the amount of memory decoherence behaves similar to the rate.
Comparing Figures \ref{fig:comparison:rate} and \ref{fig:comparison:fid_memory} reveals that indeed for both the rate and the memory decoherence,
both setups perform comparably well for small $q_\text{link}$, $N=5$ and $q_\text{BSM} = 1$ (and small $1 - p_\text{mem}$).
For the rate, increasing $N$ is in favor of Protocol \ref{prot:bipartite}.
Similarly, the amount of memory decoherence seems to scale more favourably with $N$ for Protocol \ref{prot:bipartite} than for Protocol \ref{prot:factory},
although the difference is not as pronounced as for the rate.
The effect of $q_\text{BSM}$, however, is reversed between the rate and memory decoherence.
While the amount of memory decoherence suffered in Protocol \ref{prot:factory} is unaffected by decreasing $q_\text{BSM}$,
it does affect the performance of Protocol \ref{prot:bipartite}.
The reason for this, is that while Protocol \ref{prot:factory} is reset upon a failed BSM,
the same is not true for Protocol \ref{prot:bipartite}.
This makes Protocol \ref{prot:bipartite} more resilient to failing BSMs in terms of rate, but less so in terms of fidelity.
\\

\begin{figure}[h]
    \includegraphics[width=\columnwidth]{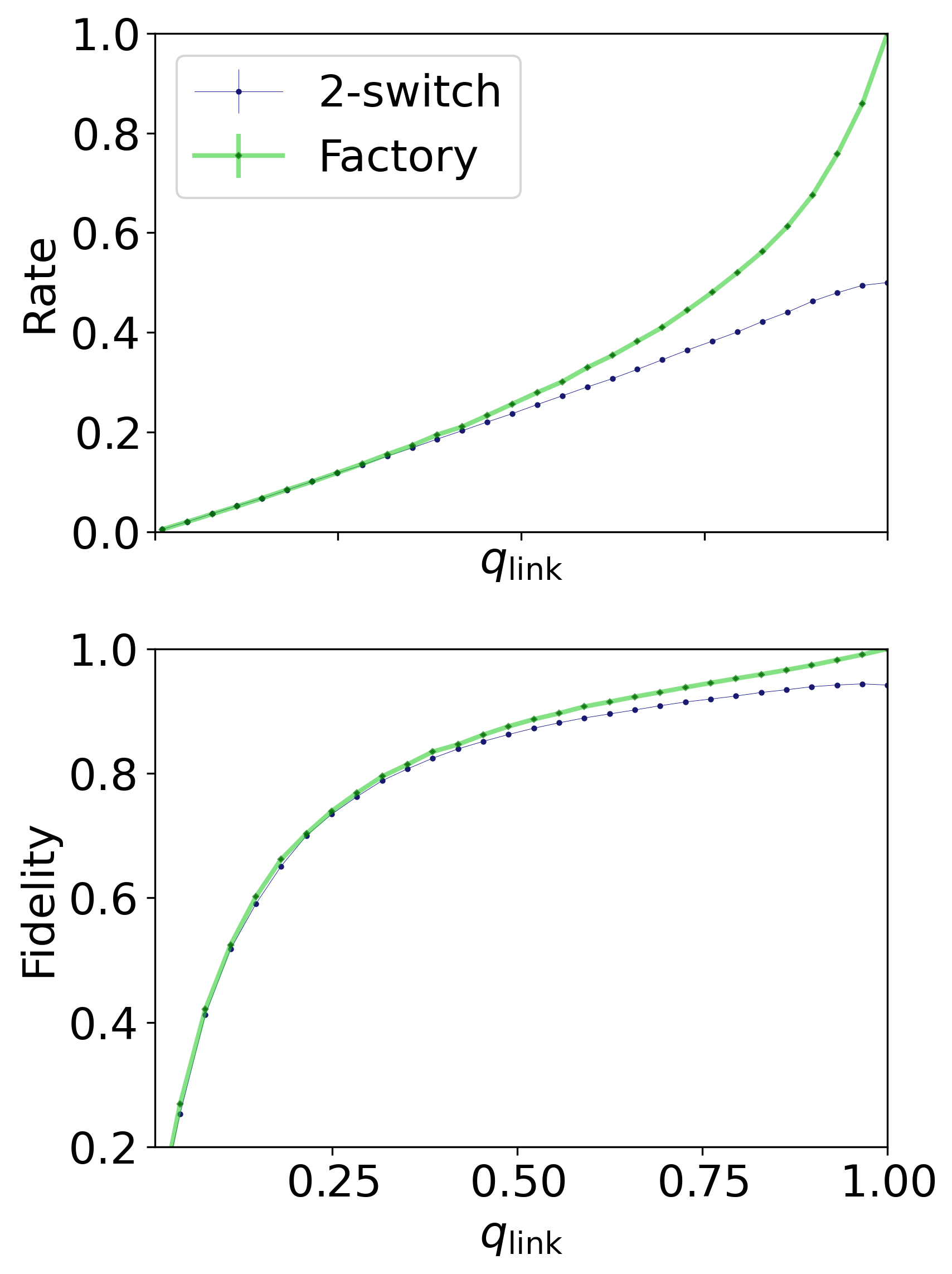}
    \caption{
       Sensitivity of both the
       rate and
       fidelity of Protocols \ref{prot:factory} (``Factory'') and \ref{prot:bipartite} (``2-switch'') to
       the success probability of Bell-state distribution $q_\text{link}$,
       when the only source of noise is memory decoherence ($p_\text{link} = p_\text{BSM} = p_\text{GHZ} = 1$).
       The other parameters are set to $p_\text{mem}=1-10^{-2}$, $N=5$ and $q_\text{BSM} = 1$.
       We see that while both protocols have similar performance for $q_\text{link} \ll 1$,
       Protocol \ref{prot:factory} wins out both in terms of rate and fidelity for $q_\text{link} \sim 1$.
       The rate is dimensionless as the round time $\Delta t$ has been set to 1.
    }
    \label{fig:comparison:qlink_large_range}
\end{figure}

Finally, we observe what happens to both the rate and the memory decoherence if $q_\text{link}$ is increased beyond the $q_\text{link} \ll 1$ regime we have studied so far.
It is seen in Figure \ref{fig:comparison:qlink_large_range} that the similarity in performance for $N=5$ and $q_\text{BSM} = 1$ observed for small values of $q_\text{link}$
disappears for larger values;
here, Protocol \ref{prot:factory} outperforms Protocol \ref{prot:bipartite} with respect to both metrics.
We note that for $q_\text{link}=1$, the rate of Protocol \ref{prot:factory} becomes one, as it takes exactly one round to distribute all $N$ Bell states.
On the other hand, the rate of Protocol \ref{prot:bipartite} becomes approximately one half, as it takes one round to distribute $N$ Bell states, and then another round to distribute the remaining $N-2$ Bell states.
This also explains the difference in fidelity for large values of $q_\text{link}$.
Note that Protocol \ref{prot:bipartite} had the advantage of using quantum connections more efficiently for small $q_\text{link}$
because an excess number of Bell states can be distributed during one protocol execution to be used during the next.
However, this advantage largely disappears for large values of $q_\text{link}$.
When all Bell states required to create a GHZ state are generated in quick succession,
there is not much  ``spare time'' during which these excess Bell states can be generated.
We remark that for $q_\text{link} \sim 1$, the classical-communication time $t_\text{cl}$ could have a large effect on both the rate and the amount of memory decoherence.
We have assumed it to be zero because for $q_\text{link} \ll 1$, the classical communication time becomes negligible compared to the time required to distribute a Bell state successfully.
This might or might not be true for larger values of $q_\text{link}$.
Therefore, we cannot draw definitive conclusions about the relative performance between the two protocols
for large values of $q_\text{link}$ from Figure \ref{fig:comparison:qlink_large_range}.

\section{Physical Implementation} \label{sec:implementation}

In this section, we discuss different ways factory nodes capable of creating GHZ states could be physically realized.
First, we discuss how they could be implemented using trapped ions in Section \ref{subsec:TI},
and then we discuss in Section \ref{subsec:NV} how they could be implemented using nitrogen-vacancy centers in diamond.

\subsection{Trapped Ions} \label{subsec:TI}

The first physical implementation we discuss is based on trapped ions \cite{harocheExploringQuantumAtoms2006}.
In an ion trap, charged atoms are suspended in an electromagnetic field.
The energy levels of the ions can be used to define qubits,
and these qubits can be manipulated by driving them with laser pulses.
Trapped ions have properties that would make them suitable to implement a factory node,
such as long coherence times \cite{wangSingleIonQubit2021, bermudezAssessingProgressTrappedIon2017, bruzewiczTrappedionQuantumComputing2019},
high-fidelity state preparation and readout \cite{hartyHighFidelityPreparationGates2014, roosDesignerAtomsQuantum2006, myersonHighFidelityReadoutTrappedIon2008},
and a good optical interface \cite{vogellDeterministicQuantumState2017, krutyanskiyLightmatterEntanglement502019, borneEfficientIonphotonQubit2020, meranerIndistinguishablePhotonsTrappedion2020, schuppInterfaceTrappedIonQubits2021, connellIonPhotonicFrequencyQubit2021, walkerImprovingIndistinguishabilitySingle2020}
that has allowed for the generation of entanglement with remote nodes
\cite{stephensonHighRateHighFidelityEntanglement2020, slodickaAtomAtomEntanglementSinglePhoton2013, moehringEntanglementSingleatomQuantum2007}.
\\

One quantum gate that can be executed on trapped ions is the M\o{}lmer-S\o{}rensen (MS) gate \cite{sorensenEntanglementQuantumComputation2000, schindlerQuantumInformationProcessor2013}.
This gate affects all qubits in the trap, and can be used to map maximally entangled GHZ-like states to computational-basis states.
In combination with single-qubit Z-basis measurements, the MS gate can therefore be used to execute a GHZ-basis measurement on all qubits.
We note that throughout this paper we have assumed the factory node creates a GHZ state locally,
and then executes BSMs between qubits of the GHZ state and qubits that are entangled to qubits at the end nodes.
However, the same result is acquired (i.e., the creation of a GHZ state shared between the end nodes) when executing a GHZ-basis measurement on the qubits that are entangled to the end nodes,
given that appropriate Pauli corrections are performed at the end nodes based on the outcome of the measurement.
\\

\begin{figure}[!]
    \includegraphics[width=0.7\linewidth]{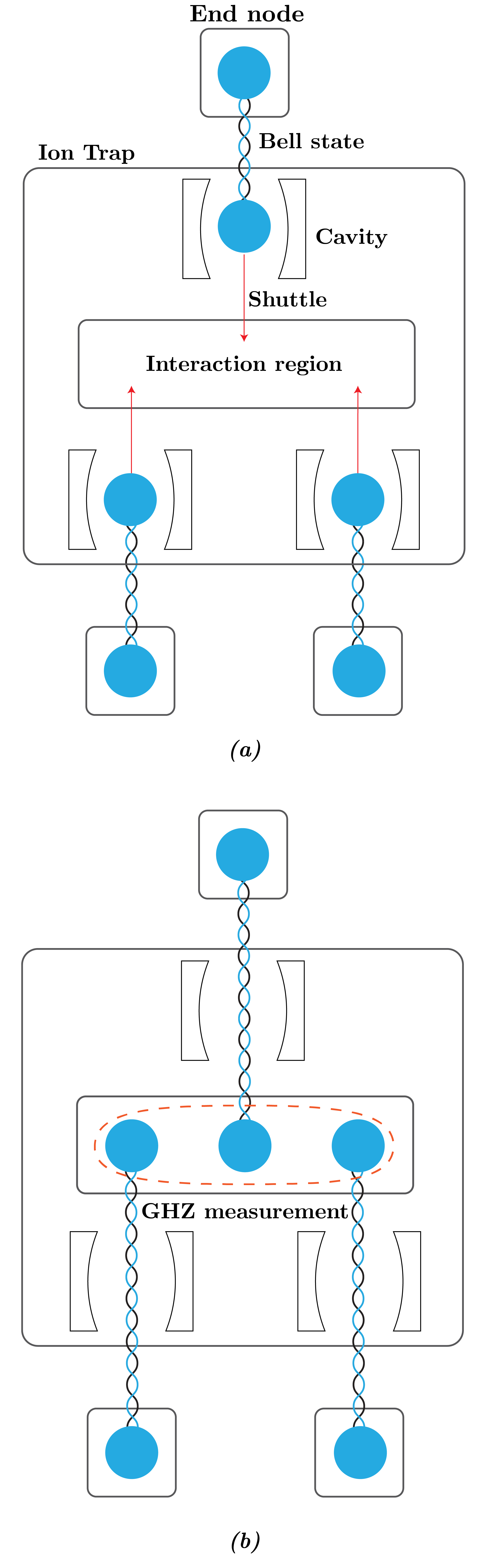}
    \caption{
        Implementation example of a factory node capable of distributing GHZ states based on trapped ions.
        \textbf{(a)}
        Single ions in cavities provide optical interfaces, allowing for Bell-state distribution with all $N=3$ end nodes.
        After all ions are entangled, they are shuttled to an interaction region.
        \textbf{(b)}
        At the interaction region, a GHZ measurement is executed using an MS gate and single-qubit measurements,
        which has the effect of creating a GHZ state shared by the end nodes.
    }
    \label{fig:implementation:ion_trap}
\end{figure}

We note that an additional challenge when using trapped ions to realize a factory node is that $N$ different ionic qubits in the same device need to participate in simultaneous Bell-state distribution with end nodes.
One potential method to allow for a good photonic interface with individual ions is to use shuttling techniques
\cite{kielpinskiArchitectureLargescaleIontrap2002, sangouardQuantumRepeatersBased2009, monroeScalingIonTrap2013, pfisterQuantumRepeaterNode2016, pinoDemonstrationTrappedionQuantum2021, leeIonShuttlingMethod2021, kaushalShuttlingbasedTrappedionQuantum2020}.
This way, ions could be physically moved to separate cavities, where they can be made to emit entangled photons suitable for Bell-state distribution.
After ions have been successfully entangled, they can be shuttled to an interaction region where the GHZ-basis measurement is executed.
This setup is illustrated in Figure \ref{fig:implementation:ion_trap}.
Potentially, different ion species could be used for generating and storing entanglement,
such that for each task the species can be selected with the most favourable properties \cite{santraQuantumRepeatersBased2019, dharaMultiplexedQuantumRepeaters2022}.
\\

\subsection{Nitrogen-Vacancy Centers} \label{subsec:NV}

The second physical implementation of factory nodes we discuss is based on nitrogen-vacancy (NV) centers in diamond \cite{rufResonantExcitationPurcell2021, pompiliRealizationMultinodeQuantum2021, kalbEntanglementDistillationSolidState2017, humphreysDeterministicDeliveryRemote2018, hensenLoopholefreeBellInequality2015, bernienHeraldedEntanglementSolidstate2013, rozpedekNeartermQuantumrepeaterExperiments2019}.
An NV center provides an electronic communication qubit that can be used as optical interface,
and is surrounded by Carbon-13 nuclear spins that can be used as memory qubits.
NV centers were used to perform the first loophole-free Bell test \cite{hensenLoopholefreeBellInequality2015},
have been used to demonstrate entanglement distillation between remote nodes \cite{kalbEntanglementDistillationSolidState2017},
and have recently been used to construct the first three-node quantum network \cite{pompiliRealizationMultinodeQuantum2021}.
\\

A downside to NV centers is that they only provide a single communication qubit.
Although entanglement can in principle be stored in $N$ memory qubits,
$N$ Bell states cannot be distributed simultaneously,
which is a prerequisite for Protocol \ref{prot:factory}.
If the time required to perform a single attempt at Bell-state distribution with a remote node, $\Delta t$,
is much larger than the time it takes to emit an entangled photon and transfer a state to a carbon atom,
temporal multiplexing could potentially be used to perform $N$ entangling attempts during a single round \cite{vandamMultiplexedEntanglementGeneration2017}.
After Bell states have been established with all $N$ end nodes,
a GHZ-basis measurement can be executed within the NV center \cite{vandamMultipartiteEntanglementGeneration2019}.
\\

\begin{figure}[!]
    \includegraphics[width=0.8\linewidth]{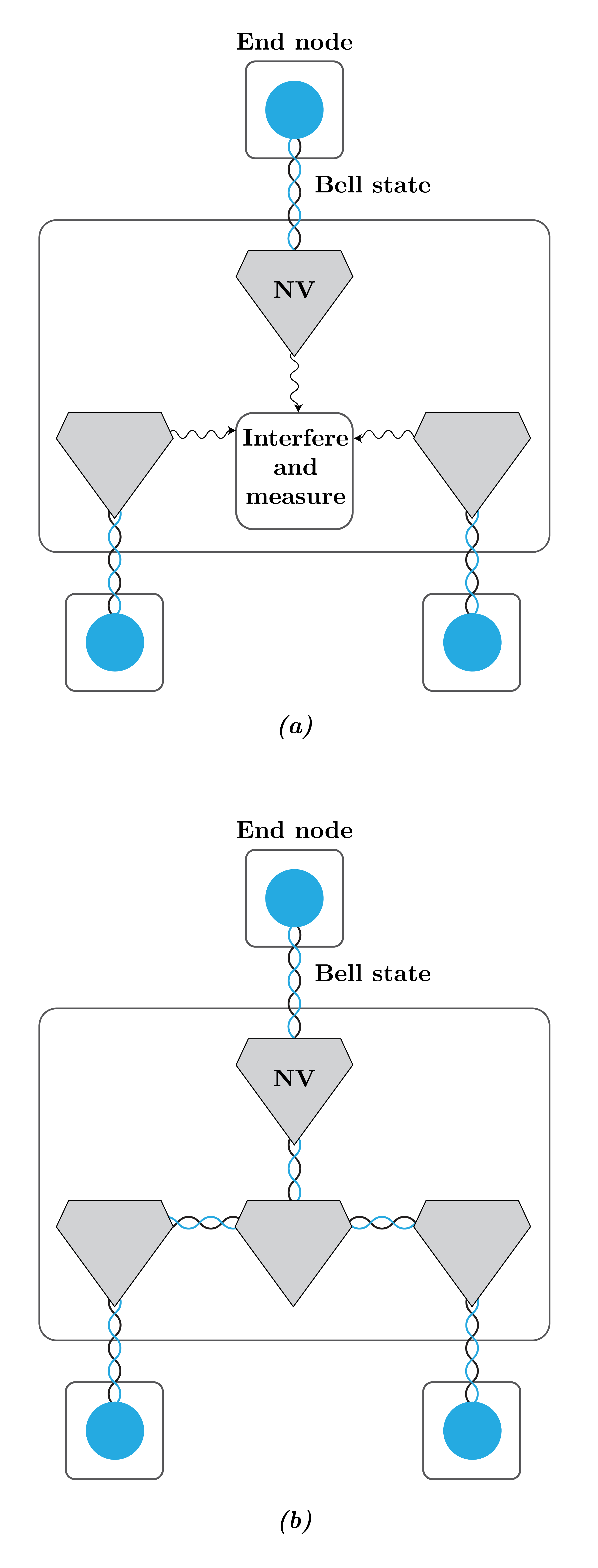}
    \caption{
        Implementation examples of factory nodes capable of distributing GHZ states based on NV centers in diamond.
        Within the factory node, $N=3$ NV centers distribute and store entanglement with the end nodes.
        When all these NV centers are entangled, a GHZ state is distributed between them,
        after which each executes a BSM to teleport the GHZ state to the end nodes.
        \textbf{(a)}
        The GHZ state can be distributed between the NV centers by emitting entangled photons, interfering these photons, and measuring them.
        \textbf{(b)}
        The GHZ state can be distributed between the NV centers by first creating Bell states between all $N$ NV centers and one additional NV center.
        Then, a GHZ measurement is executed at this NV center.
    }
    \label{fig:implementation:NV}
\end{figure}

If temporal multiplexing is not feasible, however,
a factory node could be realized from $N$ separate NV centers.
Each NV center can then be dedicated to creating and storing Bell states with a single end node.
When all Bell states are in place, a GHZ state needs to be distributed between the $N$ NV centers,
after which deterministic BSMs can be executed.
We here discern two methods of generating this GHZ state.
The first is to interfere and measure entangled photons emitted by all $N$ NV centers \cite{capraravivoliHighfidelityGreenbergerHorneZeilingerState2019, wangSchemesGenerationMultipartite2009a}.
This is illustrated in Figure \ref{fig:implementation:NV} (a).
However, the success probability of such schemes drops exponentially with $N$,
and thus many attempts may be needed to generate a single GHZ state.
Apart from having a negative influence on the rate of GHZ-state distribution for large $N$,
this can also be expected to severely degrade the fidelity of the final GHZ state,
as the memory qubits undergo decoherence each time the communication qubit is interfaced with \cite{kalbDephasingMechanismsDiamondbased2018}.
An alternative method that circumvents this exponential scaling,
is to add one more NV center to the factory node.
After all Bell states are in place, each of the $N$ outward facing NV centers can generate a Bell state with the extra NV center.
Then, the extra NV center can execute a GHZ-basis measurement on the entangled qubits it has stored,
thereby creating a GHZ state between the $N$ outward-facing NV centers.
Because Bell states can be generated with each outward-facing NV center sequentially,
the number of required attempts will scale linearly with $N$.
This can be thought of as a ``factory within a factory'' approach,
and is illustrated in Figure \ref{fig:implementation:NV} (b).
Using a single NV center as a factory within a factory could be feasible even when using a single NV center as the entire factory node is not.
The reason for this is that Bell-state distribution between NV centers located within the same node can happen at smaller time scales than with remote end nodes.

\section{Conclusion} \label{sec:conclusion}

In this paper, we have studied the distribution of multipartite entangled states in networks through local preparation of the target state at a factory node,
and subsequent quantum teleportation of the state to a set of end nodes.
We have presented two main results.
First, we have derived analytical results for the rate and fidelity of GHZ-state distribution on a symmetrical star-shaped network,
with a factory node at the center.
Second, we have compared the rate and fidelity to what is achievable on the same setup without a factory node,
using a 2-switch that is only capable of executing BSMs instead.
\\

From the comparison,
we found that the use of a factory node provides more resilience to noise in Bell states that are distributed between the central node and end nodes.
Furthermore, when BSMs at the central node are not deterministic,
using a factory node provides better protection against memory decoherence.
We note that two additional advantages of using a factory node are that it only requires the end nodes to store a single qubit,
while using a 2-switch requires more quantum capabilities of the end nodes,
and that it can be used to distribute any multipartite target state using the same method,
while the 2-switch protocol is specific to GHZ states.
However, the results are not all in favor of the factory node.
The 2-switch attains exponentially higher rates when BSMs are probabilistic,
is less sensitive to noise in BSMs,
and both the rate and (to lesser extent) the sensitivity to memory decoherence scale more favourably with the number of end nodes.
We note that no thorough search for an optimal protocol utilizing a 2-switch has been performed,
and doing so could boost performance even further.
For example, it might be possible to increase performance by incorporating cutoff times in the protocol, that is, by discarding Bell states when they have undergone too much memory decoherence \cite{liEfficientOptimizationCutoffs2021, santraQuantumRepeaterArchitecture2019, kozlowskiDesigningQuantumNetwork2020a, rozpedekParameterRegimesSingle2018}.
Cutoff times are expected to increase the fidelity, but at the cost of having a smaller rate.
However, it must be noted that we have also not optimized the factory-node protocol.
Also for this protocol e.g. cutoff times could be introduced.
As discussed in Section \ref{sec:introduction},
various protocols and network architectures that have been proposed in earlier work make use of factory nodes.
We conclude that when hardware limitations are present,
depending on the nature and severity of those limitations,
it could be worthwhile to consider other types of central nodes instead.
\\

One of our motivations for studying the factory node is to allow for assessment of proposed schemes involving factory nodes in the presence of hardware limitations.
We consider the analytical results presented in this paper a first step towards better assessment.
However, we have made various assumptions that limit the scope of applicability.
Here, we discuss how some of these assumptions could be removed.
First, all the results in this paper assume the star-shaped network is symmetric,
meaning that noise parameters are the same for each end node (same coherence time, same Bell-state fidelity, and same quality of BSMs),
and that attempts at Bell-state distribution take the same amount of time and have the same success probability for each end node.
With respect to the calculation of fidelity,
the assumption of same noise parameters can straightforwardly be removed within the framework of the analysis presented in this paper.
In Section \ref{sec:fidelity}, when evaluating Eq. \eqref{eq:factory_fidelity},
an average should be taken over all possible orderings in which end nodes generate a Bell state with the factory node.
Because of the assumption of symmetry, we were able to avoid performing such an average explicitly,
but in principle there is nothing preventing us from doing so.
Then, each of the terms in this average can be evaluated using the Eqs. \eqref{eq:app:G_final} and \eqref{eq:app:G_bound}
(or Eqs. \eqref{eq:factory_fidelity_approx} and \eqref{eq:factory_fidelity_bound} in case $p_\text{mem}$ is the same for each qubit in the network).
On the other hand,
it is a key assumption in the results of Appendix \ref{app:exp_val} that the success probability of Bell-state distribution is the same for each connection.
Removing this assumption, therefore, would be less straightforward and could provide an interesting subject for future research.
The same holds for the assumption that the attempt durations are the same for each connection.
\\

Second, all the results in this paper are specific to the distribution of GHZ states.
However, Protocol \ref{prot:factory} could also be used to distribute other states,
as long as they can be prepared locally
and consist of exactly one qubit per end node.
The analytical results for the rate that are presented in Section \ref{sec:rate} are applicable for the distribution of any such state,
as the time that each step takes in Protocol \ref{prot:factory} does not depend on the specific quantum state,
nor does the success probability of the teleportation procedure.
For the analytical fidelity results that are presented in Section \ref{sec:fidelity},
we note that the final distributed state will be equal to the target state but with the individual qubits depolarized with the parameters $p_i$ given by Eq. \eqref{eq:fidelity:p_i},
and the full state depolarized with a parameter that was called $p_\text{GHZ}$ in the GHZ case (analogously to Eq. \eqref{eq:fidelity:state}).
The fidelity of this state as a random variable is a weighted sum over products of depolarizing parameters (analogously to Eqs. \eqref{eq:fidelity:fidelity_not_yet_rearranged} and \eqref{eq:fidelity:coefficients}).
Here, the weights depend on the fidelity of the state after specific sets of qubits undergo depolarizing errors.
The expected values of these products of depolarizing parameters can be evaluated using Eqs. \eqref{eq:factory_fidelity_approx} and \eqref{eq:factory_fidelity_bound}.
Therefore, the only ingredient missing to determine the lower bound or leading-order expression for the fidelity in case of a different target state,
are the weights that appear in the fidelity.
We note that in case the target state is not invariant under qubit permutations, the symmetry of the setup is broken.
In that case, an explicit average should be taken over the different orders in which Bell states can be distributed, as discussed above.
\\

% A final assumption that we are not aware how to circumvent, is that entanglement generation needs to occur according to discrete attempts with a constant success probability
% (i.e. the time required to distribute a Bell state is geometrically distributed).
% This is, for example, not the case when entanglement is generated using nested quantum-repeater schemes \cite{briegelQuantumRepeatersRole1998, duanLongdistanceQuantumCommunication2001}.
% \\

The leading-order expressions and lower bounds presented in this paper are accurate when
the success probability per attempt at Bell-state distribution ($q_\text{link}$) is small,
and when the probability of losing a qubit to the environment when storing it in memory during a single attempt ($1 - p_\text{mem}$) is small.
When the first assumption holds, the second typically also holds;
otherwise, qubits need to be stored in memory during many attempts as new states are generated,
and if the probability of losing the qubit is large already for a single attempt,
then the final distributed state will not be entangled.
The parameter regime of small $q_\text{link}$ but large $1 - p_\text{mem}$ is therefore not very interesting to study.
E.g. for heralded entanglement generation,
the success probability per attempt is expected to be small because of photon (attenuation) losses.
However, there are also physical setups for which the assumption does not hold,
such as quantum-repeater chains making use of error correction \cite{munroQuantumCommunicationNecessity2012, muralidharanUltrafastFaultTolerantQuantum2014, inside_quantum_repeaters, borregaardOneWayQuantumRepeater2020, rozpedekQuantumRepeatersBased2021, azumaAllphotonicQuantumRepeaters2015, pantRatedistanceTradeoffResource2017, fukuiAllOpticalLongDistanceQuantum2021}
or massive multiplexing \cite{sinclairSpectralMultiplexingScalable2014, guhaRatelossAnalysisEfficient2015, seriQuantumStorageFrequencyMultiplexed2019},
for which the success probability is close to one.
For such setups, the approximations presented in this paper are not applicable,
although we have found that our leading-order expression for the fidelity is remarkably accurate for large values of $q_\text{link}$.
Additionally, we note that setups for which the quantum connections are near deterministic can be approximated by assuming they are fully deterministic.
In this case, the protocol becomes easy to analyze, as no probabilities need to be accounted for.
\\

Now, we discuss how the techniques presented in this paper can be used to study the performance of quantum-network protocols different from the one we have studied.
An entanglement switch is a central node that is able to generate Bell states shared with $k$ end nodes,
and executes local GHZ-state measurements on groups of $n$ entangled qubits.
As remarked in Section \ref{sec:introduction}, the factory-node setup studied in this paper is equivalent to an entanglement switch with $n=k$.
A possible extension of the calculations in this paper is to apply them also to entanglement switches for which $n<k$.
In Appendix \ref{app:rate},
we present a leading-order expression for the maximum switching rate for any value of $n$ when there is a single qubit of buffer memory per end node.
However, it would be especially interesting to study the fidelity of states produced by the entanglement switch,
as there are almost no known results about this.
Such an extension of the fidelity calculation, assuming a symmetric star-shaped network and one qubit of buffer memory per end node,
could be realized by repeating the calculation in Section \ref{sec:fidelity} and replacing the parameter $N$ (the number of end nodes, equal to $k$) by $n < N$
in Eq. \eqref{eq:factory_fidelity},
but not replacing it in Eq. \eqref{eq:factory_fidelity_approx} (which is needed to evaluate Eq. \eqref{eq:factory_fidelity}).
Evaluating this expression and verifying it (against a Monte Carlo simulation) is beyond the scope of this paper.
\\

Another possible extension of the work done in this paper,
is the approximation of the rate and fidelity of Bell states distributed by specific types of quantum-repeater chains.
In the factory-node setup, there are $N$ Bell states that are distributed according to geometric distributions.
Entangled states that are established need to be stored in memory until all states are distributed,
after which they are transformed into some target state through BSMs.
If any of the BSMs fails, the protocol is restarted.
The target state is a GHZ state.
Now consider a quantum-repeater chain consisting of $N$ elementary links,
where entanglement swapping (i.e. BSMs) is only executed after entangled states have been distributed on all links.
If any of the BSMs fail, all entanglement is discarded and Bell-state distribution starts anew.
This is then exactly the same scenario as for the factory node, only the target state is not a GHZ state but a bipartite state.
For the rate of such a repeater chain, analytical results similar to ours already exist \cite{coopmansImprovedAnalyticalBounds2022, schmidtMemoryassistedLongdistancePhasematching2020, shchukinWaitingTimeQuantum2019}.
\\

The fidelity of Bell states distributed by such a repeater protocol can however also be analyzed using the techniques presented in this paper.
The expression for the state's fidelity in terms of different depolarizing parameters (Eq. \eqref{eq:factory_fidelity} for the factory node) will look different
(simpler, as all depolarizing noise can be ``moved'' to a single qubit),
but the same type of expected values will need to be evaluated,
allowing for the direct use of Eqs. \eqref{eq:factory_fidelity_approx} and \eqref{eq:factory_fidelity_bound} to obtain a leading-order expression and a lower bound respectively.
Examples of repeater protocols where swapping is only performed after all links are present are
schemes that use error correction to protect against operational errors in the repeater nodes \cite{jiangQuantumRepeaterEncoding2009},
such as the ones studied for NV centers in \cite{jingQuantumRepeatersEncoding2021}.
In \cite{jingQuantumRepeatersEncoding2021}, it is remarked that accounting for depolarizing noise in individual memories is no easy task,
and the authors instead assume each qubit decoheres an amount of time equal to the average waiting time.
In contrast, our techniques, although approximate, do account for the depolarizing noise in each individual qubit.
A similar approach to \cite{jingQuantumRepeatersEncoding2021} is taken in \cite{schmidtMemoryassistedLongdistancePhasematching2020},
where the case of all swaps occurring only in the end is considered to calculate analytical bounds on the decoherence suffered when swaps are performed earlier.
This approximation provides a lower bound on the fidelity by Jensen's inequality.
An interesting direction for further study is to compare the tightness of Jensen's inequality to the lower bound presented in this paper.

\section{Data Availability}

The data presented in this paper has been made available at \url{https://doi.org/10.4121/19235937} \cite{factory-data}.
Scripts that generate all the plots presented in this paper can also be found here.

\section{Code Availability}

All the code used to evaluate the analytical results presented in this paper,
and to perform NetSquid simulations of Protocol \ref{prot:factory} and Protocol \ref{prot:bipartite},
has been made available at \url{https://gitlab.com/softwarequtech/netsquid-snippets/netsquid-factory} \cite{netsquid-factory}.

\section*{Acknowledgements}

We thank Álvaro Gómez Iñesta, Gayane Vardoyan and Tim Coopmans for feedback on the manuscript.
This work was supported by NWO Zwaartekracht QSC 024.003.037, ARO MURI (W911NF-16-1-0349) and NSF (OMA-1936118, EEC-1941583, OMA-2137642).

\bibliography{Analysis_of_Multipartite_Entanglement_Distribution_using_a_Central_Quantum-Network_Node}{}
\bibliographystyle{ieeetr}

%%%%%%%%%%%%%%%%%%%%%%%%%%%%%%%%%%%%%%%%%%%%
%%%%%%%%%%%%% APPENDIX
%%%%%%%%%%%%%%%%%%%%%%%%%%%%%%%%%%%%%%%%%%%%

\onecolumngrid

\appendix

\section{Verification of Analytical Expressions for Rate and Fidelity} \label{app:verification}

In this appendix we verify the analytical results for the rate and fidelity of Protocol \ref{prot:factory},
as presented in Section \ref{sec:estimates},
against Monte Carlo simulations of the protocol.
These simulations have been performed using the quantum-network simulator NetSquid \cite{coopmansNetSquidNETworkSimulator2021}
and user-contributed NetSquid snippets \cite{netsquid-magic, netsquid-netconf}.
The simulation code can be found in the public repository \cite{netsquid-factory}.
Just like in Section \ref{sec:comparison},
we use $\Delta t = 1$ to make the results independent of specific time scales,
each data point is the result of 10,000 simulated executions of the protocol,
and error bars represent the standard deviation of the mean.
Often, the error bars are smaller than the marker size, making them hard to see.
\\

\begin{figure}[h]
    \includegraphics[width=\linewidth]{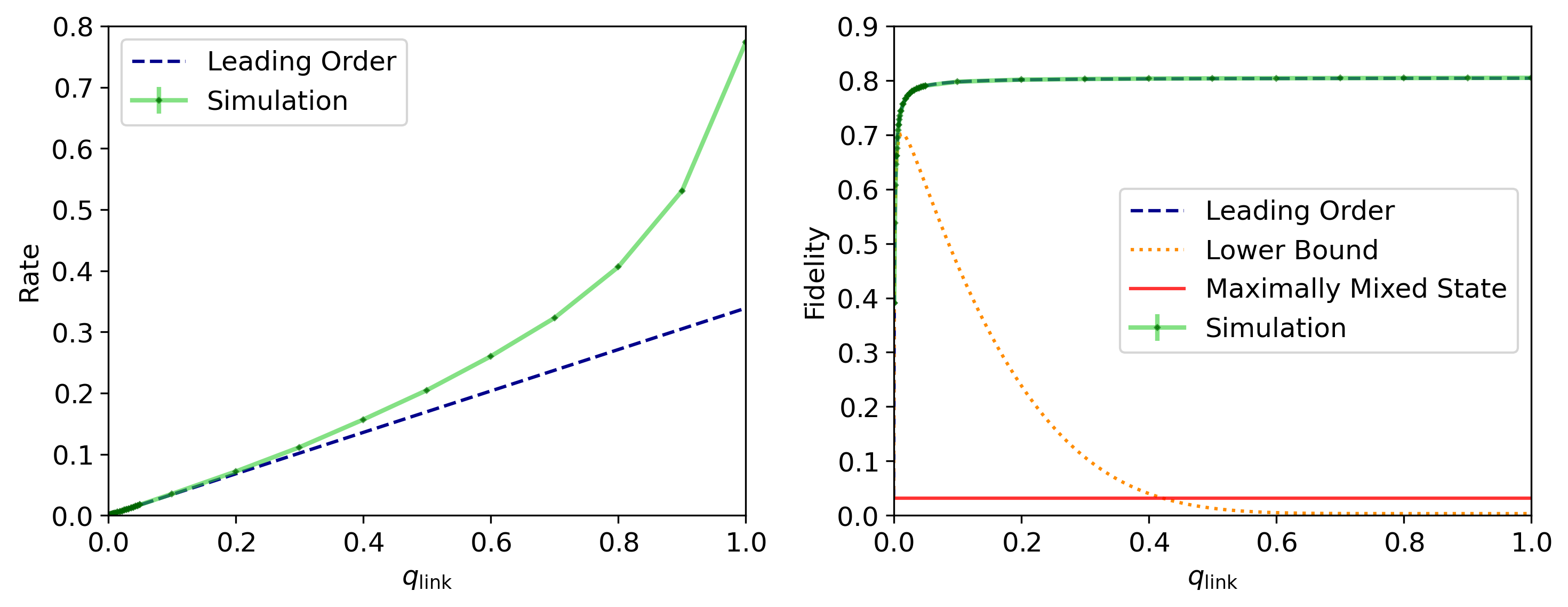}
	\caption
	{
        Comparison between simulation results and analytical expressions for the performance of Protocol \ref{prot:factory} for different values of $q_\text{link}$.
        On the left, the simulated rate is compared to the leading-order expression in Eq. \eqref{eq:factory_rate}.
        On the right, the simulated fidelity is compared to the leading-order expression and lower bound from Section \ref{sec:fidelity}.
        The parameters are $N=5$, $q_\text{BSM} = 0.95$, $p_\text{BSM} = p_\text{link} = 1 - 10^{-2}$ and $p_\text{mem} = 1 - 10^{-4}$.
        GHZ states are locally prepared with a fidelity of 0.9,
        which corresponds to $p_\text{GHZ} \approx 0.872$.
        We see that there is close agreement between analytical results for small values of $q_\text{link}$.
        As $q_\text{link}$ is increased up to a value of one, deviations in the rate grow up to a factor of $\sim 2$ while the leading-order estimate for the fidelity remains accurate.
        The lower bound for the fidelity is tight for approximately $q_\text{link} \leq 0.05$ (which is hard to see in this figure) but not for larger values,
        eventually even dropping below the fidelity of the maximally mixed state.
        The rate is dimensionless as the round time $\Delta t$ has been set to 1.
        Note that the lines showing analytical results and simulation results can sometimes be hard to distinguish because of their overlap.
	}
	\label{fig:verification:rate_fidelity_qlink}
\end{figure}

There are three parameters that can influence the rate of GHZ-state distribution.
These are the success probability of Bell-state distribution $q_\text{link}$, the number of end nodes $N$ and the BSM success probability $q_\text{BSM}$.
First, we examine the influence of $q_\text{link}$ on the accuracy of the leading-order expression for the rate (Eq. \eqref{eq:factory_rate}).
On the left in Figure \ref{fig:verification:rate_fidelity_qlink}
we verify that the difference between the leading-order expression and its simulated value becomes negligible for $q_\text{link} \ll 1$.
For larger values of $q_\text{link}$ it is much larger, with a maximum deviation of a factor $\sim 2$ for $q_\text{link} = 1$.
While not shown here, we have checked that the leading-order expression is accurate for small values of $q_\text{link}$ for the number of end nodes $3 \leq N \leq 8$ (larger values become computationally demanding to simulate).
The corresponding data can be found in our data repository \cite{factory-data}.
Finally, we note that our treatment of the effect of $q_\text{BSM}$ on the rate in Section \ref{sec:estimates} is exact.
Therefore, we do not explicitly investigate the influence of this parameter on the accuracy of the leading-order result here.
However, we do note that the leading-order result is accurate for at least one nontrivial value of $q_\text{BSM}$,
as the parameter was set to 0.95 for Figure \ref{fig:verification:rate_fidelity_qlink}.
\\

On the right in Figure \ref{fig:verification:rate_fidelity_qlink},
we do the same but for the fidelity,
but apart from the leading-order expression (obtained from combining Eq. \eqref{eq:factory_fidelity} with Eq. \eqref{eq:factory_fidelity_approx})
we also include the lower bound (obtained from combining Eq. \eqref{eq:factory_fidelity} with Eq. \eqref{eq:factory_fidelity_bound}).
Again we see close agreement for the leading-order expression for small values of $q_\text{link}$.
Remarkably, it remains highly accurate even for $q_\text{link} \sim 1$.
The lower bound does not attain the same level of agreement.
While it is tight for very small values of $q_\text{link}$,
the lower bound on the fidelity starts decreasing at $q_\text{link} \approx 0.015$,
even though the fidelity itself is a monotonically increasing function.
Consequently, the bound is very loose already for $q_\text{link} \gtrapprox 0.015$.
\\

\begin{figure}[h]
	\includegraphics[width=\linewidth]{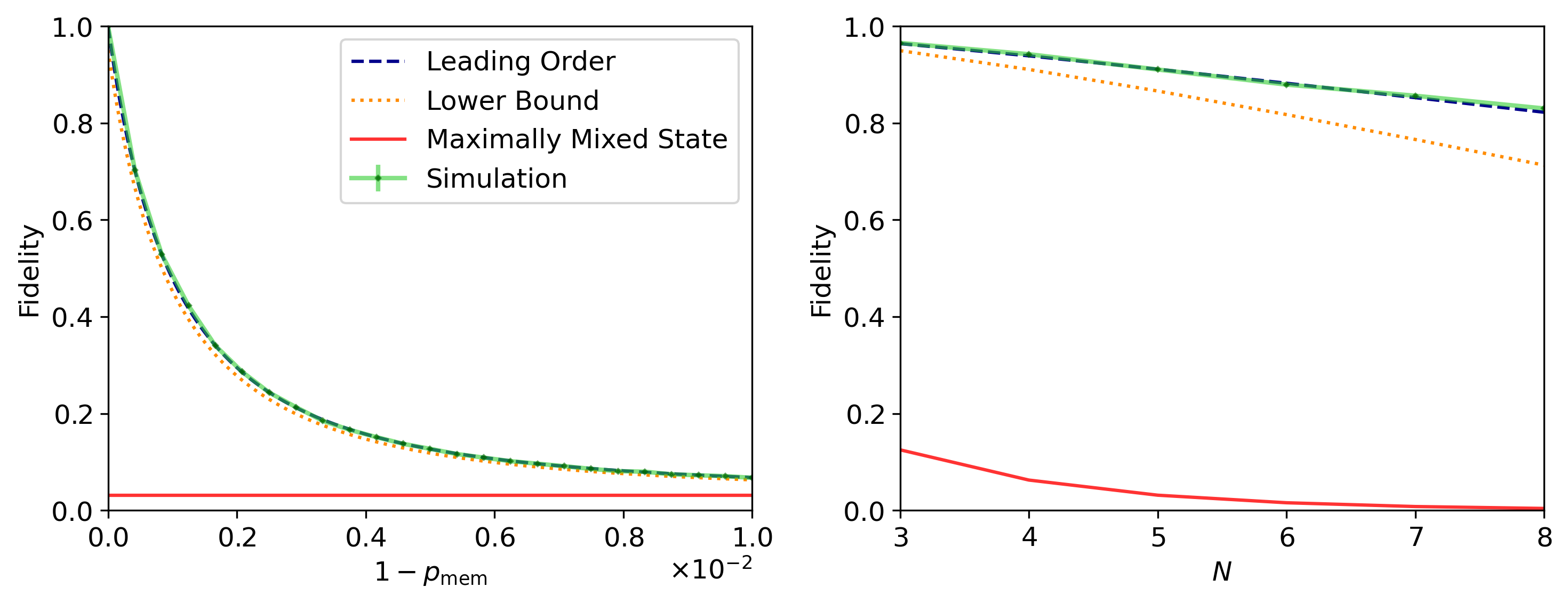}
	\caption
	{
        Comparison between simulation result and the analytical leading-order expression and lower bound from Section \ref{sec:fidelity}
        for the fidelity of Protocol \ref{prot:factory}.
        The parameters, when they are not varied over,  are $q_\text{link} = 0.01$, $N=5$, $q_\text{BSM} = 1$, $p_\text{mem} = 1 - 10^{-4}$ and $p_\text{BSM} = p_\text{link} = p_\text{GHZ} = 1$.
        On the left, we see that when $q_\text{link}$ is sufficiently small,
        the lower bound is tight and the leading-order expression remains accurate as $1 - p_\text{mem}$ becomes large,
        even as the fidelity becomes close to that of a maximally mixed state.
        On the right, we see that while the lower bound is never very tight,
        the leading-order expression remains accurate up to at least $N = 8$.
        Note that the lines showing the leading-order result, lower bound and the simulation result can be hard to distinguish because of their overlap.
	}
	\label{fig:verification:fidelity_pmem_and_N}
\end{figure}

On the left in Figure \ref{fig:verification:fidelity_pmem_and_N},
the fidelity is considered as a function of $p_\text{mem}$,
for a small value of $q_\text{link}$ (0.01).
Both the leading-order expression and lower bound remain remarkably close as $1 - p_\text{mem}$ grows,
up to the point where the fidelity becomes close to that of a maximally-mixed state.
This seems to suggest that as long as $q_\text{link}$ is small,
the analytical expressions are accurate for all values of $p_\text{mem}$ that allow for the generation of useful entanglement.
We note that the other noise parameters,
$p_\text{GHZ}$, $p_\text{BSM}$ and $p_\text{link}$,
have a much simpler effect on the fidelity
as their effect does not depend on the times at which entanglement is distributed between the factory node and the different end nodes.
This has allowed our treatment of these parameters to be exact
and therefore verification plots where these parameters are varied are not required.
We note though that in Figure \ref{fig:verification:rate_fidelity_qlink}
the accuracy of the analytical expressions is verified for nontrivial values of these parameters.
\\

Finally, on the right in Figure \ref{fig:verification:fidelity_pmem_and_N},
we consider the fidelity as a function of the number of end nodes $N$.
We observe that the leading-order expression is accurate in the range $3 \leq N \leq 8$,
while the lower bound deviates already for small values of $N$.
The lower bound becomes increasingly loose as $N$ increases.
As it is computationally demanding to simulate large quantum states,
we have not investigated the accuracy of the leading-order expression or lower bound beyond $N=8$.

\section{Deriving the Density Matrix Created by Protocol \ref{prot:factory}} \label{app:dm}

In this appendix, we formally derive the density matrix $\rho$ that is shared after executing Protocol \ref{prot:factory}.
To this end, we first define three relevant Hilbert spaces.
Let $\mathcal H_A$ be the space spanned by the $N$ qubits used by the factory node to create GHZ states locally.
Let $\mathcal H_B$ be the space spanned by the $N$ qubits used by the factory node to store Bell states shared with end nodes.
Finally, let $\mathcal H_C$ be the space spanned by the $N$ qubits at the $N$ different end nodes.
Then, Protocol \ref{prot:factory} does the following.
First, a state $\sigma_A \otimes \tau_{BC}$ is prepared,
where $\sigma$ is a noisy $N$-qubit GHZ state,
and where $\tau$ is a noisy entangled state between $2N$ qubits.
Specifically, it contains depolarizing noise due to noise in the distribution of Bell states and storage of those Bell states in noisy memory.
Secondly, noisy BSMs are executed between the qubits of $\mathcal H_A$ and $\mathcal H_B$.
The measurement outcomes are sent to the end nodes, where Pauli corrections are performed in accordance with the measurement outcomes.
The final state on $\mathcal H_C$ shared between the end nodes is $\rho_C$.
\\

The four Bell states are defined by 
\begin{equation}
\ket{\phi_{ij}} = (\mathbb 1 \otimes X^i Z^j) \ket{\phi_{00}} = \pm (X^i Z^j \otimes \mathbb 1) \ket{\phi_{00}}.
\end{equation}
for $i, j = 0, 1$.
As is apparent from this equation, the Bell states have the special property that it does not matter (up to a global sign) on which of the two qubits the Pauli operator $X^i Z^j$ acts.
This means that Pauli operators in the system can be ``moved'' through Bell states: $(P \otimes \mathbb 1) \ket{\phi_{ij}} = \pm (\mathbb 1 \otimes P) \ket{\phi_{ij}}$ for any Pauli operator $P$.
We combine this with the fact that the single-qubit depolarizing channel is a Pauli channel. That is, its Kraus operators are Pauli operators.
The consequence is that also single-qubit depolarizing noise can be moved through Bell states.
We can make use of this in the following way:

\begin{enumerate}

    \item
    When a BSM is executed between a pair of qubits (one in $\mathcal H_A$, one in $\mathcal H_B$), we use the measurement operators (which are projectors onto the Bell states) to move all the single-qubit depolarizing noise from $\mathcal H_A$ to $\mathcal H_B$.

    \item
    Now, because before the measurement every qubit in $\mathcal H_B$ is (up to single-qubit depolarizing noise) in the state $\ket{\phi_{00}}$ with a qubit in $\mathcal H_C$, we move all single-qubit depolarizing noise and the operator $X^iZ^j$ in the definition of each measurement operator from  $\mathcal H_B$ to $\mathcal H_C$.

\end{enumerate}

At $\mathcal H_C$ the operators $X^iZ^j$ from the measurement operators cancel exactly against the Pauli corrections that are applied at Step 5 of Protocol \ref{prot:factory}, which are chosen to match the measurement outcome.
Therefore, all measurement operators effectively become the same projector on $\ket{\phi_{00}}$, and each BSM can therefore be modelled as a projection of two qubits on the state $\ket{\phi_{00}}$.
Additionally, as the probability of a measurement outcome occuring is determined by the corresponding measurement operator and all outcomes effectively have the same measurement opertor, each of the four outcomes must occur with equal probability $\tfrac 1 4$.
This means that the normalization factor in the post-measurement state is given by $4$.
We define the maximally entangled state $\ket \omega$ as the tensor product of $N$ copies of $\ket {\phi_{00}}$, i.e.,
\begin{equation}
	\ket{\omega}  \equiv \ket{\phi_{00}}^{\otimes N} = \frac 1 {2^{N/2}} \sum_{i\in \{0, 1\}^{\otimes 2^N}} \ket i \otimes \ket i.
\end{equation}
Then, we can write the post-measurement state on $\mathcal H_C$ (and thus the final state produced by the protocol) as
\begin{equation} \label{eq:fidelity:BSM}
\rho_C = 2^{2N} \bra{\omega}_{AB} \sigma_A \otimes \tau_{BC} \ket{\omega}_{AB}.
\end{equation}
Furthermore, the effect of moving all the single-qubit depolarizing channels to the system $\mathcal H_C$ results in the pre-measurement states $\sigma$ and $\tau$ to effectively become
\begin{equation}
\sigma = p_{\text{GHZ}} \ketbra{\text{GHZ}} + \frac {1} {2^N}(1 - p_{\text{GHZ}}) \mathbb{1},
\end{equation}
\begin{equation}\label{eq:fidelity:tau}
\tau_{BC} = \mathcal{E_C} \Big( \ketbra \omega_{BC} \Big),
\end{equation}
where $\mathcal E$ is a quantum channel applying single-qubit depolarizing noise to $N$ different qubits.
This quantum channel accounts for the noisy BSMs, the noisy distributed Bell states, and noise due to the storage of Bell states in memory.
As can be seen, the noise in the GHZ state prepared within the factory is the only source of noise that is not contained in the channel $\mathcal E$.
Instead, this source of noise is contained by the expression for $\sigma$.
\\

Now, we notice that the state $\tau$ is exactly the Choi state \cite{choiCompletelyPositiveLinear1975, jamiolkowskiLinearTransformationsWhich1972} of the quantum channel $\mathcal E$.
Additionally, Eq. \eqref{eq:fidelity:BSM} is exactly the expression for the effect of a quantum channel in terms of its Choi state \cite{khatri2020a}.
Therefore, we can immediately conclude that
\begin{equation}
\rho = \mathcal{E} (\sigma).
\end{equation}
Using the fact that the maximally-mixed component of $\sigma$ will remain maximally mixed by the effect of $\mathcal E$, we can write
\begin{equation}
\rho = p_{\text{GHZ}} \mathcal E \Big(\ketbra{\text{GHZ}}\Big) + \frac{1 - p_{\text{GHZ}}}{2^N} \mathbb{1}.
\end{equation}
The final remaining step towards determining $\rho$ is thus evaluating the quantum channel $\mathcal E$.
\\

Because depolarizing channels have the property
\begin{equation}
\mathcal D_{\mathcal H_A, p_1} \circ \mathcal D_{\mathcal H_A, p_2} = \mathcal D_{\mathcal H_A, p_1p_2},
\end{equation}
all the depolarizing noise that has been moved to the qubits of $\mathcal H_C$ can be combined into a single depolarizing channel per qubit,
giving
\begin{equation} \label{eq:fidelity:ghz_depolar_channels}
\mathcal E \Big(\ketbra{\text{GHZ}}\Big) = \mathcal D_{\mathcal H_1, p_1} \circ \mathcal D_{\mathcal H_2, p_2} \circ \dots \circ \mathcal D_{\mathcal H_N, p_N} \Big(\ketbra{\text{GHZ}}\Big).
\end{equation}
Here, $\circ$ indicates the composition (i.e., subsequent application) of the channels and $\mathcal H_i$ denotes the Hilbert space of the qubit at the $i^\text{th}$ end node.
The combined depolarizing parameter $p_i$ accounts for noise due to one BSM, one noisy distributed Bell state and memory decoherence at both the factory node and the end node itself,
and is given by Eq. \eqref{eq:fidelity:p_i}.
Each depolarizing channel $\mathcal D_{\mathcal H_i, p_i}$ gives one term proportional to $p_i$ where nothing happens to the $\mathcal H_i$ subspace,
and one term proportional to $1 - p_i$ where $\mathcal H_i$ is traced out of the GHZ state and then put into the state $\mathbb 1_i / 2$.
Thus, evaluating Eq. \eqref{eq:fidelity:ghz_depolar_channels} comes down to accounting for all different combinations of terms.
Tracing out one qubit from a GHZ state results in
\begin{equation} \label{eq:fidelity:trace_from_GHZ}
\Tr_i \Big( \ketbra{\text{GHZ}} \Big)_{1, 2, \dots, k} = \frac 1 2 \mathcal P_{1, 2, \dots, i-1, i+1, \dots, k},
\end{equation}
where $\mathcal P$ is the classically correlated, unnormalized state defined in Eq. \eqref{eq:fidelity:P}.
Tracing out a qubit from $\mathcal P$ yields
\begin{equation} \label{eq:fidelity:trace_from_P}
\Tr_{i} \mathcal P_{1, 2, \dots, k} = \mathcal P_{1, 2, \dots, i-1, i+1, \dots, k},
\end{equation}
unless $k=1$, in which case
\begin{equation} \label{eq:fidelity:trace_from_1}
\Tr_1 \mathcal P_1 = \Tr_1 \mathbb 1_1 = 2.
\end{equation}

Now, we define the set $\mathcal N = \{1, 2, \dots, N\}$ as the set of all qubit indices.
Working out the combinatorics, we find
\begin{equation}
\begin{aligned}
\rho = \frac{1 - p_{\text{GHZ}}}{2^N} \mathbb 1_{\mathcal N} + p_{\text{GHZ}} \Bigg[ & \prod_{i\in \mathcal N} p_i \big( \ketbra{\text{GHZ}} \big)_{\mathcal N} + \prod_{i \in \mathcal N} \frac{1 - p_i}{2} \mathbb 1 _{\mathcal N} \\
&+ \frac 1 2 \sum_{\substack{U \subset \mathcal N \\ 1 < |U| < N}} \left( \prod_{i \in U } \frac{1 - p_i}{2} \prod_{j \in \mathcal N \setminus U} p_j \right) \mathbb 1_U \otimes \mathcal P_{\mathcal N \setminus U} \Bigg].
\end{aligned}
% \begin{aligned}
% \rho = \frac{1 - p_{\text{GHZ}}}{2^N} \mathbb 1_{\mathcal N} + p_{\text{GHZ}} \Big[ &\prod_{i\in \mathcal N} p_i \big( \ketbra{\text{GHZ}} \big)_{\mathcal N} + \prod_{i \in \mathcal N} \frac{1 - p_i}{2} \mathbb 1 _{\mathcal N} \\
% &+ \frac 1 2 \sum_{\substack{U \subseteq \mathcal N \\ 1 < |U| < N}} \prod_{i \in U } p_i \prod_{j \in \mathcal N \setminus U} \frac{1 - p_j}{2} \mathcal P_U \otimes \mathbb 1_{\mathcal N \setminus U} \Big].
% \end{aligned}
\end{equation}
Note that due to the factors appearing when taking traces in Eqs. \eqref{eq:fidelity:trace_from_GHZ}, \eqref{eq:fidelity:trace_from_P}, and \eqref{eq:fidelity:trace_from_1},
the terms where more than 0 but less than $N$ of the qubits are traced out effectively have an ``extra'' factor of $\tfrac 1 2$.

\section{Coefficients of Fidelity Function} \label{app:coefficients}

In this appendix, we derive the coefficients in the expression for the fidelity of GHZ states distributed by Protocol \ref{prot:factory}.
That is, we show that Eq. \eqref{eq:fidelity:fidelity_not_yet_rearranged} can be rewritten into the form of Eq. \eqref{eq:factory_fidelity},
with the coefficients $A_{|U|}$ given by Eq. \eqref{eq:fidelity:coefficients}.
\\

First, we collect products of $p_i$'s such that we may write
\begin{equation}\label{eq:coefficients:prop_B}
\sum_{U \subseteq \mathcal N} 2^{\delta_{|U|, 0} + \delta_{|U|, N} - 1}  \expectationvalue{\prod_{i \in U}  \frac{1 - p_i}{2} \prod_{j \in \mathcal N \setminus U} p_j} = \sum_{U \subseteq \mathcal N} B_U  \expval{\prod_{i \in U} p_i}
\end{equation}
for some constants $B_U$.
To find these constants, we start by expanding
\begin{equation}
\prod_{i \in W} \frac{1 - p_i}{2} = \left( \frac 1 2 \right) ^ {|W|} \sum_{V \subseteq W} (-1)^{|V|} \prod_{i \in V} p_i,
\end{equation}
giving
\begin{equation} \label{eq:coefficients:after_expanding_1-p}
\sum_{W \subseteq \mathcal N} 2^{\delta_{|W|, 0} + \delta_{|W|, N} - 1}  \expectationvalue{\prod_{i \in W}  \frac{1 - p_i}{2} \prod_{j \in \mathcal N \setminus W} p_j} = \sum_{W \subseteq \mathcal N}  2^{\delta_{|W|, 0} + \delta_{|W|, N} - 1 - |W|} \sum_{V \subseteq W} (-1)^{|V|} \expectationvalue{\prod_{i \in V \cup (\mathcal N \setminus W)} p_i}.
\end{equation}
We now equate Eqs. \eqref{eq:coefficients:prop_B} and \eqref{eq:coefficients:after_expanding_1-p}.
Each is the expected value of a polynomial in the independent random variables $p_i$.
They are equal if the coefficients of all terms in the polynomial are equal.
Therefore, we determine $B_U$ by collecting all parts of the sum in Eq. \eqref{eq:coefficients:after_expanding_1-p} that are proportional to $\expectationvalue{\prod_{i \in U} p_i}$ and thus contribute to the same term.
Writing as a shorthand $\overline W = \mathcal N \setminus W$, this gives
\begin{equation} \label{eq:coefficients:A_using_delta}
B_U = \sum_{W \subseteq \mathcal N} 2^{\delta_{|W|, 0} + \delta_{|W|, N} - 1 - |W|} \sum_{V \subseteq W} (-1)^{|V|} \delta_{V \cup \overline W, U},
\end{equation}
where we are slightly abusing notation by using the Kronecker delta for two sets.
It is defined by
\begin{equation}
\delta_{U, V} =
\begin{cases}
1 &\text{for } U = V, \\
0 &\text{otherwise}, \\
\end{cases}
\end{equation}
where $U$ and $V$ are sets.
The delta function ensures that we are adding together exactly those coefficients of \eqref{eq:coefficients:A_using_delta} that contribute to the right term of the polynomial.
\\

We note that the equation $V \cup \overline W = U$ implies that $\overline W \subseteq U$.
Therefore, the Kronecker delta will always be zero when this condition does not hold, allowing us to refine the summation limit and write
\begin{equation}
B_U = \sum_{\substack{W \\ \overline W \subseteq U}} 2^{\delta_{|W|, 0} + \delta_{|W|, N} - 1 - |W|} \sum_{V \subseteq W} (-1)^{|V|} \delta_{V \cup \overline W, U}.
\end{equation}
The Kronecker delta now limits the sum to values of $V$ and $W$ where $V \cup \overline W = U$ holds.
Because $V \subseteq W$ for all terms in the sum, it always holds that $V \cap \overline W = \emptyset$, i.e. there is no overlap between the two sets.
Therefore, the equation $V \cup \overline W = U$ implies that $V = U \setminus \overline W$.
Additionally, because $\overline W \subseteq U$ for all terms in the sum, the equation $V = U \setminus \overline W$ implies that $V \cup \overline W = U$.
It follows that the two equations are equivalent given the conditions imposed on $V$ and $W$ by the summation limits,
and we can safely rewrite the Kronecker delta function to obtain
\begin{equation}
B_U = \sum_{\substack{W \\ \overline W \subseteq U}} 2^{\delta_{|W|, 0} + \delta_{|W|, N} - 1 - |W|} \sum_{V \subseteq W} (-1)^{|V|} \delta_{V, U \setminus \overline W}.
\end{equation}
Since $U \setminus \overline W$ contains only elements not in $\overline W$, and since $W$ contains all elements in $\mathcal N$ that are not in $\overline W$,
it follows that $U \setminus \overline W \subseteq W$.
If this were not always the case, it could be the case for some $W$ that the sum over $V \subseteq W$ contains no terms for which the delta function is nonzero.
But since it is the case, for every $W$ there is exactly one value of $V$, namely $V = U \setminus \overline W$,
for which the delta function has a nonzero value.
For this value, $|V| = |U| - |\overline W|$, and therefore the equation becomes
\begin{equation}
B_U = \sum_{\substack{W \\ \overline W \subseteq U}} 2^{\delta_{|W|, 0} + \delta_{|W|, N} - 1 - |W|} (-1)^{|U| - |\overline W|}.
\end{equation}

To further resolve the equation, we note that when the cardinality of $\overline W$ is equal to $|\overline W| = i$,
there are exactly $|U|$ choose $i$ different ways $\overline W$ can be chosen from $U$.
Since only the cardinalities of $\overline W$ and $W$ (with $|W| = N - i$) appear in the sums,
this allows us to write
\begin{equation}
B_U = \sum_{i=0}^{|U|} \binom{|U|}{i} 2^{\delta_{N - i, 0} + \delta_{N - i, N} - 1 - N + i} (-1)^{|U| - i} =  \sum_{i=0}^{|U|}  \binom{|U|}{i} 2^{\delta_{i, N} + \delta_{i, 0} - 1 - N + i} (-1)^{|U| - i}.
\end{equation}
Now, we make a change of variable, $i \to |U| - i$.
Conveniently, the binomial coefficient is invariant under this transformation, giving
\begin{equation}
B_U = \sum_{i=0}^{|U|} \binom{|U|}{i} 2^{\delta_{|U| - i, N} + \delta_{|U| - i, 0} - 1 - N + |U| - i} (-1)^i = \left(\frac 1 2 \right)^{N + 1 - |U|} \sum_{i=0}^{|U|} \binom{|U|}{i} 2^{\delta_{|U|, N} \delta_{i,0} + \delta_{i, |U|}} \Big(\frac{-1}{2}\Big)^i.
\end{equation}
By the binomial theorem,
\begin{equation}
\sum_{i=0}^{|U|} \binom{|U|}{i} \left(\frac{-1}{2}\right)^i = \left(1 - \frac 1 2\right)^{|U|} = \left(\frac 1 2 \right)^{|U|}.
\end{equation}
By adding the contributions from when the delta functions are nonzero separately on top of that, we find
\begin{equation}
B_U = \left(\frac 1 2 \right)^{N + 1 - |U|} \left\{ \left( \frac 1 2 \right)^{|U|} + \delta_{|U|, N} + \left(\frac {-1}{2} \right)^{|U|}\right\},
\end{equation}
which can be rewritten as (using the fact that $N - |U| = 0$ whenever the remaining delta function is nonzero)
\begin{equation}
B_U = \left( \frac 1 2 \right)^{N + 1} \left( 1 + (-1) ^{|U|} \right) + \frac 1 2 \delta_{|U|, N}.
\end{equation}
Noticing furthermore that the value of $B_U$ only depends on the cardinality of the set $U$, we write
\begin{equation}
B_{|U|} =
\begin{cases}
\frac 1 {2 ^ N} + \frac 1 2 \delta_{|U|, N} \hspace{0.1 cm} &\text{if $|U|$ is even,} \\
\frac 1 2 \delta_{|U|, N}  &\text{if $|U|$ is odd.}
\end{cases}
\end{equation}

Now, we can derive the coefficients $A_{|U|}$ in Eq. \eqref{eq:factory_fidelity}.
To this end, we substitute Eq. \eqref{eq:fidelity:p_i} into Eq. \eqref{eq:coefficients:prop_B} to find
\begin{equation}
\begin{aligned}
\sum_{U \subseteq \mathcal N} 2^{\delta_{|U|, 0} + \delta_{|U|, N} - 1}  \expectationvalue{\prod_{i \in U}  \frac{1 - p_i}{2} \prod_{j \in \mathcal N \setminus U} p_j} &= \sum_{U \subseteq \mathcal N} B_{|U|} \left(p_\text{link} p_\text{BSM}^2\right)^{|U|} \expectationvalue{\prod_{i \in U} \left(p_\text{mem}^2\right)^{\Delta n_i}} \\
&=  \sum_{U \subseteq \mathcal N} A_{|U|}\expectationvalue{\prod_{i \in U} \left(p_\text{mem}^2\right)^{\Delta n_i}}
\end{aligned}
\end{equation}
where $A_{|U|}$ is exactly as defined in Eq. \eqref{eq:fidelity:coefficients}.
Therefore, Eq. \eqref{eq:factory_fidelity} indeed follows from Eq. \eqref{eq:fidelity:fidelity_not_yet_rearranged}.

\section{Expected Values for Memory Decoherence} \label{app:exp_val}

In this appendix, we derive both a leading-order expression and a lower bound for the effect of memory decoherence on the fidelity of GHZ states produced using Protocol \ref{prot:factory}.
These results allow us to write down a leading-order expression for the fidelity of states produced using this protocol (Eq. \eqref{eq:factory_fidelity_approx}),
and a lower bound (\eqref{eq:factory_fidelity_bound}).
To this end, we first derive more general results for the case where the decoherence rate is different for each quantum memory.
\\

\subsection{Indices}

Trying to establish a Bell state happens according to discrete rounds, with the probability of succeeding during each round being $q_\text{link}$ for all end nodes.
When all Bell states are in place, a GHZ state is generated locally and then teleported by the factory node towards the end nodes after which, in case all BSMs are successful, the protocol terminates.
While the BSM success probability influences the rate with which GHZ states can be distributed (see Section \ref{sec:rate}),
it will not influence the fidelity,
since all states are discarded whenever a BSM fails and the protocols starts again from the beginning.
Therefore, without loss of generality, we will henceforth assume BSMs are deterministic.
In that case, each execution of the protocol is uniquely defined by which end node established a Bell state during which attempt.
This can be described by assigning indices $i \in \mathcal N$ to the different end nodes (where $\mathcal N = \{1, \cdots, N\}$ as before), and denoting the round during which end node $i$ established a Bell state by $n_i$.
\\

For any given realization of the protocol, an ordering can be imposed on the indices in correspondence with the order in which the different Bell states were distributed.
We denote the ordered index corresponding to end node $i$ by $d_i$,
and they have the property
\begin{equation} \label{eq:app:order_of_d}
n_{i} \geq n_{j} \,\,\,\text{ if }\,\,\, d_i > d_j.
\end{equation}
for $i, j = 1, 2, ..., N$.
That means that if $d_5 = 1$, end node with label 5 was the first end node to share a Bell state with the factory node,
while if $d_1 = N$, end node with label 1 was the last to do so.\\

What we want to calculate, are expected values including only the waiting times of a specific subset of the end nodes.
We denote this subset $V \subseteq \mathcal N$, with $|V| \equiv M$,
and define the indices $v_1, v_2, \cdots, v_M$ as the ordered elements of the subset $V$.
That is, $V = \{v_1, v_2, ..., v_M\}$ and
\begin{equation} \label{eq:app:order_of_v}
d_{v_{i+1}} > d_{v_i}
\end{equation}
for $i = 1, 2, \cdots, M-1$.
To simplify our notation, we now introduce the symbols
\begin{equation}
\begin{aligned}
c_i \equiv d_{v_i},\\
m_i \equiv n_{v_i}.
\end{aligned}
\end{equation}
We note that Eq. \eqref{eq:app:order_of_d} and \eqref{eq:app:order_of_v} together imply that
\begin{equation}
m_{i+1} \geq m_i
\end{equation}
for $i = 1, 2, \cdots, M-1$.
\\

An example of the values these different indices can take, let us consider the case $N=4$.
For a specific realization of the protocol,
it might be that the end node with index 2 shared a Bell state with the factory node first during $n_2=3$, then 3 during $n_3=5$, then 1 at $n_1 = 10$ and finally 4 at $n_4=17$.
In that case, $d_2 = 1$, $d_3 = 2$, $d_1 = 3$, and $d_4 = 4$.
Now, if we take $V = \{1, 3\}$, then $v_1 = 3$ and $v_2 = 1$.
This gives, $c_1 = d_3 = 2$ and $c_2 = d_1 = 3$, which correctly satisfies $c_2 > c_1$.
Furthermore, $m_1 = n_3 = 5$ and $m_2 = n_1 = 10$.

\subsection{Probability Building Blocks} \label{sec:app:prob_building_blocks}

At the start of Protocol \ref{prot:factory},
there are $N$ quantum connections simultaneously distributing Bell states between the factory node and end nodes $1, 2, \cdots, N$.
Each of these will follow a geometric distribution.
That is,
\begin{equation}
\text{Pr}\Big(\text{Bell state $i$ is successfully distributed during round $n$}\Big) = q_\text{link} (1 - q_\text{link})^{n-1},
\end{equation}
for $i = 1, 2, ..., N$, and $n = 1, 2, 3, ...$ .
\\

Now, we introduce some probabilities based on this that will be useful later on:
\begin{equation} \label{eq:app:P_i/N}
\begin{aligned}
P_{i/N}(n) \equiv \text{Pr}\Big(&\text{during round $n$, the $i^\text{th}$ Bell state is distributed, given that there were zero before round 1,}
\\ &\text{and distribution takes place on $N$ quantum connections}\Big),
\end{aligned}
\end{equation}
\begin{equation}
\begin{aligned}
P'_{i/N}(n) \equiv \text{Pr}\Big(&\text{after round $n$, exactly $i$ Bell states are distributed, given that there were zero before round 1,}\\
&\text{and distribution takes place on $N$ quantum connections;} \\
&\text{the $i^\text{th}$ Bell state was established during round $n$}\Big).
\end{aligned}
\end{equation}
Note that the difference between $P_{i/N}(n)$ and $P'_{i/N}(n)$ is that the first also includes the probability for the case that, during round $n$, more Bell states are simultaneously established than was required to reach $i$.
The first of these two is a properly normalized probability distribution,
and has the random variable $n_{i/N}$ associated to it,
representing the number of rounds needed to distribute $i$ Bell states using $N$ quantum connections.
A special case is the variable $n_{1/N}$, as it is a geometrically distributed random variable.
The reason for this is that the probability that the first Bell state is distributed during round $n$,
is equal to the probability that all quantum connections failed up until round $n$,
and that not all quantum connections fail during round $n$.
That is,
\begin{equation}\label{eq:app:P1/N}
P_{1/N}(n) = [1 - (1-q_{\text{link}})^N] (1-q_{\text{link}})^{N(n-1)},
\end{equation}
which is geometric with $1 / \expval{n_{1/N}} = 1 - (1 - q_\text{link})^N$.
\\

Furthermore, we define
\begin{equation}
\begin{aligned}
P^j_{i/N}(n) \equiv \text{Pr}\Big(&\text{after round $n$, exactly $i$ Bell states are distributed, given that there were zero before round 1,}\\
&\text{and distribution takes place on $N$ quantum connections;} \\
&\text{$j$ of those $i$ Bell states were distributed during round $n$}\Big).
\end{aligned}
\end{equation}
Here, $j\leq i\leq N$, and $j \geq 1$.
This allows us to be more specific about the number of success events during the last round.
Since for $P'_{i/N}(n)$ the number of success events at round $n$ can be any number larger than zero (and, of course, smaller or equal to $i$),
we can write down the relation
\begin{equation}\label{eq:app:P'}
P'_{i/N}(n) = \sum_{l = 1}^i P_{i/N}^l(n).
\end{equation}
Similar, since $P_{i/N}(n)$ is the same as $P'_{i/N}(n)$ but also includes to possibility that ``too many'' successes occurred during round $n$,
bringing the number of entangled states above $i$,
we can write
\begin{equation} \label{eq:app:expansion_of_P_i/N}
P_{i/N}(n) = \sum_{k=0}^{N-i} \sum_{l=1}^i P_{(i+k)/N}^{k+l}(n) = P'_{i/N}(n) + \sum_{k=1}^{N-i} \sum_{l=1}^i P_{(i+k)/N}^{k+l}(n).
\end{equation}
Note however that both equations only hold for $i>0$.
\\

It is possible to derive a recursive relation for $P^j_{i/N}(n)$.
We can express the probability as
\begin{equation}
\begin{aligned}
P^j_{i/N}(n) =& \binom{N-(i-j)}{j} \text{Pr}\Big(\text{during round $n$, out of $N-(i-j)$ quantum connections}\\
&\hspace{2.8cm}\text{trying to establish a Bell state, exactly $j$ succeed}\Big) \\
&\times \text{Pr}\Big(\text{after round $n-1$, there were $i-j$ Bell states}\Big).
\end{aligned}
\end{equation}
The first probability is simply $q_\text{link}^j (1-q_\text{link})^{N-i}$.
The second probability depends on what $i-j$ is.
If it is zero, it is simply the probability that there have been no success events up to and including round $n-1$, i.e. $(1-q_\text{link})^{N(n-1)}$.
If $i-j \neq 0$, we must distinguish between the different cases in which the final Bell state is established during different rounds.
This gives
\begin{equation}
\begin{aligned}
&\text{Pr}\Big(\text{after round $n-1$, there were $i-j$ Bell states}\Big) \\
&= \sum_{n'=1}^{n-1} P'_{(i-j)/N}(n') \times \text{Pr}\Big(\text{none out of $N-(i-j)$ active quantum connections distribute a}\\
& \hspace{4cm}\text{Bell state after round $n'$ up to round $n-1$} \Big) \\
&= \sum_{n'=1}^{n-1} P'_{(i-j)/N}(n') (1-q_\text{link})^{[N-(i-j)][(n-1)-n']}.
\end{aligned}
\end{equation}

Now, we note that the definition of $P'_{i/N}(n)$ is somewhat ambiguous for $i=0$ and $n=0$.
Therefore, we here define it explicitly for these values,
in such a way that we can extend the above relation to the cases $j=i$ and $n=1$.
The definition is as follows:
\begin{equation} \label{eq:app:P'_0/N}
P'_{0/N}(n) \equiv \delta_{n,0}.
\end{equation}
This allows us to extend the above sum to include $n'=0$, which gives exactly what we need for $j=i$ and vanishes anyway for $j<i$, i.e.
\begin{equation}\label{eq:P^j_i/N}
P^j_{i/N}(n) = \binom{N-i+j}{j} \sum_{n'=0}^{n-1} q_{\text{link}}^j (1-q_{\text{link}})^{(n-n')(N-i +j)-j}P'_{(i-j)/N}(n').
\end{equation}
We can rewrite this equation into a form that makes it easier to deal with later on.
Using Eq. \eqref{eq:app:P1/N} we can write
\begin{equation}\label{eq:calc_ti:Pji/N}
P^j_{i/N}(n) = \binom{N-i+j}{j} \frac{q_{\text{link}}^j(1-q_{\text{link}})^{N-i}}{1- (1-q_{\text{link}})^{N-i+j}} \sum_{n'=0}^{n-1} P_{1/(N-i+j)}(n-n') P'_{(i-j)/N}(n').
\end{equation}
Furthermore, to turn this into a true recursion relation, we also fill in Eq. \eqref{eq:app:P'} to find
\begin{equation}\label{eq:app:P^j_i/N_recursion}
P^j_{i/N}(n) = \binom{N-i+j}{j} \frac{q_{\text{link}}^j(1-q_{\text{link}})^{N-i}}{1- (1-q_{\text{link}})^{N-i+j}} \sum_{n'=0}^{n-1} P_{1/(N-i+j)}(n-n')  \sum_{l = 1}^i P_{i/N}^l(n).
\end{equation}
However, we must be aware of the fact that this equation only covers the $i>0$ cases.
If $i=j=0$, there are no Bell states distributed at all, and thus we can also not split up the success events as we did in our arguing above.
Analogues to $P'_{0/N}(n) = \delta_{n,0}$, we define $P^0_{0/N}(n) = \delta_{n,0}$.
Furthermore, while $P^j_{i/N}(n)$ is technically undefined for $j = 0$ and $i>0$, we define it to be zero for later convenience.
Note that therefore $P^0_{i/N}(n)$ for $i>0$ is not equal to the probability that there are $i$ Bell states after round $n$, of which there where $0$ distributed during round $n$,
since this would be a nonzero quantity.
\\

Finally, we will abuse notation to write
\begin{equation}
\sum_{n=0}^\infty n P^j_{i/N}(n) = \expval{n^j_{i/N}},
\end{equation}
even though $P^j_{i/N}(n)$ is not a normalized probability distribution and thus $n^j_{i/N}$ is not a well-defined random variable.

\subsection{Probability Distribution of Links}

Now, we introduce the probability distribution
\begin{equation}
P(m_1=m_1', m_2=m_2', \cdots, m_M=m_M'),
\end{equation}
which is the probability that, if Protocol \ref{prot:factory} is executed once,
and labels are defined and ordered as described above,
that $m_i$ has the value $m_i'$ for each $i = 1, 2, \cdots, M$.
Below, we will use this probability distribution to write down expected values of the type we need to account for memory decoherence.
First, we will investigate what the probability distribution looks like.
\\

Then, what is the probability that Bell state $c_i$ is distributed at round $m_i$?
Consider the fact that Bell state $c_{i-1}$ was distributed at round $m_{i-1}$.
During this round, many Bell states could have been distributed simultaneously, as multiple quantum connections are attempting to distribute them in parallel.
However, assume for the moment that only Bell state $c_{i-1}$ was distributed at round $m_{i-1}$.
In that case, the probability that $c_i$ succeeds during round $m_i$ is equal to the probability that $c_i - c_{i-1}$ Bell states are distributed
using $N - c_{i-1}$ parallel quantum connections in $m_i - m_{i-1}$ rounds,
which is the probability $P_{(c_i - c_{i-1})/(N-c_{i-1})}(m_i - m_{i-1})$ defined above.
Now assume that there were in fact multiple successes during round $m_{i-1}$.
Specifically, let it be such that there were so many successes that after round $m_{i-1}$,
the number of distributed Bell states is $c_{i-1} + k_{i-1}$.
That is, $k_{i-1}$ is the ``overshoot'' during round $m_{i-1}$.
Then, we can distinguish two different cases.
In the first case, $k_{i-1} < c_i - c_{i-1}$, and Bell state number $c_i$ is not yet distributed after round $m_{i-1}$.
We can then repeat the logic above: the probability of distributing Bell state $c_i$ during round $m_i$ is $P_{(c_i - c_{i-1} - k_{i-1})/(N-c_{i-1} - k_{i-1})}(m_i - m_{i-1})$.
However, in the second case, $k_{i-1} \geq c_i - c_{i-1}$; the overshoot is so large that Bell state $c_i$ was already distributed during round $m_{i-1}$,
and the probability can be written as the Kronecker delta function $\delta_{m_i, m_{i-1}}$.
\\

Using this logic, the probability distribution can be completely characterized using $P^j_{i/N}(n)$-type probabilities that were defined above.
For each $c_i$, we can put a Heaviside step function $\theta(c_i - c_{i-1} - k_{i-1} - 1)$ to account for the case where the overshoot was small enough to ensure $m_i \neq m_{i-1}$,
and $\theta(c_{i-1} + k_i-1 - c_i)$ when they are the same.
The Heaviside step function is defined as
\begin{equation}
\theta(x) =
\begin{cases}
0 \text{ if } x < 0, \\
1 \text{ if } x \geq 0.
\end{cases}
\end{equation}
There are just two additional aspects we need to consider.
First of all, the number of successes during round $m_{i-1}$ is not necessarily equal to $k_{i-1}$; $k_{i-1}$ is just the overshoot.
It could e.g. be the case that $c_{i-1} = 6$ and $k_{i-1} = 3$.
That means that after $m_{i-1}$, the number of distributed Bell states is 9.
But it says nothing about the number of Bell states before that round.
It could e.g. be 4, in which case there were 5 successes during round $m_{i-1}$.
We denote the number of ``additional'' successes that did not go into the overshoot by $l_{i-1}$.
Thus, the number of successes during round $m_{i-1}$ is $l_{i-1} + k_{i-1}$.
In the example, $l_{i-1} = 2$.
Secondly, we need to consider the fact that if $k_{i-1}$ is large enough that $m_i = m_{i-1}$,
then the overshoot $k_i$ must be equal to $k_{i-1} - (c_i - c_{i-1})$,
which can be accounted for using a Kronecker delta.
Combining all this into a single equation, we find
\begin{equation} \label{eq:app:prob_connections}
\begin{aligned}
&\text{Pr}(m_1=m_1', m_2=m_2', \cdots, m_M=m_M')\\
&=\prod_{i=1}^{M}  \sum_{k_i = 0}^{N-c_i}  \Big[ \theta(c_i-c_{i-1}-k_{i-1}-1)\sum_{l_i=1}^{c_i - c_{i-1} - k_{i-1}}  P^{k_i+l_i}_{(c_i+k_i - c_{i-1}-k_{i-1})/(N-c_{i-1}-k_{i-1})}(m_i' - m_{i-1}')\\
&+ \theta(c_{i-1}+k_{i-1} -c_i) \delta_{k_i, c_{i-1} + k_{i-1} - c_i} \delta_{m_i' ,m_{i-1}'} \Big]\\
&=\prod_{i=1}^{M}\Big[  \sum_{k_i = 0}^{N-c_i} \sum_{l_i=-k_i}^{c_i - c_{i-1} - k_{i-1}} \Big( \theta(l_i-1) + \delta_{k_i, c_{i-1} + k_{i-1} - c_i} \Big)\\
&\times P^{k_i+l_i}_{(c_i+k_i - c_{i-1}-k_{i-1})/(N-c_{i-1}-k_{i-1})}(m_i' - m_{i-1}')\Big],
\end{aligned}
\end{equation}
where we set $m_0' \equiv c_0 \equiv k_0 \equiv 0$ by definition to allow for the more compact form of the equation.

\subsection{Expected Value}

In order to calculate the expected values for the amount of decoherence in quantum memory,
what we need is a probability distribution not for at what time each Bell state was distributed,
but for how long each Bell state had to sit in memory before Protocol \ref{prot:factory} terminated.
Luckily, the second can be easily obtained from the first.
First, we define $n_f$ to be the round during which the final Bell state is distributed.
Then, we define $\Delta m_i = n_f - m_i$ as the number of rounds Bell state $v_i$ waits in memory until all Bell states are distributed.
The probability distribution we are then interested in is
\begin{equation}
\text{Pr}(\Delta m_1 = \Delta m_1', \Delta m_2 = \Delta m_2', \cdots, \Delta m_M = \Delta m_M'),
\end{equation}
which can be written as
\begin{equation}
\begin{aligned}
& \text{Pr}(\Delta m_1 = \Delta m_1', \Delta m_2 = \Delta m_2', \cdots, \Delta m_M = \Delta m_M')\\
&= \sum_{n_f'=1}^{\infty}  \prod_{i=1}^M \Big(\sum_{m_i'=1}^{\infty} \delta_{n_f'-m_i', \Delta m_i'} \Big) \text{Pr}(m_1=m_1', m_2=m_2', \cdots, m_M=m_M', n_f = n_f').
\end{aligned}
\end{equation}

The latter probability distribution is the one from Eq. \ref{eq:app:prob_connections},
except for the additional condition $n_f = n_f'$.
However, this condition can be easily incorporated by extending the set $V$ of end nodes under consideration slightly,
such that we include $v_{M+1}$ which corresponds to the last Bell state that is distributed.
That is,
\begin{equation}
c_{M+1} = N,
\end{equation}
and $m_{N+1} = n_f$.
In that case, we can directly use Eq. \ref{eq:app:prob_connections} to write down
\begin{equation}
\begin{aligned}
& \text{Pr}(\Delta m_1 = \Delta m_1', \Delta m_2 = \Delta m_2', \cdots, \Delta m_M = \Delta m_M')\\
&= \sum_{m_{M+1}'=1}^{\infty}  \prod_{i=1}^M \Big(\sum_{m_i'=1}^{\infty} \delta_{n_f'-m_i', \Delta m_i'} \Big) \text{Pr}(m_1=m_1', m_2=m_2', \cdots, m_M=m_M', m_{M+1} = m_{M+1}')\\
&= \sum_{m_{M+1}'=1}^{\infty}   \prod_{i=1}^M \Big(\sum_{m_i'=1}^{\infty} \delta_{m_{M+1}' - m_i', \Delta m_i'}\Big) \prod_{i=1}^{M+1} \Big[ \sum_{k_i = 0}^{N-c_i} \sum_{l_i=-k_i}^{c_i - c_{i-1} - k_{i-1}} \Big( \theta(l_i-1) + \delta_{k_i, c_{i-1} + k_{i-1} - c_i} \Big)\\
&\times P^{k_i+l_i}_{(c_i+k_i - c_{i-1}-k_{i-1})/(N-c_{i-1}-k_{i-1})}(m_i' - m_{i-1}') \Big].
\end{aligned}
\end{equation}
First, we resolve the Kronecker delta functions.
If $\Delta m_i' = m_{M+1}' - m_i'$, then $m_i' - m_{i-1}' = \Delta m_{i-1} - \Delta m_i$.
Therefore, if we define $\Delta m_{M+1}' \equiv 0$ and write $\Delta m_0' = m_{M+1}'$,
we find
\begin{equation} \label{eq:app:prob_mem}
\begin{aligned}
& \text{Pr}(\Delta m_1 = \Delta m_1', \Delta m_2 = \Delta m_2', \cdots, \Delta m_M = \Delta m_M')\\
&= \sum_{\Delta m_0' = 0}^\infty \prod_{i=1}^{M+1}  \Bigg[ \sum_{k_i = 0}^{N-c_i} \sum_{l_i=-k_i}^{c_i - c_{i-1} - k_{i-1}} \Big( \theta(l_i-1) + \delta_{k_i, c_{i-1} + k_{i-1} - c_i} \Big) \\
&\times  P^{k_i+l_i}_{(c_i+k_i - c_{i-1}-k_{i-1})/(N-c_{i-1}-k_{i-1})}(\Delta m_{i-1}' - \Delta m_i') \Bigg].
\end{aligned}
\end{equation}

Now, we will use these results to calculate the expected value
\begin{equation}\label{eq:app:G_def}
\begin{aligned}
G(r_1, r_2, \cdots, r_M) &\equiv \expectationvalue{\prod_{i=1}^M (1 - r_i)^{\Delta m_i}}\\
&=\prod_{i=1}^M \Big[ \sum_{\Delta m_i'=0}^\infty (1 - r_i)^{\Delta m_i'} \Big] \text{Pr}(\Delta m_1 = \Delta m_1', \Delta m_2 = \Delta m_2', \cdots, \Delta m_M = \Delta m_M')\\
\end{aligned}
\end{equation}
Here, the $r_i$ are some numbers between zero and one.
The fidelity of GHZ states created by Protocol \ref{prot:factory} is expressed as a sum over such expected values in Eq. \eqref{eq:factory_fidelity}.
Therefore, if we are able to evaluate Eq. $\eqref{eq:app:G_def}$, we are able to evaluate the fidelity using the substitution $r_i = 1 - p_\text{mem}^2$ for all $i$
(i.e. $r_i$ becomes the probability that a quantum state is lost in memory per round of Bell-state distribution).
We make two remarks about the expected value $G$.
First, an evaluation of $G$ is a more general result than what we need to calculate the fidelity,
as here we allow each $r_i$ to take a different value.
As discussed in Section \ref{sec:conclusion}, this makes such a result suitable to study asymmetric quantum networks.
Second, in the definition of $G$, a product over quantities of the form $1 - r_i$ appears.
We could just as well make the redefinition $r_i \to 1 - r_i$.
This would make the definition of $G$ more compact, and would lead to the perhaps more natural mapping $r_i = p_\text{mem}^2$ in order to calculate the fidelity.
However, we are ultimately interested in the regime $1 - p_\text{mem}^2 \ll 1$, where the probability of losing a quantum state when storing it in memory for a single round is small.
This translates here to $r_i \ll 1$.
Therefore, if we want to calculate the fidelity to leading order in $1 - p_\text{mem}^2$, we need to evaluate $G$ to leading order in the variables $r_i$.
This is easier to do than working to leading order in $1 - r_i$.
\\

First of all, we substitute Eq. \eqref{eq:app:prob_mem} into Eq. \eqref{eq:app:G_def}.
By defining $r_0 \equiv 0$, we can conveniently write the result as
\begin{equation}
\begin{aligned}
&G(r_1, r_2, \cdots, r_M) \\
&= \sum_{\Delta m_0', \Delta m_1', \cdots, \Delta m_M'=0}^\infty \prod_{i=1}^{M+1} \Bigg[ \sum_{k_i = 0}^{N-m_i} \sum_{l_i=-k_i}^{m_i - m_{i-1} - k_{i-1}} \Big( \theta(l_i-1) + \delta_{k_i, m_{i-1} + k_{i-1} - m_i} \Big) \\
&\times (1-r_{i-1})^{\Delta n_{i-1}}  P^{k_i+l_i}_{(m_i+k_i - m_{i-1}-k_{i-1})/(N-m_{i-1}-k_{i-1})}(\Delta m_{i-1}' - \Delta m_i') \Bigg],
\end{aligned}
\end{equation}
To evaluate it, we can make use of the fact that probability $P^j_{i/N}(n)$ is only nonzero for $n \geq 0$, and that the sum only contains terms for which $\Delta m_i' \geq 0$.
Thus, for some number $0<a<1$,
\begin{equation}
\begin{aligned}
&\sum_{\Delta m_{i-1}'=0}^\infty a^{\Delta m_{i-1}'}  P^j_{i/N}(\Delta m_{i-1}' - \Delta m_i') \\
&= \sum_{\Delta m_{i-1}' = \Delta m_i'}^\infty a^{\Delta m_{i-1}'}  P^j_{i/N}(\Delta m_{i-1}' - \Delta m_i') \\
&= \sum_{n=0 }^\infty a^{n + \Delta m_i'}  P^j_{i/N}(n)\\
&= \expval{ a^{n^j_{i/N}} } a^{\Delta m_i'}.
\end{aligned}
\end{equation}
This shows that the summation over $\Delta m_i'$ cannot be resolved independently from the summation over $\Delta m_{i-1}'$.
However, the summation over $\Delta m_{i-1}'$ can be safely performed before the summation over $\Delta m_i'$, as shown above.
Thus, our strategy is to sum over the $\Delta m_i'$'s in the order of their index
(i.e. $\Delta m_0'$ first, $\Delta m_M'$ last).
For $\Delta m_0'$, we get
\begin{equation}
\expval{(1-r_0)^{n^{k_1 + l_1}_{m_1 + k_1 - m_0 - k_0/(N-m_0-k_0)}}} (1-r_0)^{\Delta m_1'}.
\end{equation}
Before performing the sum over $\Delta m_1'$,
we must remember to also include the $(1-r_0)^{\Delta m_1'}$ that came out of the sum over $\Delta m_0'$ and thus we get
\begin{equation}
\expval{[(1-r_0)(1-r_1)]^{n^{k_2 + l_2}_{m_2 + k_2 - m_1 - k_1/(N-m_1-k_1)}}} [(1-r_0)(1-r_1)]^{\Delta m_2'}.
\end{equation}
Then, for the sum over $\Delta m_3'$, we should not forget to add the $[(1-r_0)(1-r_1)]^{\Delta m_2'}$ to the $(1-r_2)^{\Delta m_2'}$ already present.
And so on. The result is
\begin{equation}\label{eq:app:G_as_sum}
\begin{aligned}
&G(r_1, r_2, ..., r_M) \\
& = \prod_{i=1}^{M+1} \Bigg[ \sum_{k_i = 0}^{N-c_i} \sum_{l_i=-k_i}^{c_i - c_{i-1} - k_{i-1}} \Big( \theta(l_i-1) + \delta_{k_i, c_{i-1} + k_{i-1} - c_i} \Big) \\
&\times \Bigg<  \Big( \prod_{j=0}^{i-1} (1 - r_j) \Big)^{n^{k_i+l_i}_{(c_i + k_i - c_{i-1} - k_{i-1})/(N-c_{i-1}-k_{i-1})}}\Bigg> \Bigg]\\
& = \prod_{i=1}^{M+1} \Bigg[ \sum_{k_i = 0}^{N-c_i} \sum_{l_i=-k_i}^{c_i - c_{i-1} - k_{i-1}} \Big( \theta(l_i-1) + \delta_{k_i, c_{i-1} + k_{i-1} - c_i} \Big) \\
&\times \Big<  \big( 1 - \bar r_{i-1} \big)^{n^{k_i+l_i}_{(c_i + k_i - c_{i-1} - k_{i-1})/(N-c_{i-1}-k_{i-1})}}\Big> \Bigg],
\end{aligned}
\end{equation}
where we defined
\begin{equation}
\bar r _i = 1 - \prod_{j=0}^i (1-r_j).
\end{equation}

The next step is to calculate the expected values of the form encountered in the above equation.
That is, we need to calculate
\begin{equation}\label{eq:app:<(1-r^n)>}
\Big< (1-r)^{n^j_{i/N}} \Big> = \sum_{n=0}^\infty P^j_{i/N} (n) (1-r)^n.
\end{equation}
We can use equation \eqref{eq:app:P^j_i/N_recursion} to write down the recursive relation
\begin{equation}
\begin{aligned}
&\Big< (1-r)^{n^j_{i/N}} \Big>\\
 &= \binom{N-i+j}{j} \frac{q_{\text{link}}^j (1-q_{\text{link}})^{N-i}}{1- (1-q_{\text{link}})^{N-i+j}}  \sum_{n=0}^\infty \sum_{n'=0}^{n-1}(1-r)^n P_{1/(N-i+j)}(n-n') \sum_{l=0}^{i-j} P^l_{(i-j)/N}(n') \\
&= \binom{N-i+j}{j} \frac{q_{\text{link}}^j (1-q_{\text{link}})^{N-i}}{1- (1-q_{\text{link}})^{N-i+j}} \sum_{l=0}^{i-j} \sum_{\Delta n=1}^\infty\sum_{n'=0}^\infty  (1-r)^{n' + \Delta n} P_{1/(N-i+j)}(\Delta n) P^l_{(i-j)/N}(n') \\
&= \binom{N-i+j}{j} \frac{q_{\text{link}}^j (1-q_{\text{link}})^{N-i}}{1- (1-q_{\text{link}})^{N-i+j}} \expval{(1-r)^{n_{1/(N-i+j)}}} \sum_{l=0}^{i-j} \Big< (1-r)^{n^l_{(i-j)/N}} \Big>.
\end{aligned}
\end{equation}
Since $n_{1/N}$ is geometric with $1 / \expval{n_{1/N}} = 1 - (1 - q_\text{link})^N$,
and since
\begin{equation}\label{eq:app:a^x}
\expval{a^x} = aq / (1 - a[1-q])
\end{equation}
for any geometric variable $x$ with $1 / \expval{x} = q$ and $0 < a < 1$,
we can write
\begin{equation}\label{eq:<F>:recursive}
\Big< (1-r)^{n^j_{i/N}} \Big>= \binom{N-i+j}{j} \frac{q_{\text{link}}^j (1-q_{\text{link}})^{N-i}(1-r)}{1- (1-r)(1-q_{\text{link}})^{N-i+j}} \sum_{l=0}^{i-j}\Big< (1-r)^{n^l_{(i-j)/N}} \Big>.
\end{equation}
for $i>0$.

\subsection{Recursive Relation} \label{sec:app:recursive}

We will now proceed in the limit $q_{\text{link}}, r \ll 1$, since this is the regime that we are mostly interested in, and since this allows for some convenient approximations.
Throwing out higher-order terms in both $r$ and $q_{\text{link}}$, we get
\begin{equation}
\Big< (1-r)^{n^j_{i/N}} \Big> \approx \binom{N-i+j}{j} \frac{q_{\text{link}}^j }{r + (N-i+j)q_{\text{link}}} \sum_{l=0}^{i-j}\Big< (1-r)^{n^l_{(i-j)/N}} \Big>.
\end{equation}
Now, we will argue that to leading order in $q_\text{link}$ and $r$, we only need to consider the term for which $l=1$,
making it much easier to resolve the recurrence relation.
\\

Let us for the moment represent $\expval{(1-r)^{n^b_{a/N}}}$ schematically by the tuple $(a,b)$.
Then, any $(i,j)$ is expressed as a sum over $(i-j,l_1)$'s, for $l_1 = 0, 1 , ...,i-j$.
In turn each $(i-j,l_1)$ will be a sum over $(i-j-l_1,l_2)$'s for $l_2 = 0, 1, ..., i-j-l_1$.
Therefore, each term in the sum can be represented by a sequence
\begin{equation}
\text{term in sum} = \Big( (a_0, b_0), (a_1, b_1), (a_2, b_2), \cdots \Big)
\end{equation}
following the rule $a_{i+1} = a_i - b_i$ and the boundary condition $a_0 =i, b_0 =j$.
Now, since
\begin{equation}\label{eq:app:(1-r)^n^0_i/N}
\Big< (1-r)^{n^0_i/N} \Big> = \sum_{n=0}^\infty P^0_{i/N} (n) (1-r)^n = \sum_{n=0}^\infty \delta_{i,0} \delta_{n,0} (1-r)^n = \delta_{i,0},
\end{equation}
tuples of the form $(a, 0)$ can only occur in the sequence if $a = 0$.
That means $b_i > 0$ for each tuple where $a_i \neq 0$.
As a result, $a_{i+1} \leq a_i - 1$ unless $a_i = 0$.
Furthermore, the $(0, 0)$ term itself does not contain a reference other $(a, b)$; it simply has the value one.
Thus, the recurrence relation terminates when $a_i = 0$ is reached.\\

As a consequence, we can rewrite the sequence above as
\begin{equation}
\text{term in sum} = \Big( (i, j), (i - j, l_1), (i - j - l_1, l_2), \cdots, (i - j - \sum_{i=1}^{K-1} l_i, l_K), (0, 0) \Big),
\end{equation}
for some $l_i > 0$ for $i = 1, 2, \cdots, K$ and for some value $K$.
This sequence can be thought of as a ``path'' from $(i, j)$ to $(0, 0)$.
Each path is uniquely defined by a sequence $(l_1, l_2, \cdots, l_K)$,
and each such sequence uniquely defines a path as long as it satisfies the condition
\begin{equation}
\sum_{i=1}^K l_i = i - j.
\end{equation}
Note that as $l_i \geq 1$, this automatically imposes $K \leq i - j$.
We denote the set of all sequences $(l_1, l_2, \cdots, l_K)$ that define a path from $(i, j)$ to $(0, 0)$ by $\mathcal L_{i,j}$,
which allows us to expand the recurrence relation as
\begin{equation}
\begin{aligned}
\Big< (1-r)^{n^j_{i/N}} \Big> &\approx \sum_{ (l_1,...,l_{K} )\in \mathcal L_{i,j} } \binom{N-i+j}{j} \frac{q_{\text{link}}^j }{r + (N-i+j)q_{\text{link}}}  \Big< (1-r)^{n^0_{0/N}} \Big>\\
&\times \prod_{k = 1}^{K}\binom{N-(i-j-\sum_{a=1}^{k-1} l_a) +l_k}{l_k} \frac{q_{\text{link}}^{l_k} }{r + (N-(i-j-\sum_{a=1}^{k-1} l_a) +l_k)q_{\text{link}}} \\
&= \sum_{ (l_1,...,l_K )\in \mathcal L_{i,j}} \frac{q_{\text{link}}^{j + \sum_{k=1}^{K} l_k }}{r + (N-i+j)q_{\text{link}}} \prod_{k = 1}^{K} \frac{1 }{r + (N-(i-j-\sum_{a=1}^{k-1} l_a) +l_k)q_{\text{link}}}\\
&\times  \binom{N-i+j}{j} \prod_{k = 1}^{K}\Big[ \binom{N-(i-j-\sum_{a=1}^{k-1} l_a) +l_k}{l_k}     \\
&= \sum_{ (l_1,...,l_K )\in \mathcal L_{i,j} } \frac{q_{\text{link}}^i}{\mathcal O \Big((r+ q_{\text{link}})^{K}\Big)} \times \mathcal O (r^0q_{\text{link}}^0).
\end{aligned}
\end{equation}

For $r, q_\text{link} \ll 1$, this sum will be dominated by paths that have the largest $K$.
As explained above, the maximum value that $K$ can take is $i - j$.
Furthermore, there is exactly one path that realizes this value,
which is defined by $l_a = 1$ for $a = 1, 2, \cdots, i-j$.
When we keep only this path in the above equation, we find
\begin{equation} \label{eq:app:resolving_recursive_relation}
\begin{aligned}
\Big< (1-r)^{n^j_{i/N}} \Big> &\approx \binom{N-i+j}{j}  \frac{q_{\text{link}}^i}{r + (N-i+j)q_{\text{link}}}\\
& \times \prod_{k=1}^{i-j}  \binom{N- (i-j-(k-1)) + 1}{1} \frac 1 {r+ (N- (i-j-(k-1)) + 1)q_{\text{link}}} \\
&= \binom{N-i+j}{j}  \frac{q_{\text{link}}^{j-1}}{N-i+j} \prod_{k=N-i+j}^{N}   \frac {kq_{\text{link}}} {r+ kq_{\text{link}}}.
\end{aligned}
\end{equation}
We must note that all of the above is only valid for $i>0$, since the recursive relation \eqref{eq:<F>:recursive} is not applicable for $i=0$.
In order to also incorporate equation \eqref{eq:app:(1-r)^n^0_i/N}, we write
\begin{equation} \label{eq:app:(1-r)^n^j_i/N}
\Big< (1-r)^{n^j_{i/N}} \Big> \approx \Big( \theta(j-1) + (r+Nq_{\text{link}}) \delta_{i,0} \Big)\binom{N-i+j}{j}  \frac{q_{\text{link}}^{j-1}}{N-i+j} \prod_{k=N-i+j}^{N}   \frac {kq_{\text{link}}} {r+ kq_{\text{link}}}.
\end{equation}

How can we interpret the dominance of terms corresponding to the ``longest path''?
What it means is that realizations of Protocol \ref{prot:factory} for which multiple successes occur during the same round occur with suppressed probability,
as shown by the fact that we only include $P^j_{i/N}$'s for which $j=1$.
This can also be intuitively expected:
if for each quantum connection the probability of distributing a Bell state per round is very small ($q_\text{link} \ll 1$),
there will be a large spread in the rounds during which the different Bell states are distributed.
It will then be very unlikely that two Bell states are distributed during the exact same round.
However, when $r$ is large (close to 1), the quantity $(1 - n)^n$ will decrease very quickly with $n$.
The average will then have much larger weight for small $n$ than for large $n$.
However, these terms with small $n$ are exactly those that are excluded by the large spread implied by $q_\text{link} \ll 1$.
In fact, if $r = 1 - \epsilon$ with $\epsilon \ll 1$, the only linear term in the average is the one corresponding to $n = 1$,
which implies all Bell states being distributed collectively during the first round ( $P^j_{i/N}$ with $j=i$).
This explains why neglecting simultaneous successes requires both $q_\text{link}$ and $r$ to be small.
\\

Finally, before we move on, we are interested to know whether equation \eqref{eq:app:(1-r)^n^j_i/N} also holds for $r=0$.
This does not follow from the above, because the use of equation \eqref{eq:app:a^x} required $0<r<1$.
For $r=0$, Eq. \eqref{eq:app:<(1-r^n)>} yields
\begin{equation}
\Big<(1-r)^{n^j_{i/N}} \Big| _{r=0} \Big> = \sum_{n=0}^\infty P^j_{i/N}(n).
\end{equation}
Using again equation \eqref{eq:app:P^j_i/N_recursion} we find
\begin{equation}
\begin{aligned}
\Big< (1-r)^{n^j_{i/N}}\Big|_{r=0}\Big> &= \binom{N-i+j}{j} \frac{q_{\text{link}}^j (1-q_{\text{link}})^{N-i}}{1- (1-q_{\text{link}})^{N-i+j}} \sum_{n=0}^\infty \sum_{n'=0}^{n-1} P_{1/(N-i+j)}(n-n')P'_{(i-j)/N}(n') \\
&= \binom{N-i+j}{j} \frac{q_{\text{link}}^j (1-q_{\text{link}})^{N-i}}{1- (1-q_{\text{link}})^{N-i+j}} \sum_{\Delta n=0}^\infty  P_{1/(N-i+j)}(\Delta n) \sum_{l=0}^{i-j}\sum_{n'=0}^{\infty}P^l_{(i-j)/N}(n') \\
&= \binom{N-i+j}{j} \frac{q_{\text{link}}^j (1-q_{\text{link}})^{N-i}}{1- (1-q_{\text{link}})^{N-i+j}} \sum_{l=0}^{i-j} \Big< (1-r)^{n^l_{(i-j)/N}}\Big|_{r=0}\Big>
\end{aligned}
\end{equation}
which is exactly recursive relation \eqref{eq:<F>:recursive} but with $r=0$.
Because
\begin{equation}
\Big< (1-r)^{n^0_i/N}\Big|_{r=0} \Big> = \delta_{i,0},
\end{equation}
the recursive relation expresses any $(i,j)$ in terms of $(0,0)$'s, and these are expressed the same for both $r=0$ and $0<r<1$.
Because both the recursive relation and the final term $(0, 0)$ can be written the same,
we conclude that it does not matter whether $r$ is set to zero before or after resolving the recursion relation.
Therefore,
\begin{equation}
\Big< (1-r)^{n^j_i/N}|_{r=0} \Big> = \Big< (1-r)^{n^j_i/N}\Big|_{0<r<1} \Big> \Big|_{r=0}.
\end{equation}
Thus, equation \eqref{eq:app:(1-r)^n^j_i/N} is valid for $0\leq r \ll 1$.
This means that we do not need to treat the $r_0$ that we defined to be zero before any differently from the other $r_i$'s when calculating $G((r_1, r_2, \cdots, r_M)$,
and our results are still valid if $r_i = 0$ for some $0 < i < M$.
\\

\subsection{Counting Orders}

Now, we can in principle substitute Eq. \eqref{eq:app:(1-r)^n^j_i/N} into Eq. \eqref{eq:app:G_as_sum}.
However, if we limit ourselves to leading order in $q_\text{link}$ and the various $r_i$ variables
(which we denote as all being of order $\mathcal O(r)$),
this allows us to disregard part of the summation.
In this section, we count orders to find that only $l_i=1$ and $k_i=0$ terms contribute to $G$ at leading order.
This allows us to more easily calculate $G$ to leading order in the next section.
\\

First of all, note that
\begin{equation}
\bar r _i \equiv 1 - \prod_{j=0}^i (1-r_j) = \sum_{j=0}^i r_j + \mathcal O (r^2).
\end{equation}
Therefore, each $\bar r_i$ is of order $\mathcal O(r)$.
Furthermore, from Eq. \eqref{eq:app:(1-r)^n^j_i/N} we see that
\begin{equation}
\Big< (1-r)^{n^j_{i/N}} \Big> = \Big( \theta(j-1) + \delta_{i,0} \mathcal O (r+q_\text{link}) \Big)\mathcal O \Big( \frac{q_\text{link}^i}{(r+q_\text{link})^{i-j+1}} \Big).
\end{equation}
Substituting this into equation \eqref{eq:app:G_as_sum} yields
\begin{equation}
\begin{aligned}
&G(r_1, r_2, \cdots, r_M) \\
&= \prod_{i=1}^{M+1} \Bigg[ \sum_{k_i=0}^{N-c_i} \sum_{l_i = -k_i}^{c_i - c_{i-1} - k_{i-1}} \Big( \theta(l_i -1) + \delta_{k_i,c_{i-1}+k_{i-1}-c_i} \Big)\\
&\times  \Big( \theta(k_i + l_i -1) + \delta_{c_i + k_i - c_{i-1} - k_{i-1},0}   \mathcal O (r+q_\text{link}) \Big)\\
&\times\mathcal O \Big( \frac {q_\text{link}^{c_i +k_i - c_{i-1} - k_{i-1}}} {(r+q_\text{link})^{c_i - c_{i-1} - k_{i-1} - l_i + 1}} \Big) \Bigg] \\
&= \prod_{i=1}^{M+1} \Bigg[ \sum_{k_i=0}^{N-c_i} \sum_{l_i = -k_i}^{c_i - c_{i-1} - k_{i-1}} \Big( \theta(l_i -1) + \delta_{k_i,c_{i-1} + k_{i-1}-c_i}   \mathcal O (r+q_\text{link}) \Big)\\
&\times\mathcal O \Big( \frac {q_\text{link}^{c_i +k_i - c_{i-1} - k_{i-1}}} {(r+q_\text{link})^{c_i - c_{i-1} - k_{i-1} - l_i + 1}} \Big) \Bigg].
\end{aligned}
\end{equation}
Here, we have used the fact that for every term in the sum, $k_i \geq 0$, and thus $\theta(l_i-1) \theta(k_i+l_i-1) = \theta(l_i - 1)$.
Furthermore, the delta functions are the same, and squaring it gives the same delta function again.
Cross terms $\theta \times \delta$ vanish, because the only term for which the delta function does not vanish has $l_i = -k_i \leq 0$, making the step function vanish.
Now, we make use of the identity
\begin{equation}
\prod_i^N \left(\sum_{x_{i}} f(x_i)\right) = \sum_{x_1} \sum_{x_2} \cdots \sum_{x_N} f(x_1) f(x_2) \cdots f(x_N) = \prod_{i} \Bigg(\sum_{x_i}\Bigg) \prod_i \Bigg(f(x_i) \Bigg)
\end{equation}
to split the product in  ``three parts'' and hence collect part of the order counting in a way that is very convenient, giving
% \begin{equation}
% \begin{aligned}
% &G(r_1, r_2, \cdots, r_M) \\
% &= \prod_{i=1}^{M+1} \Big[ \sum_{k_j=0}^{N-c_j}\Big] \mathcal O \Bigg( \left( \frac {q_\text{link}}{r+q_\text{link}} \right)^{\sum_{j=1}^{M+1} (c_i + k_i - c_{i-1} - k_{i-1})} \Bigg)\\
% &\times \prod_{i=1}^{M+1}  \sum_{l_i = -k_i}^{c_i - c_{i-1} - k_{i-1}}  \Big( \theta(l_i -1) + \delta_{k_i,c_{i-1}+k_{i-1}-c_i} \mathcal O (r+q_\text{link}) \Big) \mathcal O \Big( (r+q_\text{link})^{l_i+k_i-1} \Big).
% \end{aligned}
% \end{equation}
\begin{equation}
\begin{aligned}
&G(r_1, r_2, \cdots, r_M) \\
&= \prod_{i=1}^{M+1} \Bigg[ \sum_{k_i=0}^{N-c_i}\Bigg] \prod_{i=1}^{M+1} \Bigg[\mathcal O \Bigg( \left( \frac {q_\text{link}}{r+q_\text{link}} \right)^{c_i + k_i - c_{i-1} - k_{i-1}} \Bigg)\Bigg]\\
&\times \prod_{i=1}^{M+1} \Bigg[ \sum_{l_i = -k_i}^{c_i - c_{i-1} - k_{i-1}}  \Big( \theta(l_i -1) + \delta_{k_i,c_{i-1}+k_{i-1}-c_i} \mathcal O (r+q_\text{link}) \Big) \mathcal O \Big( (r+q_\text{link})^{l_i+k_i-1} \Big)\Bigg] \\
&= \prod_{i=1}^{M+1} \Big[ \sum_{k_i=0}^{N-c_i}\Big] \mathcal O \Bigg( \left( \frac {q_\text{link}}{r+q_\text{link}} \right)^{\sum_{i=1}^{M+1} \left(c_i + k_i - c_{i-1} - k_{i-1}\right))} \Bigg)\\
&\times \prod_{i=1}^{M+1}  \sum_{l_i = -k_i}^{c_i - c_{i-1} - k_{i-1}}  \Big( \theta(l_i -1) + \delta_{k_i,c_{i-1}+k_{i-1}-c_i} \mathcal O (r+q_\text{link}) \Big) \mathcal O \Big( (r+q_\text{link})^{l_i+k_i-1} \Big).
\end{aligned}
\end{equation}
This then allows us to make use of
\begin{equation}
\sum_{i=1}^{M+1} (c_i + k_i - c_{i-1} - k_{i-1}) = c_{M+1} +k_{M+1} - c_{0} - k_{0} = N,
\end{equation}
since we manually defined $m_0 = k_0 = 0$ and $m_{M+1} = N$,
and since the sum over $k_{M+1}$ only runs over $k_{M+1} = 0$
(the last success cannot ``overshoot'' as all Bell states are already in place).
Thus, this quantity is the same for every term and can safely be taken out of the sum.
\\

Now, working out the $\theta$ and $\delta$ parts separately, we get
\begin{equation}
\begin{aligned}
&G(r_1, r_2, \cdots, r_M) \\
&= \mathcal O \Bigg( \left( \frac {q_\text{link}}{r+q_\text{link}} \right)^{N} \Bigg)\prod_{i=1}^{M+1} \Big[ \sum_{k_i=0}^{N-c_i}\Big] \prod_{i=1}^{M+1} \Big[ \\
&  \sum_{l_i = 1}^{c_i - c_{i-1} - k_{i-1}} \mathcal O \Big( (r+q_\text{link})^{l_i + k_i-1} \Big) + \delta_{k_i, c_{i-1}+k_{i-1}-c_i}  \Big].
\end{aligned}
\end{equation}
The part where we sum over $l_i$ now is clearly dominated by the term for which $l_i$ is lowest, since a larger $l_i$ means a larger order in $r+q_\text{link}$.
Since this is $l_i = 1$, we find
\begin{equation}
\begin{aligned}
&G(r_1, r_2, \cdots, r_M) \\
&= \mathcal O\Bigg( \left( \frac {q_\text{link}}{r+q_\text{link}} \right)^{N} \Bigg)\prod_{i=1}^{M+1} \Big[ \sum_{k_i=0}^{N-c_i}\Big] \prod_{i=1}^{M+1} \Bigg[ \\
&  \theta(c_i - c_{i-1} - k_{i-1}-1) \mathcal O \Big( (r+q_\text{link})^{ k_i} \Big) + \delta_{k_i,c_{i-1}+k_{i-1}-c_i}  \Bigg],
\end{aligned}
\end{equation}
where the step function is due to the summation over $l_i$ being empty and hence zero for $c_i - c_{i-1} - k_{i-1} < 1$.
This quantity will be dominated by terms which are products of $\delta$'s, and of $\theta$'s with $k_i=0$, since these terms do not carry an additional $\mathcal O(r+q_\text{link})$.
Now note that the Kronecker $\delta$ function $\delta_{k_i,c_{i-1}+k_{i-1}-c_i}$ enforces $k_{i-1} \geq c_i - c_{i-1} > 0$.
This implies two things.
Firstly, it implies that any term that contains a $\theta(c_i - c_{i-1} - k_{i-1}-1)$ for $i=j$ but a $\delta_{k_i,c_{i-1}+k_{i-1}-c_i} $ for $i = j+1$ will be of higher order in $r + q_\text{link}$.
Secondly, because $k_0 = 0$ by definition, it implies that all nonzero terms of the sum must ``start'' with a $\theta$, i.e. include a $\theta(c_i - c_{i-1} - k_{i-1}-1)$ for $i=1$.
Together, these two implications mean any leading terms cannot contain a $\delta$;
they only contain $\theta$'s.
The only leading term with only $\theta$'s is the one for which all $k_i$'s are 0.
Combining this with what we found for the $l_i$'s,
we can conclude that the leading contribution to $G$ has $l_i=1$ and $k_i = 0$ for $i = 0, 1, 2, ..., M+1$.
This can again be interpreted as neglecting the possibility that multiple Bell states are distributed simultaneously.

\subsection{Calculating $G$}

Now, we are ready to calculate $G$ to leading order.
Only keeping $l_i=1$, $k_i=0$ in Eq. \eqref{eq:app:G_as_sum}
and then filling in Eq. \eqref{eq:app:(1-r)^n^j_i/N},
we find
\begin{equation} \label{eq:app:working_out_G}
\begin{aligned}
&G(r_1, r_2, \cdots, r_M) \\
& \approx \prod_{i=1}^{M+1}  \Big<  \big( 1 - \bar r_{i-1} \big)^{n^{1}_{(c_i - c_{i-1})/(N-c_{i-1})}}\Big> \\
& \approx  \prod_{i=1}^{M+1} \Bigg[ \Big( \theta(1-1) + (\bar r_{i-1}+Nq_\text{link}) \delta_{c_i-c_{i-1},0} \Big) \\
&\times \binom{N-c_i+1}{1}  \frac{1}{N-c_i+1} \prod_{k=N-c_i+1}^{N-c_{i-1}}   \frac {k q_\text{link}} {\bar r_{i-1}+ kq_\text{link}}\Bigg]\\
&=  \prod_{i=1}^{M} \prod_{k=c_i+1}^{c_{i+1}}   \frac {(N+1-k) q_\text{link}} {\bar {r}_i+ (N+1-k)q_\text{link}}\\
&\approx  \prod_{i=1}^{M} \prod_{k=c_i+1}^{c_{i+1}}   \frac {(N+1-k) q_\text{link}} {\sum_{j=1}^i r_{j}+ (N+1-k)q_\text{link}}.
\end{aligned}
\end{equation}
Here, we have used the fact that $r_0 \equiv 0$ (and thus $\bar r_{1} = 0$) to drop the lowest term in the product.
This can also be rewritten as
\begin{equation} \label{eq:app:G_final}
G(r_1, r_2, \cdots, r_M) \approx \prod_{k=1}^N \frac{ (N+1-k)q_\text{link}}{\sum_{%i \text{ s.t. }
c_i < k} r_i + (N+1-k)q_\text{link}}.
\end{equation}

\subsection{Lower Bound}

Apart from the leading-order approximation of the function $G$ derived above,
we can also derive a lower bound.
At the core of the approximation lies the fact that, to leading order in $q_\text{link}$ and $r$,
we are able to ignore all events for which multiple Bell states are distributed during the same round.
That function $G$ obtained by ignoring these events is an average over a sub-normalized probability distribution,
and thus provides a lower bound on the real function.
In turn, using a lower bound of the function $G$ to evaluate the fidelity (Eq. \eqref{eq:factory_fidelity}) gives a lower bound on the real fidelity.
Even so, the result Eq. \eqref{eq:app:G_final} is not necessarily a lower bound on the function $G$.
The reason for this is that,
in order to work consistently at leading order,
we have thrown out some additional terms that are not linked to ignoring multiple simultaneous successes.
Some of these terms would lower the function G if they were kept,
and thus Eq. \eqref{eq:app:G_final} is only a lower bound if the effect of throwing out these terms is smaller than the effect of throwing out events corresponding to multiple simultaneous successes.
We do not know if this is generally the case.
\\

In this section, we derive a lower bound by repeating the above calculation without throwing out these additional terms.
That means that we are not working at leading order,
but just deriving a lower bound by throwing out all contributions to $G$ due to multiple distributed Bell states during the same round.
We start by lower-bounding the expected value $\expval{(1-r)^{n^j_{i/N}}}$.
To this end, we use the recursive relation Eq. \eqref{eq:<F>:recursive}.
Because the factor in front of the summation is a positive quantity,
and because each term of the sum is ultimately expressed in terms of $\expval{(1-r)^{n^0_0/N}} = 1$ (see Eq. \eqref{eq:app:(1-r)^n^0_i/N}),
we can conclude that
\begin{equation}
\expval{(1-r)^{n^j_{i/N}}} \geq 0.
\end{equation}
Because of this, Eq. \eqref{eq:<F>:recursive} tells us
\begin{equation}
\Big< (1-r)^{n^j_{i/N}} \Big> \geq \binom{N-i+j}{j} \frac{q_{\text{link}}^j (1-q_{\text{link}})^{N-i}(1-r)}{1- (1-r)(1-q_{\text{link}})^{N-i+j}} \Big< (1-r)^{n^1_{(i-j)/N}} \Big>.
\end{equation}
This inequality can be applied recursively until reaching
\begin{equation}
\Big< (1-r)^{n^1_{1/N}} \Big> = \frac 1 N \frac{q_\text{link}  (1 - q_\text{link})^{N-1} (1 - r)}{1 - (1-r)(1-q_\text{link})^N}.
\end{equation}
This is exactly the ``leading order path'' discussed in Section \ref{sec:app:recursive} and yields,
in analogue to Eq. \eqref{eq:app:resolving_recursive_relation},
\begin{equation}
\begin{aligned}
\Big< (1-r)^{n^j_{i/N}} \Big> &\geq \binom{N-i+j}{j}\frac{q_{\text{link}}^j (1-q_{\text{link}})^{N-i}(1-r)}{1- (1-r)(1-q_{\text{link}})^{N-i+j}} \\
& \times \prod_{k=1}^{i-j}  \binom{N - i + j + k}{1} \frac{q_{\text{link}} (1-q_{\text{link}})^{N-i+j+k-1}(1-r)}{1- (1-r)(1-q_{\text{link}})^{N-i+j+k}}.
\end{aligned}
\end{equation}
We will now focus on the case $j=1$,
since this will ultimately be the only type of term occurring in the lower bound for $G$
(after all, $j>1$ would correspond to distributing multiple Bell states during the same round).
We then find
\begin{equation}
\begin{aligned} \label{eq:app:bound_on_(1-r)^n}
\Big< (1-r)^{n^1_{i/N}} \Big> &\geq \prod_{k=0}^{i-j}  \left( N-i+k+1 \right) \frac{q_{\text{link}} (1-q_{\text{link}})^{N-i+k}(1-r)}{1- (1-r)(1-q_{\text{link}})^{N-i+k+1}} \\
&= \prod_{k={N-i+1}}^N \frac{kq_\text{link} (1 - q_\text{link})^{k-1} (1 - r)}{1 - (1-r) (1-q_\text{link})^k}.
\end{aligned}
\end{equation}

Now, we can use Eq. \eqref{eq:app:bound_on_(1-r)^n} in combination with Eq. \eqref{eq:app:G_as_sum} to bound $G$.
Because all terms in the sum of Eq. \eqref{eq:app:G_as_sum} are positive,
we can write (analogously to Eq. \eqref{eq:app:working_out_G})
\begin{equation}
\begin{aligned}
G(r_1, r_2, \cdots, r_M) &\geq \prod_{i=1}^{M+1}  \Big<  \big( 1 - \bar r_{i-1} \big)^{n^{1}_{(c_i - c_{i-1})/(N-c_{i-1})}}\Big> \\
&\geq \prod_{i=1}^{M+1} \prod_{k=N-c_i+1}^{N-c_{i-1}}  \frac{kq_\text{link} (1 - q_\text{link})^{k-1} (1 - \bar r_{i-1})}{1 - (1-\bar r_{i-1}) (1-q_\text{link})^k} \\
&= \prod_{i=0}^{M} \prod_{k=c_i+1}^{c_{i+1}}  \frac{(N+1-k)q_\text{link} (1 - q_\text{link})^{N-k} (1 - \bar r_i)}{1 - (1-\bar r_i) (1-q_\text{link})^{N+1-k}}.
\end{aligned}
\end{equation}
This can be rewritten as
\begin{equation} \label{eq:app:G_bound}
G(r_1, r_2, \cdots, r_M) \geq \prod_{k=1}^M \frac{(N+1-k)q_\text{link} (1 - q_\text{link})^{N-k} \prod_{c_i < k} (1 - r_i)}{1 - (1-q_\text{link})^{N+1-k}\prod_{c_i < k} (1 - r_i) }.
\end{equation}

\section{Expected Value of Distribution Time} \label{app:rate}

In this appendix, we use the tools developed in Appendix \ref{app:exp_val} to prove the equation
\begin{equation} \label{eq:app_rate:result}
\expval{n_{i/N}} \approx \frac 1 {q_\text{link}} \sum_{k = N + 1 - i}^{N} \frac 1 {k}
\end{equation}
is true up to leading order in $q_\text{link}$.
Here, $n_{i/N}$ is the number of rounds required to distribute $i$ Bell states over $N$ quantum connections.
That is, it is the $i^\text{th}$ largest value out of $\{n_1, n_2, ..., n_N\}$,
where we remind the reader that each $n_j$ is a geometrically-distributed random variable with mean $\tfrac 1 {q_\text{link}}$.
Additionally, we provide the upper bound
\begin{equation} \label{eq:app_rate:result_upper_bound}
\expval{n_{i/N}} \leq \sum_{k=N+1-i}^{N} \frac{1}{1 - (1 - q_\text{link}) ^ {k}}.
\end{equation}

We note that it directly follows from Eq. \eqref{eq:app_rate:result} that
\begin{equation} \label{eq:app_rate:max_of_geoms}
\expectationvalue{n_{N/N}} \equiv \expectationvalue{n_\text{all}} \equiv \expectationvalue{\max \{n_1, n_2, ..., n_N\}} \approx  \frac {H_N}{q_\text{link}},
\end{equation}
where $H_N$ is the $N^\text{th}$ harmonic number,
is valid up to leading order in $q_\text{link}$.
This is a well-known result \cite{coopmansImprovedAnalyticalBounds2022, shchukinWaitingTimeQuantum2019, schmidtMemoryassistedLongdistancePhasematching2020}.
Additionally, Eq. \eqref{eq:app_rate:result_upper_bound} can be used to upper bound $\expectationvalue{n_{N/N}}$.
However, the bound is less tight than the existing bound given in Eq. \eqref{eq:rate:bound}.
\\

We now explain the intuition behind Eq. \eqref{eq:app_rate:result}.
If $k > 1$ connections try to establish entanglement,
the first success will occur sooner than when only one connection is trying.
For one connection, the time it takes is on average $\tfrac 1 {q_\text{link}}$
(this is the expected value of the geometric distribution).
But when there are $k$ connections trying, there is a ``boost factor'';
entanglement is generated exactly $k$ times faster,
and therefore the time required is on average only $\tfrac 1 {kq_\text{link}}$.
In the limit $q_\text{link} \to 0$, it is very unlikely that multiple Bell states are distributed during the same round,
and therefore one can repeatedly use this argument to go from success to success.
The rest of this appendix is dedicated to proving Eq. \eqref{eq:app_rate:result},
thereby making the intuitive argument exact.

\subsection{Exact Recursion Relation}

The random variable $n_{i/N}$ follows the probability distribution $P_{i/N}$ defined in Eq. \eqref{eq:app:P_i/N}.
Key to deriving Eq. \eqref{eq:app_rate:result},
is to determine the difference between $\expval{n_{(i+1)/N}}$ and $\expval{n_{i/N}}$,
as it allows us to write a recursion relation.
To this end, we first take the difference between their probability distributions.
Using Eq. \eqref{eq:app:expansion_of_P_i/N} yields
\begin{equation}
\begin{aligned}
P_{(i+1)/N} - P_{i/N} &= \sum_{k=0}^{N-i-1} \sum_{l=1}^{i+1} P_{(k+i+1)/N}^{k+l} - \sum_{k=0}^{N-i} \sum_{l=1}^i P_{(k+i)/N}^{k+l} \\
&= \sum_{k=1}^{N-i} P^{k}_{(k+i)/N} - \sum_{l=1}^i P^l_{i/N} \\
&= \sum_{k=1}^{N-i} P^{k}_{(k+i)/N} - P'_{i/N}.
\end{aligned}
\end{equation}
From linearity of the average, it then follows directly that
\begin{equation}\label{eq:app_rate:difference_1}
\expval{n_{(i+1)/N}} - \expval{n_{i/N}} = \sum_{k=1}^{N-i} \expval{n^k_{(k+1)/N}} - \expval{n'_{i/N}}.
\end{equation}

To evaluate Eq. \eqref{eq:app_rate:difference_1},
we first give an expression for $\expval{n^k_{(k+1)/N}}$.
We use Eq. \eqref{eq:app:P^j_i/N_recursion} to write
\begin{equation}
\expval{n^j_{i/N}}= \binom{N-i+j}{j} \frac{q_\text{link}^j(1-q_\text{link})^{N-i}}{1- (1-q_\text{link})^{N-i+j}} \sum_{n=1}^\infty \sum_{n'=0 }^{n-1} nP_{1/(N-i+j)}(n-n') P'_{(i-j)/N}(n').
\end{equation}
This can be calculated by making the change of variables $n = n'+\Delta n$
and using the fact that $P_{1/(N-i+j)}(n)$ is a normalized probability distribution,
giving
\begin{equation}\label{eq:calc_ti/N:t^j_i/N}
\begin{aligned}
\expval{n^j_{i/N}} &= \binom{N-i+j}{j} \frac{q_\text{link}^j(1-q_\text{link})^{N-i}}{1- (1-q_\text{link})^{N-i+j}} \sum_{n'=0}^\infty \sum_{\Delta n = 1}^{\infty} (n'+\Delta n)P_{1/(N-i+j)}(\Delta n) P'_{(i-j)/N}(n') \\
&= \binom{N-i+j}{j} \frac{q_\text{link}^j(1-q_\text{link})^{N-i}}{1- (1-q_\text{link})^{N-i+j}} \Big( \expval{n_{1/(N-i+j)}} T_{(i-j)/N} + \expval{n'_{(i-j)/N}}\Big).
\end{aligned}
\end{equation}
Here, we have defined
\begin{equation}\label{key}
T_{i/N} \equiv \sum_{n=0}^\infty P'_{i/N} (n),
\end{equation}
which is the total probability mass of the sub-normalized probability distribution $P'_{i/N}$
(and therefore always smaller than one).
Then, resolving the summation in Eq. \eqref{eq:app_rate:difference_1} yields
\begin{equation}
\sum_{k=1}^{N-i} \expval{n^k_{(k+1)/N}} = \Big( \expval{n_{1/(N-i)}} T_{i/N} + \expval{n'_{i/N}}\Big)\sum_{k=1}^{N-i}\binom{N-i}{k} \frac{q_\text{link}^k(1-q_\text{link})^{N-i-k}}{1- (1-q_\text{link})^{N-i}}.
\end{equation}
To deal with the final summation, we use the binomial theorem to write
\begin{equation}
\sum_{k=0}^{N-i}\binom{N-i}{k}q_\text{link}^k(1-q_\text{link})^{N-i-k} = \Big( q_\text{link} + (1-q_\text{link}) \Big) ^{N-i} = 1.
\end{equation}
Therefore,
\begin{equation}
\sum_{k=1}^{N-i}\binom{N-i}{k}q_\text{link}^k(1-q_\text{link})^{N-i-k} = 1 - q_\text{link}^0 (1-q_\text{link})^{N-i-0} = 1 - (1 - q_\text{link})^{N-i}
\end{equation}
(note that the lower limit of the summation is one here as opposed to zero).
From this, we conclude conveniently that
\begin{equation}\label{key}
\sum_{k=1}^{N-i}\binom{N-i}{k} \frac{q_\text{link}^k(1-q_\text{link})^{N-i-k}}{1- (1-q_\text{link})^{N-i}} = 1.
\end{equation}
This brings Eq. \eqref{eq:app_rate:difference_1} into the form
\begin{equation}\label{eq:app_rate:difference_2}
\expval{n_{(i+1)/N}} - \expval{n_{i/N}} = \expval{n_{1/(N-i)}} T_{i/N}.
\end{equation}
This recursive relation can be written down in a closed form,
as long as we leave the $T_{i/N}$ explicit.
We then find
\begin{equation}
\expval{n_{i/N}} = \expval{n_{1/N}} + \sum_{k=1}^{i-1} T_{k/N} \expval{n_{1/{N-k}}}.
\end{equation}
It was remarked in Section \ref{sec:app:prob_building_blocks} that $\expval{n_{1/N}}$ is geometrically distributed with $1 / \expval{n_{1/N}} = 1 - (1 - q_\text{link})^N$.
Therefore, we can also write this result at
\begin{equation}\label{eq:app_rate:result_exact}
\expval{n_{i/N}} = \frac 1 {1 - (1 - q_\text{link}) ^ N} + \sum_{k=1}^{i-1} \frac{T_{k/N}}{1 - (1 - q_\text{link}) ^ {N-k}}.
\end{equation}

\subsection{Upper Bound}

Now, we use Eq. \eqref{eq:app_rate:result_exact} to derive an upper bound on $\expval{n_{i/N}}$.
Because $T_{i/N}$ is the total probability mass of a sub-normalized probability function,
we have $T_{i/N} \leq 1$.
From this, it follows directly that Eq. \eqref{eq:app_rate:result_upper_bound} is true.

\subsection{Leading Order}

Finally, we use Eq. \eqref{eq:app_rate:result_exact} to show that Eq. \eqref{eq:app_rate:result} is valid up to leading order in $q_\text{link}$.
Because, to leading order,
\begin{equation}
\frac 1 {1 - (1 - q_\text{link})^N} \approx \frac 1 {N q_\text{link}},
\end{equation}
to leading order we can write Eq. \eqref{eq:app_rate:result_exact} as
\begin{equation}
\expval{n_{i/N}} \approx \frac 1 {q_\text{link}} \Big(\frac 1 N + \sum_{k=1}^{i-1} \frac{T_{k/N}}{N-k} \Big).
\end{equation}
This exactly reduces to Eq. \eqref{eq:app_rate:result} if we can show that $T_{k/N} \approx 1$ to leading order in $q_\text{link}$.
\\

To calculate $T_{i/N}$, we use yet another recursion relation.
First, using Eq. \eqref{eq:app:P'}, we can write (for $i \geq 1$)
\begin{equation}
T_{i/N} = \sum_{l=1}^i \sum_{n=1}^\infty P^l_{i/N}(n).
\end{equation}
Then, using Eq. \eqref{eq:app:P^j_i/N_recursion},
making once more the change in variables $n \to n' + \Delta n$,
and making use of the normalization of $P_{1/N}(n)$,
\begin{equation}\label{eq:app_rate:Ti}
\begin{aligned}
T_{i/N} &= \sum_{l=1}^i \binom{N-i+l}{l} \frac{q_\text{link}^l (1-q_\text{link})^{N-i}}{1- (1-q_\text{link})^{N-i+l}} \sum_{n=1}^\infty  \sum_{n'=0}^\infty P_{1/(N-i+l)}(n-n') P'_{(i-l)/N}(n')\\
&=\sum_{l=1}^i \binom{N-i+l}{l} \frac{q_\text{link}^l (1-q_\text{link})^{N-i}}{1- (1-q_\text{link})^{N-i+l}} \sum_{\Delta n=1}^\infty P_{1/(N-i+l)}(\Delta n) \sum_{n'=0}^\infty  P'_{(i-l)/N}(n') \\
&=\sum_{l=1}^i \binom{N-i+l}{l} \frac{q_\text{link}^l (1-q_\text{link})^{N-i}}{1- (1-q_\text{link})^{N-i+l}} T_{(i-l)/N}.
\end{aligned}
\end{equation}
This recursion relation can be completely resolved if $T_{0/N}$ is known.
From the definition of $P'_{0/N}$ (Eq. \eqref{eq:app:P'_0/N}), we have
\begin{equation}
T_{0/N} = \sum_{n=0}^\infty \delta_{n,0} = 1.
\end{equation}

Now we will resolve the recursion relation to leading order in $q_\text{link}$.
We note that
\begin{equation}
\frac{(1-q_\text{link})^{N-i}}{1-(1-q_\text{link})^ {N-i+l}} = \frac 1 {(N-i+l)q_\text{link}} + \mathcal O (q_\text{link}^ 0),
\end{equation}
and therefore
\begin{equation}
T_{i/N} = \Big( 1 + \mathcal O (q_\text{link}) \Big) T_{(i-1)/N} + \sum_{l=2}^ i \mathcal O (q_\text{link}) T_{(i-l)/N}.
\end{equation}
Thus,
\begin{equation}
T_{i/N} \approx T_{(i-1)/N}
\end{equation}
to leading order.
This holds for every $i \geq 1$ until we hit $T_{0/N} = 1$.
Therefore,
\begin{equation}
T_{i/N} \approx 1
\end{equation}
up to leading order in $q_\text{link}$.
This is exactly what we needed to show,
and therefore we can conclude that Eq. \eqref{eq:app_rate:result} is indeed valid up to leading order in $q_\text{link}$.

\end{document}